\documentstyle[psfig]{l-aa}

\begin{document}
 
\thesaurus{13
            (11.01.2;  
             11.17.3;  
             13.18.1)} 

\title{A New Sample of Faint Gigahertz Peaked Spectrum Radio Sources}

\author{I.A.G. Snellen \inst{1, 2},
         R.T. Schilizzi \inst{1, 3},
         A.G. de Bruyn \inst{4, 5},
         G.K. Miley \inst{1},
         R.B. Rengelink \inst{1},
         H.J.A. R\"ottgering \inst{1},
         M.N. Bremer \inst{1, 6}} 
 
\institute{ \inst{1} Leiden Observatory, P.O. Box 9513, 2300 RA, Leiden, The
Netherlands \\
\inst{2} Institute of Astronomy, Madingley Road, Cambridge CB3 0HA, United Kingdom\\
\inst{3} Joint Institute for VLBI in Europe, Postbus 2, 7990 AA, Dwingeloo
The Netherlands\\
\inst{4} Netherlands Foundation for Research in Astronomy, Postbus 2, 7990 AA, 
Dwingeloo, The Netherlands\\
\inst{5} Kapteyn Institute, Postbus 800, 9700 AV, Groningen, 
The Netherlands\\
\inst{6} Institut d'Astrophysique de Paris, 98bis Boulevard Arago, 75014 Paris, France} 

\maketitle
 
\markboth{Snellen et al.: A New Sample of Faint GPS Radio Sources }{}

\begin{abstract}
The Westerbork Northern Sky Survey (WENSS) has been used to select a
sample of Gigahertz Peaked Spectrum (GPS) radio sources at flux densities
one to two orders of magnitude lower than bright GPS sources investigated in
earlier studies. Sources with inverted spectra at frequencies above 
$325$ MHz have
been observed with the WSRT\footnote{The Westerbork Synthesis Radio telescope (WSRT) is operated by 
the Netherlands Foundation for Research in Astronomy with financial support
from the Netherlands Organisation for Scientific Research (NWO).}
at 1.4 and 5 GHz and with the VLA\footnote{The Very Large Array (VLA) is operated by the U.S. National Radio Astronomy Observatory which is operated by the Associated Universities, Inc. under cooperative
agreement with the National Science Foundation.} 
at 8.6 and 15
GHz to select genuine GPS sources. This has resulted in a sample of 47 GPS
sources with peak frequencies ranging from $\sim$500 MHz to $>$15 GHz, and
peak flux densities ranging from $\sim$40 to $\sim$900 mJy. 
Counts of GPS sources in our sample as a function of flux density have been 
compared with counts of large scale sources from WENSS scaled to 2 GHz, the 
typical peak frequency of our GPS sources. The counts can be made similar if
the number of large scale sources at 2 GHz is divided by 250, and their flux 
densities increase by a factor of 10. On the scenario that all GPS sources
evolve into large scale radio sources, these results show that the lifetime
of a typical GPS source is $\sim 250$ times shorter than a typical large scale
radio source, and that the source luminosity must decrease by a factor of 
$\sim 10$ in evolving from GPS to large scale radio source.
However, we note that the redshift distributions of GPS and large scale radio 
sources are different and that this hampers a direct and straightforward 
interpretation of the source counts. Further modeling of radio source 
evolution combined with cosmological evolution of the radio luminosity function
for large sources is required.
\end{abstract}

\section{Introduction}

Gigahertz Peaked Spectrum (GPS) radio sources are a class of extragalactic
radio source characterized by a spectral peak near 1 Gigahertz in frequency
(e.g. Spoelstra et al. 1985)
The spectral peak in these compact luminous objects is believed to be due to
synchrotron self absorption caused by the high density of the synchrotron
emitting electrons in the radio source. GPS sources are interesting objects,
both as Active Galactic Nuclei (AGN) and as cosmological probes. It has been
suggested that they are young radio sources ($<10^4$ yr) which evolve into
large radio sources (Fanti et al. 1995, Readhead et al. 1996, 
O'Dea and Baum 1997), and studying them
would then provide us with important information on the early stages of radio
source evolution. 
Alternatively GPS sources may be compact because a particularly dense
environment prevents them from growing larger (e.g. O'Dea et al. 1991).

Important information about the nature of GPS radio sources comes from the 
properties of their optical counterparts. The
galaxies appear to be a homogeneous class of giant ellipticals with old
stellar populations (Snellen et al. 1996a, 1996b, O'Dea et al. 1996)
and are thus useful probes of galaxy
evolution with little or no contamination from the active nucleus 
in the optical. GPS quasars have a different redshift distribution to their 
galaxy counterparts ($2<z<4$, O'Dea et al 1991) and their radio morphologies 
are also quite unlike those of GPS galaxies.
The relationship between GPS quasars and galaxies, if any, 
remains uncertain (Stanghellini et al. 1996).

Previous work on GPS sources has concentrated on the radio-bright members
of the class, with $S_{5GHz}>1$ Jy (Fanti et al. 1990, O'Dea et al. 1991, 
Stanghellini et al. 1996, de Vries et al. 1997). 
We are carrying out investigations of GPS sources at fainter flux density
levels, in order to compare their properties with their 
radio bright counterparts. This enables us to investigate the properties of
GPS sources as a function of radio luminosity, redshift, and rest frame
peak frequency.
The selection of a sample at intermediate flux densities was described in
Snellen et al. (1995a). This paper describes and discusses the selection 
of an even fainter sample from the Westerbork Northern Sky Survey
(WENSS, Rengelink et al. 1997).

\section {Selection of GPS Sources}
\subsection{The Westerbork Northern Sky Survey}

The Westerbork Northern Sky Survey (WENSS) is being carried
out at 325 and 609 MHz (92 and 49 cm) with the Westerbork Synthesis Radio
Telescope (WSRT). At 325 MHz, WENSS covers the complete sky north of $30^\circ$ to a
limiting flux density of approximately 18 mJy ($5 \sigma$). At 609 MHz, about a
quarter of this area, concentrated at high galactic latitudes, has been
surveyed to a limiting flux density of approximately 15 mJy ($5 \sigma$). 
The systematic errors in flux density in WENSS were found to be
$\sim 5\%$ (Rengelink et al. 1997). The
survey was conducted in mosaicing mode which is very efficient in terms of
observing time. In this mode, the telescope cycles through 80 evenly spaced
field centres, during each of a number of $12^h$ syntheses with different
spacings of array elements. The visibilities are sufficiently well sampled for
all 80 fields that it is possible to reconstruct the brightness distribution
in an area of the sky, $\sim$100 square degrees, which is many times larger
than the primary beam of the WSRT. Individual fields are referred to as {\it
mosaics}, and have a resolution (FWHM of the restoring beam) of $54''
\times 54''$ cosec $\delta$ at 325 MHz and $28''\times 28''$ cosec $\delta$ at
609 MHz. From the combined mosaics, maps are made with a uniform sensitivity and
regular size, which are called {\it frames}. The 325 MHz frames are $6^\circ
\times 6^\circ $ in size and positioned on a regular $5^\circ \times 5^\circ $
grid over the sky, which coincides with the position grid of the new Palomar
Observatory Sky Survey (POSS II, Reid et al. 1991) plates.
A detailed description of WENSS is given by Rengelink et al. (1997)

\subsection{Selection of a Sample of Candidate GPS Sources.}

\begin{figure}[!t]
\centerline{
\psfig{figure=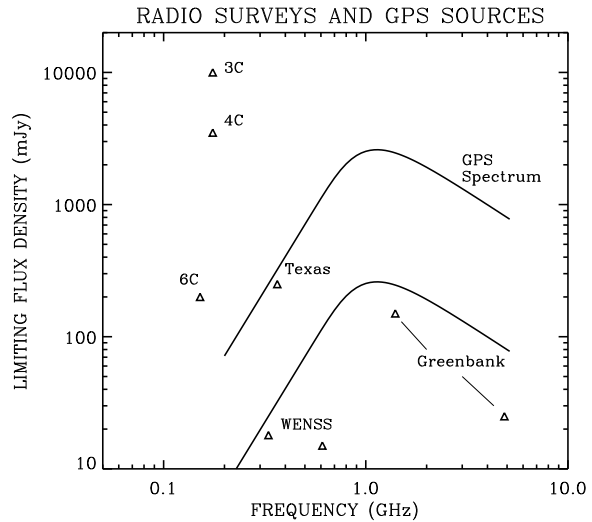,width=8cm}}
\caption{\label{surveys}
Overview of the major radio surveys in the northern sky:
the Greenbank Surveys (Condon and Broderick 1985, Gregory and Condon 1991),
 the Texas Survey (Douglas et al. 1996), and the Cambridge 3C, 4C, and 6C 
surveys. 
The curves represent the spectra of a homogeneous synchrotron self absorbed
radio source, with a peak frequency of 1 GHz and peak flux density of 
300 mJy (lower curve) and 3000 mJy (upper curve). 
Samples of GPS sources can be constructed using WENSS flux density measurements
in the optically thick part of their spectra which are more than an order of magnitude fainter
than samples selected using the Texas Survey.}
\end{figure}

A deep low frequency radio survey such as WENSS is crucial for selecting a
sample of faint GPS sources. It is the inverted spectrum at low frequencies
which distinguishes them from other types of radio sources. Figure
\ref{surveys} shows the major large-sky radio surveys in the northern sky with
theoretical spectra of homogeneous synchrotron self absorbed radio sources
(eg. Moffet 1975) superimposed, which have spectral peak frequencies of 1 GHz.
Samples of GPS 
sources can be
constructed using WENSS flux density measurements in the optically thick part
of their spectra which are ten times fainter than samples selected using the
Texas Survey (Douglas 1996).

When we selected our sample, only a small part of the WENSS
region had been observed and the data reduced to the point of providing 
source lists. The 325 MHz WENSS data used to select GPS sources are from two
regions of the sky; one at $15^{\rm h} < \alpha < 20^{\rm h}$ and $58^\circ< \delta <
75^\circ$, which is called the {\it mini-survey} region (Rengelink et al. 1997),
and the other at $4^{\rm h}00^{\rm m} < \alpha < 8^{\rm h}30^{\rm m}$ and $58^\circ< \delta < 75^\circ$,
where $\alpha$ is right ascension and $\delta$ is declination.
These were the first two regions observed, reduced and analysed for WENSS. 
The mini-survey region,
which is roughly centered on the North Ecliptic Pole, was chosen as the first
area for analysis because it coincides  with the NEP-VLA survey at 1.5 GHz
(Kollgaard et al. 1994), the deep 7C North Ecliptic Cap survey (Lacy et al.
1995, Visser et al. 1995), the deepest part of the ROSAT All Sky survey
(Bower et al., 1996) and the IRAS survey (Hacking and
Houck, 1987). The high declination of the two regions is very convenient for
VLBI experiments, because their locations are circumpolar for almost all the 
major EVN
and VLBA radio telescopes. At the time of selection WENSS 609 MHz data 
was available for only about one third of the mini-survey region. 
The regions for which both 325 and 609 MHz source lists were available cover 119
square degrees of the sky. The regions for which only 325 MHz data were
available cover 216 square degrees in the mini-survey region and 306 square
degrees in the other region.

These source lists were correlated with those from the Greenbank 5 GHz (6 cm)
survey (Gregory and Condon 1991, Gregory et al. 1996), which has a limiting
flux density of 25 mJy ($5 \sigma$). For the faintest sources the new Greenbank source list
(Gregory et al. 1996) was used, which is based on more data. 
Candidate GPS sources were selected on the basis of a
positive spectral index $\alpha$ between 325 MHz and 5 GHz, where the spectral
index is defined by $S \sim \nu^{\alpha}$. If 609 MHz data was also available,
an ``inverted'' spectrum between 325 MHz and 609 MHz was used as the selection
criterion. This in fact increased the sensitivity of the selection process to
GPS sources with low peak frequencies ($< 1 $ GHz). Note that in general for a
GPS source, the 325-609 MHz spectral index will be more positive than the
325-5000 MHz spectral index for a spectral peak in the 1 GHz range. Hence,
using the 325-609 MHz selection criterion will not miss any GPS sources which
would have been found using the 325-5000 MHz selection criterion, it will only
add extra sources with lower peak frequencies.

In total, 117 inverted spectrum sources were selected; 37 using
the 325-609 MHz selection and 82 using the 325-5000 MHz selection. 
They are listed in table \ref{canGPS}. Columns 1, and 2 give the name, 
right ascension and declination (B1950) (obtained from the VLA observations),
 columns 3, 4 and 5 the 325 MHz,
609 MHz and 5 GHz flux densities, and columns 6 and 7 give the 325-609 MHz 
and 325-5000 MHz spectral indices. The uncertainties in the 325-5000 MHz 
spectral indices range from 0.03 to 0.05 (for the faintest objects),
and the uncertainties in the 325-609 MHz spectral index range from 
0.10 to 0.40.

\subsection{Additional Radio Observations.}

An apparently inverted or peaked spectrum could be caused by variability at
any or all of the selection frequencies, due to the fact that the 325, 609 and
5000 MHz surveys were observed at different epochs. To select the genuine GPS
sources, additional quasi-simultaneous observations at other frequencies are
required to eliminate flat spectrum, variable radio sources. The 5 GHz
Greenbank survey was made in 1987, while the 325 MHz and 609 MHz  data were
taken in 1993. Furthermore, high frequency data is needed to confirm their
turnover, and measure the (steep) spectrum in the optically thin part of their
spectra. Therefore VLA observations were taken at 8.4 and 15 GHz, and WSRT
observations at 1.4 and 5 GHz. Later, after the selection process,
data at 1.4 GHz from the NRAO VLA Sky Survey (NVSS, Condon et al. 1996) 
became available and were used to supplement our spectra.

\subsubsection{WSRT Observations at 1.4 and 5 GHz}

The WSRT was used to observe the candidate GPS sources at 1.4 and 5 GHz. The
1.4 GHz observations were performed on 20 February and 10 March 1994 using
8 bands of 5 MHz between 1377.5 and 1423.5 MHz, providing a total bandwidth of
40 MHz. The sources were all observed for about 100 seconds at two to three
different hour angles. This resulted in a noise level of typically 1 mJy/beam
and a resolution of $15''\times 15''cosec \ \delta$. The results are shown in
column 8 of table \ref{canGPS}. 

In order to improve the 5 GHz Greenbank flux density measurements, observations
were carried out with the WSRT  at 4.87 GHz on May 15 1994 using a bandwidth
of 80 MHz, at a time when the WSRT was participating a VLBI session. Unfortunately only
three telescopes were equipped with 5 GHz receivers.  Only sources between 4
and 8 hours right ascension were observed, and the uncertainty in the
measured flux densities is large ($\sim 15$\%). The resulting flux densities
are listed in column 13 in table \ref{canGPS}.

\begin{table*}
\renewcommand{\arraystretch}{0.90}
\setlength{\tabcolsep}{1mm}
\begin{tabular}{|c|rrrrrr|rrr|rr|rrr|c|rrr|} \hline
Source&\multicolumn{3}{c}{R.A.(1950)}&
\multicolumn{3}{c|}{Decl.(1950)}&$S_{325}$&$S_{609}$&$S^{gb}_{5.0}$&$\alpha 
^{325}_{609}$&$\alpha ^{325}_{5000}$&
$S^{wsrt}_{1.4}$&$S^{VLA}_{8.6}$&$S^{VLA}_{14.9}$&GPS&$S^{nvss}_{1.4}$&
$S^{wsrt}_{5.0}$&$S^{merlin}_{5.0}$\\
          & h &  m &   s  &$^{\circ}$&$'$&$''$&{\tiny (mJy)}&{\tiny(mJy)}&
{\tiny(mJy)}&&    &{\tiny(mJy)}&{\tiny(mJy)}&{\tiny(mJy)}& 
&{\tiny(mJy)}&{\tiny(mJy)}&{\tiny(mJy)}\\ \hline
B0400+6042 & 4 &  0 &  7.22 & 60 & 42 & 29.0   & 81 &  & 100&    &+0.08& 180  & 85& 36 &+& 166& 73& 85           \\
B0402+6442 & 4 &  2 & 56.73 & 64 & 42 & 52.4   & 48 &  &  69&    &+0.13&  63  & 56& 45 & &  70& 32&              \\
B0406+7413 & 4 &  6 & 37.39 & 74 & 13 & 22.5   & 63 &  &  66&    &+0.02&  56  & 54& 36 & &  60& 49&              \\
B0418+6724 & 4 & 18 & 50.86 & 67 & 24 &  7.2   & 47 &  &  95&    &+0.26&  54  & 96& 79 & &  50& 46&              \\
B0436+6152 & 4 & 36 & 15.80 & 61 & 52 & 10.0   & 70 &  & 127&    &+0.22& 208  &108& 60 &+& 238&101&122           \\
B0441+5757 & 4 & 41 & 53.90 & 57 & 57 & 21.7   & 60 &  & 115&    &+0.24&  95  &128& 96 &+&  91& 91&101           \\
B0456+7124 & 4 & 56 &  0.15 & 71 & 24 & 10.1   & 88 &  & 148&    &+0.19& 121  &213&216 & & 116&181&              \\
B0507+6840 & 5 & 7 &   4.48 & 68 & 40 & 44.7   & 27 &  &  33&    &+0.07&  30  & 41& 29 & &  31& 28&              \\
B0513+7129 & 5 & 13 & 38.82 & 71 & 29 & 55.1   &121 &  & 181&    &+0.15& 236  & 96& 60 &+& 244&123&131           \\
B0515+6129 & 5 & 15 & 19.62 & 61 & 28 & 59.3   & 40 &  &  45&    &+0.04&  26  & 62& 41 & &  28& 43&              \\ 
B0518+6004 & 5 & 18 & 42.78 & 60 &  4 & 55.9   & 88 &  & 113&    &+0.09&  67  & 92&102 & & 101& 56&              \\
B0531+6121 & 5 & 31 & 55.43 & 61 & 21 & 31.0   & 18 &  &  38&    &+0.27&  19  & 44& 23 &+&  22& 33& 23           \\
B0535+6743 & 5 & 35 & 57.15 & 67 & 43 & 49.8   & 83 &  & 182&    &+0.29&  97  &235&136 &+&148&177 &168          \\
B0536+5822 & 5 & 36 &  8.26 & 58 & 22 &  3.8   & 26 &  &  27&    &+0.01&  36  & 54& 39 & & 31& 37 &             \\
B0537+6444 & 5 & 37 & 15.12 & 64 & 45 &  3.7   & 18 &  &  32&    &+0.21&  29  & 17&  9 &+& 28& 17 & 17          \\
B0538+7131 & 5 & 38 & 38.30 & 71 & 31 & 20.9   & 19 &  &  77&    &+0.51&  45  & 68& 29 &+& 48& 90 & 73          \\
B0539+6200 & 5 & 39 & 54.51 & 62 &  0 &  2.2   & 47 &  & 104&    &+0.29& 123  & 80& 66 &+&126&112 & 99          \\
B0542+7358 & 5 & 42 & 50.98?& 73 & 58 & 32.5   & 29 &  &  29&    &+0.00&  46  & 33& 24 & & 44& 34 &             \\
B0543+6523 & 5 & 43 & 40.36 & 65 & 23 & 24.6   & 26 &  &  43&    &+0.18&  65  & 47& 27 &+& 72& 49 & 43          \\
B0544+5847 & 5 & 44 &  3.18 & 58 & 46 & 55.8   & 33 &  &  42&    &+0.09&  67  & 43& 22 &+& 60& 48 & 34          \\
B0552+6017 & 5 & 52 & 35.07 & 60 & 17 & 30.1   & 15 &  &  26&    &+0.28&  44  & 11&$<3$&+& 47& 11 & 13          \\
B0556+6622 & 5 & 56 & 13.00 & 66 & 22 & 57.7   & 22 &  &  25&    &+0.05&  12  & 30& 20 & & 24& 27 &             \\
B0557+5717 & 5 & 57 & 31.76 & 57 & 17 & 19.7   & 19 &  &  29&    &+0.16&  63  & 28& 14 &+& 69& 36 & 30          \\
B0601+7242 & 6 &  1 & 57.83 & 72 & 42 & 54.6   & 16 &  &  26&    &+0.18&  14  & 18&  8 & & 14& 37 &             \\
B0601+5753 & 6 &  1 & 22.05 & 57 & 53 & 31.8   & 19 &  & 162&    &+0.83& 141  &149&138 &+&   &207 &192          \\
B0605+7218 & 6 &  5 &  6.15 & 72 & 18 & 51.1   & 24 &  & 116&    &+0.51&  80  & 39& 60 & & 60& 65 &             \\
B0607+7335 & 6 &  7 & 33.86 & 73 & 35 & 53.0   & 20 &  &  54&    &+0.36&  73  & 86& 54 & & 53& 87 &             \\
B0607+7107 & 6 &  7 & 54.42 & 71 &  8 & 14.1   & 21 &  &  25&    &+0.10&  24  & 21& 18 & & 27& 42 &             \\
B0609+7259 & 6 &  9 & 14.93 & 72 & 59 & 50.7   & 20 &  &  25&    &+0.08&  16  & 16& 11 & & 22& 23 &             \\
B0738+7043 & 7 & 38 & 37.86 & 70 & 43 &  9.2   & 47 &  &  84&    &+0.19&  34  & 18&106 & & 73&100 &             \\
B0741+7213 & 7 & 41 & 30.37 & 72 & 12 & 58.5   & 65 &  & 106&    &+0.18&  96  & 54& 55 & & 99& 75 &             \\
B0748+6343 & 7 & 48 & 27.42 & 63 & 43 & 31.7   & 24 &  &  54&    &+0.42&  39  & 69& 85 &+& 44& 87 &130          \\
B0752+6355 & 7 & 52 & 21.41 & 63 & 55 & 59.5   & 15 &  & 254&    &+1.04& 133  &298&282 &+&196&376 &303          \\
B0755+6354 & 7 & 55 & 20.66 & 63 & 54 & 25.4   & 25 &  &  38&    &+0.15&  26  & 17& 18 &+& 26& 21 & 19         \\
B0756+6647 & 7 & 56 & 47.89 & 66 & 47 & 27.8   & 44 &  & 164&    &+0.48&  90  & 80& 77 &+&104&109 & 97          \\
B0758+5929 & 7 & 58 & 13.00 & 59 & 29 & 56.9   & 97 &  & 178&    &+0.22& 203  &127&101 &+&207&185 &163          \\ 
B0759+6557 & 7 & 59 & 12.95 & 65 & 57 & 45.9   & 15 &  &  27&    &+0.22&  40  & 14&  8 &+& 48& 26 & 21          \\
B0800+6754 & 8 &  0 &  6.92 & 67 & 54 & 31.9   & 78 &  &  78&    &+0.00&  30  & 33& 32 & & 39& 38 &             \\
B0802+7323 & 8 &  2 & 32.27 & 73 & 23 & 53.3   &318 &  & 321&    &+0.00& 277  &387&445 & &307&437 &             \\
B0808+6518 & 8 &  8 &  5.14 & 65 & 18 & 10.8   & 35 &  &  42&    &+0.07&  52  & 17& 21 & & 31& 32 &             \\
B0810+6440 & 8 & 10 &  7.59 & 64 & 40 & 29.1   & 96 &  & 193&    &+0.26&  92  &132&179 & & 90&133 &             \\
B0820+7403 & 8 & 20 & 43.56 & 74 &  2 & 53.8   &104 &  & 108&    &+0.01&  81  & 65& 63 & &102& 97 &             \\
B0824+6446 & 8 & 24 & 31.62 & 64 & 46 & 27.4   & 24 &  &  35&    &+0.14&  30  & 11& 13 & & 44& 29 &             \\
B0826+7045 & 8 & 26 & 52.55 & 70 & 45 & 44.1   & 34 &  & 109&    &+0.43&  73  & 63& 56 &+& 79& 99 & 92            \\
B0827+6231 & 8 & 27 &  3.37 & 62 & 31 & 45.9   & 28 &  &  28&    &+0.00&  32  & 25& 33 & & 34& 32 &             \\
B0828+5756 & 8 & 28 & 33.49 & 57 & 56 &  5.6   & 29 &  &  32&    &+0.04&  37  & 27& 26 & & 53& 40 &             \\
B0828+7307 & 8 & 28 & 49.08 & 73 &  6 & 58.7   & 81 &  & 104&    &+0.09&  68  & 58& 68 & &102& 86 &             \\
B0830+5813 & 8 & 30 & 12.71 & 58 & 13 & 38.8   & 39 &  &  65&    &+0.29&  65  & 31& 23 &+& 59& 43 & 38             \\
B0830+6300 & 8 & 30 & 37.77 & 63 &  0 &  8.1   & 26 &  &  36&    &+0.12&  54  & 50& 54 & & 62& 65 &             \\
B0830+6845 & 8 & 30 & 59.69 & 68 & 45 & 33.1   & 53 &  &  74&    &+0.12&  83  &136&124 & & 86&159 &             \\ 
B1525+6801 &15 & 25 & 21.12 & 68 &  1 & 48.9   & 90 &  & 103&    &+0.05& 153  & 54& 29 &+&161&    & 91          \\
B1529+6741 &15 & 29 & 17.85 & 67 & 41 & 58.6   & 47 &  &  55&    &+0.06&  19  & 32& 26 & & 35&    &             \\
B1529+6829 &15 & 29 & 45.15 & 68 & 29 &  9.3   & 51 &  &  51&    &+0.00&  29  & 21& 23 & & 27&    &             \\
B1536+6202 &15 & 36 & 54.56 & 62 &  2 & 56.4   & 16 &  &  30&    &+0.23&  15  & 34& 24 & & 22&    &             \\
B1538+5920 &15 & 38 & 27.46 & 59 & 20 & 39.0   & 29 &  &  73&    &+0.34&  47  & 36& 23 &+& 45&    & 45          \\
B1539+6156 &15 & 39 & 32.28 & 61 & 56 &  1.9   & 34 &  &  40&    &+0.06&  15  & 54& 36 & & 33&    &             \\
B1542+6139 &15 & 42 &  5.03 & 61 & 39 & 20.9   & 60 &  & 129&    &+0.28&  86  &114&119 & & 90&    &             \\
B1542+6631 &15 & 42 & 54.19 & 66 & 31 & 18.2   & 54 &  &  86&    &+0.17&  44  & 81& 84 & & 51&    &             \\
B1550+5815 &15 & 50 & 55.59 & 58 & 15 & 37.5   & 86 &  & 362&    &+0.53& 157  &214&212 &+&   &    &237          \\
B1551+6822 &15 & 51 & 53.07 & 68 & 22 & 38.7   & 23 &  &  34&    &+0.14&  49  & 26& 10 &+& 55&    & 27          \\\hline
\end{tabular}
\caption{\label{canGPS} The sample of candidate GPS sources. Column 1 gives
the B1950 source name, column 2 the VLA position, column 3, 4 and 5
the flux densities from WENSS at 325 and 609 MHz and of the Greenbank Survey
at 5 GHz. Column 6 and 7 give the 325-609 MHz and the 325-5000 MHz spectral
indices, column 8, 9 and 10 the flux densities from the WSRT at 1.4 GHz, and from the VLA at 8.4 and 15 GHz.
A cross in column 11 indicates whether the source was selected in the final sample. Column 12, 13 and 14 give the NVSS 1.4 GHz, the WSRT 5 GHz and the MERLIN 
5 GHz flux densities.}
\end{table*}

\addtocounter{table}{-1}

\begin{table*}
\renewcommand{\arraystretch}{0.90}
\setlength{\tabcolsep}{1mm}
\begin{tabular}{|c|rrrrrr|rrr|rr|rrr|c|rrr|} \hline
Source&\multicolumn{3}{c}{R.A.(1950)}&
\multicolumn{3}{c|}{Decl.(1950)}&$S_{325}$&$S_{609}$&$S^{gb}_{5.0}$&$\alpha 
^{325}_{609}$&$\alpha ^{325}_{5000}$&
$S^{wsrt}_{1.4}$&$S^{VLA}_{8.6}$&$S^{VLA}_{14.9}$&GPS&$S^{nvss}_{1.4}$&
$S^{wsrt}_{5.0}$&$S^{merlin}_{5.0}$\\
          & h &  m &   s  &$^{\circ}$&$'$&$''$&{\tiny (mJy)}&{\tiny(mJy)}&
{\tiny(mJy)}&&    &{\tiny(mJy)}&{\tiny(mJy)}&{\tiny(mJy)}& 
&{\tiny(mJy)}&{\tiny(mJy)}&{\tiny(mJy)}\\ \hline
B1557+6220 &15 & 57 &  8.43 & 62 & 20 &  7.4   & 23 &   &   37 &     &+0.17& 40   & 12&  4 &+& 42& & 18           \\
B1559+5715 &15 & 59 &  5.07 & 57 & 15 & 19.2   & 27 &   &   47 &     &+0.20& 55   & 36& 39 & & 68& &               \\
B1600+5714 &16 &  0 &  8.62 & 57 & 14 & 18.8   & 29 &   &   45 &     &+0.16& 24   & 12& 76 & & 41& &               \\
B1600+7131 &16 &  0 & 57.00 & 71 & 31 & 40.2   & 26 &   &  103 &     &+0.50&311   & 37& 15 &+&308& & 85            \\
B1604+5939 &16 &  4 & 56.32 & 59 & 39 & 43.7   & 71 &   &  110 &     &+0.16&111   &130&112 & &   & &               \\
B1607+6026 &16 &  7 & 30.32 & 60 & 26 & 34.6   & 79 &   &   79 &     &+0.00& 16   & 31& 22 & & 34& &               \\
B1608+6540 &16 &  8 & 50.55 & 65 & 40 & 15.1   & 66 &   &   90 &     &+0.11& 63   & 40& 78 & & 68& &               \\
B1616+6428 &16 & 16 & 26.42 & 64 & 28 &  7.7   & 19 & 37&   62 &+1.06&+0.43& 64   & 58& 58 & & 61& &               \\
B1620+6406 &16 & 20 & 46.19 & 64 &  6 & 12.6   & 27 & 30&   25 &+0.16&$-$0.03& 41   &  6&$<3$&+& 43& & 11            \\
B1622+6630 &16 & 22 & 50.52 & 66 & 30 & 52.6   & 23 & 61&  517 &+1.55&+1.14&178   &230&176 &+&159& &230            \\
B1623+6859 &16 & 23 & 36.01 & 68 & 59 & 46.0   & 32 & 32& $<25$&+0.00&     & 11   &  3&$<3$& & 18& &               \\
B1624+6622 &16 & 24 &  7.26 & 66 & 22 &  6.7   & 38 & 42& $<25$&+0.16&     & 33   &  4&$<3$& & 26& &               \\
B1633+6506 &16 & 33 &  7.49 & 65 &  6 & 52.4   & 48 & 66&  114 &+0.51&+0.30&111   & 97& 90 & & 95& &               \\
B1639+6711 &16 & 39 & 10.76 & 67 & 11 & 47.2   & 34 & 61&   30 &+0.93&$-$0.05& 54   & 27& 19 &+& 76& & 40            \\
B1642+6701 &16 & 42 & 16.42 & 67 &  1 & 22.6   &124 & 126&  65 &+0.02&$-$0.24&121   & 43& 24 &+&126& & 56            \\
B1645+6738 &16 & 45 & 38.00 & 67 & 38 &  0.8   & 22 & 22& $<25$&+0.00&     & 28   &  8&  5 & & 28& &               \\
B1647+6225 &16 & 47 & 31.26 & 62 & 25 & 49.7   & 31 & 59&   33 &+1.02&+0.02& 69   & 13&  3 &+& 56& & 17            \\
B1655+6446 &16 & 55 & 21.09 & 64 & 46 & 21.2   & 23 & 52&   34 &+1.30&+0.14& 68   & 16&  9 &+& 61& & 23            \\
B1657+5826 &16 & 57 & 15.96 & 58 & 26 & 31.5   & 56 & 63&   17 &+0.19&$-$0.44& 44   & 18& 13 &+& 48& & 23            \\
B1711+6031 &17 & 11 & 39.99 & 60 & 31 & 45.3   & 15 & 29& $<25$&+1.05&     & 17   &  3&$<3$& & 19& &               \\
B1712+6727 &17 & 12 & 50.73 & 67 & 27 & 10.2   & 28 & 30& $<25$&+0.11&     & 28   & 23& 13 & & 31& &               \\
B1714+5819 &17 & 14 & 56.27 & 58 & 19 & 16.2   & 21 & 35& $<25$&+0.81&     & 25   &  6&  3 & & 28& &               \\
B1718+6024 &17 & 18 & 18.75 & 60 & 24 & 11.7   & 19 & 29& $<25$&+0.67&     & 20   &  3&$<3$& & 18& &               \\
B1730+6027 &17 & 30 & 15.71 & 60 & 27 & 24.9   & 41 & 83&   34 &+1.12&$-$0.13&178   &103& 89 & &161& &               \\
B1746+6921 &17 & 46 & 53.21 & 69 & 21 & 33.5   & 65 & 96&  144 &+0.62&+0.31&161   &127&100 &+&154& &139            \\
B1749+6919 &17 & 49 & 31.62 & 69 & 19 & 41.3   & 21 & 31&   18 &+0.62&$-$0.06& 27   & 10&  7 & & 32& &               \\
B1755+6905 &17 & 55 & 42.48 & 69 &  5 & 48.4   & 16 & 72&   77 &+2.40&+0.28& 84   & 51& 48 & & 78& &               \\
B1807+5959 &18 &  7 & 17.36 & 59 & 59 & 26.5   & 16 & 43&   30 &+1.52&+0.28& 47   & 37& 22 &+& 42& & 38            \\
B1807+6742 &18 &  7 & 23.43 & 67 & 42 & 22.3   & 29 & 52&$<25$ &+0.93&     & 47   & 12&  8 &+& 43& & 20            \\
B1808+6813 &18 &  8 & 25.41 & 68 & 13 & 36.1   & 33 & 37&   24 &+0.18&+0.04& 42   & 11&  8 &+& 42& & 19              \\
B1818+6445 &18 & 18 & 24.79 & 64 & 45 & 17.1   &  35 & 38 &$<25$&+0.13     &  &  24 & 27 &$<3$& & 27& &               \\
B1818+6249 &18 & 18 & 50.02 & 62 & 49 & 56.2 &  41 & 50 &$<25$&+0.32     &  &  34 &  8 &  6&  & 33& &              \\
B1819+6707 &18 & 19 & 48.42 & 67 &  7 & 20.8 & 265 &330 & 154 &+0.35   &$-0.20$  & 297 & 93 & 68&+ &311& &142           \\
B1821+6251 &18 & 21 & 20.03 & 62 & 51 & 52.6 &  30 & 38 &  31 &+0.38   &+0.01 &  32 & 28 & 19&  & 29& &              \\
B1827+6432 &18 & 27 & 55.40 & 64 & 32 & 13.7 & 152 &228 & 262 &+0.65   &+0.20 & 204 &135 &120&  &216& &              \\
B1829+6419 &18 & 29 & 16.62 & 64 & 19 & 23.1 &  62 & 95 &$<25$&+0.68     &  &  80 &  6 &  4&  & 75& &              \\
B1834+6319 &18 & 34 & 48.26 & 63 & 19 & 49.6 &  37 & 46 &$<25$&+0.35     &  &  36 &  5 &$<3$& & 36& &               \\
B1838+6239 &18 & 38 & 12.00 & 62 & 39 & 56.2 &  15 & 33 &$<25$&+1.25     &  &  54 &  7 & 6&   & 36& &             \\
B1841+6715 &18 & 41 &  7.21 & 67 & 15 & 51.2 &  36 & 94 & 163 &+1.53   &+0.55  & 142 & 98 &68& + &178& &125          \\
B1841+6343 &18 & 41 & 18.25 & 63 & 43 & 56.3 &  15 & 29 &$<25$&+1.05  &    &  41 & 10 &  6  &   & 36& &           \\
B1843+6305 &18 & 43 &  6.16 & 63 &  5 & 42.8 &  15 & 41 &  52 &+1.67   &+0.45  &  59 & 27 & 16  & + & 81& & 40        \\
B1850+6447 &18 & 59 & 27,78 & 64 & 47 & 31.6 &  49 & 70 &$<25$&+0.57  &  &  52 &  6 & $<3$&   & 55& &             \\
B1916+6817 &19 & 16 & 37.66 & 68 & 17 & 51.6 &  23 & 39 &$<25$&+0.84  &  &  20 &  6 &  4  &   & 26& &           \\
B1919+6912 &19 & 19 & 57.99 & 69 & 12 & 26.2 &  21 & 28 &  18 &+0.46  &$-0.06$  &  16 & 16 & 14 15&   & & &           \\
B1926+6111 &19 & 26 & 49.66 & 61 & 11 & 20.9 & 404 &    & 613 &  &+0.15 & 718 & 85 &678  &   & &535&           \\
B1934+7111 &19 & 34 & 41.81 & 71 & 11 & 10.4 &  92 &    & 108 &  &+0.06 & 142 &119 &102  &   &179 & &           \\
B1938+5824 &19 & 38 & 50.57 & 58 & 24 & 49.9 &  24 &    &  29 &  &+0.07 &  34 & 30 & 25  &   &21 & &           \\
B1942+7214 &19 & 42 &  2.24 & 72 & 14 & 31.9 &  81 &    & 158 &  &+0.24 & 233 &147 &110  & + &233 & &183        \\
B1944+6007 &19 & 44 & 21.42 & 60 &  7 & 40.5 &  20 &    &  79 &  &+0.50 &  12 & 62 & 32  &   &17 & &           \\
B1945+6024 &19 & 45 & 24.83 & 60 & 24 & 12.6 &  25 &    &  80 &  &+0.43 &  55 &125 &188  & + & 55& & 84        \\
B1946+7048 &19 & 46 & 12.02 & 70 & 48 & 21.6 & 234 &    & 643 &  &+0.37 & 887 &389 &268  & + & 953& &574        \\
B1951+6915 &19 & 51 & 34.02 & 69 & 15 &  6.3 &  23 &    &  32 &  &+0.12 &  24 & 40 & 35  &   & 33& &           \\
B1951+6453 &19 & 51 & 42.52 & 64 & 53 & 56.1 & 111 &    & 103 &  &+0.00 &  89 & 83 & 59  &   & 88& &           \\
B1954+6146 &19 & 54 & 11.69 & 61 & 45 & 58.1 &  66 &    & 132 &  &+0.25 &  66 &182 &152  & + & 61& &153        \\
B1958+6158 &19 & 58 & 45.58 & 61 & 58 & 27.1 &  52 &    & 140 &  &+0.36 & 111 & 96 & 84  & + & 129& &136        \\
B2006+5916 &20 & 06 & 52.14 & 59 & 16 & 43.5 &  30 &    &  37 &  &+0.08 &  36 & 21 & 18  &   & 30& &           \\
B2011+7156 &20 & 11 & 22.95 & 71 & 56 &  9.4 &  48 &    & 134 &  &+0.38 & 118 &103 &102  &   & & &           \\\hline
\end{tabular}
\caption{{\it Continued...}}
\end{table*}

\subsubsection{VLA Observations at 8.4 and 15 GHz}

The candidate GPS sources were observed with the VLA in B-configuration at 8.4
and 15 GHz on 23 July 1994. At both frequencies, the objects were observed in
a standard way using a bandwidth of $2\times 25$ MHz. 
The phases
were calibrated using standard nearby VLA phase calibrators. Total integration
times were typically 100 seconds at both frequencies, resulting in noise levels
of 0.2 and 1.0 mJy/beam respectively. 
Systematic errors in flux density of VLA observations
at these frequencies are typically about $3\%$ (eg. Carilli et al. 1991).
The data were reduced using AIPS in a standard manner,
including
several iterations of phase self-calibration. The synthesized beams
have half widths of $1.5''\times 0.8''$ and $0.8''\times 0.5 ''$ at 8.4 and 15
GHz respectively. Several candidate GPS sources had already been observed at
8.4 GHz on February 26 1994 and April 3 1994 
during the Cosmic Lens All Sky Survey (CLASS) program (eg. Myers et al. 1995); 
these were not re-observed by us at 8.4 GHz. The CLASS 8.4 GHz
observations were made using the VLA in A configuration in a standard way, also
with a bandwidth  of $2\times 25$ MHz and an average integration time of 30
seconds. The resolution of the CLASS observations was $\sim 0.2''$, and the
noise level $\sim 0.4$ mJy/beam.

The results of the VLA observations are listed in columns 9 and 10 in table
\ref{canGPS}. All of the sources were unresolved, 
except for B1608+6540, which was found
to be a quadruple gravitational lens (Snellen et al. 1995b, Myers et al. 1995,
Fassnacht et al. 1996)

\subsubsection{The NRAO VLA Sky Survey at 1.4 GHz}

Observations for the 1.4 GHz NRAO VLA Sky Survey (NVSS, Condon et al. 1996)
began in September 1993 and are planned to cover the sky north of 
declination $-40^{\circ}$ (82\% of the celestial sphere). 
Data in our regions of interest
were  taken on 1 November 1993 for the region $4^h00^m<R.A.<8^h00^m$, and on 2
April 1995 for the region between $15^h00^m<R.A.<20^h00^m$. The noise level in
an image is typically 0.5 mJy/beam and the resolution is $45''$.

\subsubsection{MERLIN Observations at 5 GHz}

The final sample of genuine GPS sources, as selected in section 2.4,
 was observed with MERLIN at 5 GHz
on 15 and 16 May 1995 during our global VLBI measurements. The sources were 
observed in three to four ``snapshots'' of 13 minutes each, resulting in 
a noise level of typically 0.3 mJy/beam, and a resolution of 0.04$''$.
All the sources were unresolved.
The results are listed in
column 14 of table \ref{canGPS}. Note that these observations were obtained
after, and therefore not used for, the final selection.

\subsection{Selection of the Genuine GPS Sources}

The genuine GPS sources were selected using the 325 MHz and, if available, 
the 609 MHz WENSS data, 1.4 GHz WSRT data, 5 GHz Greenbank data
and 8.4 and 15 GHz VLA data. The WSRT 5 GHz data were not used for selection because they were only available for a part of the sample. The NVSS data was not
used for selection, being not available at the time of source selection.   
However, both the 5 GHz WSRT and 1.4 GHz 
NVSS data were used for variability studies (see section \ref{varsec}).

The selection criteria were as follows:
\begin{itemize}
\item[1.] The spectrum must decrease monotonically 
below the frequency with the highest flux density, taking into account an 
assumed uncertainty of 10\% in flux density. 
\item[2.] The spectrum must decrease monotonically 
above the frequency with the highest flux density,taking into account an 
assumed uncertainty of 10\% in flux density. 
\item[3.] The Full Width Half Maximum (FWHM) defined by the logarithm 
of the spectrum must be less than 2 decades in frequency. 
A spectral index
of 0.5 is assumed below 325 MHz, and a spectral index of -0.5 above
15 GHz. 
\item[4.] The Greenbank 5 GHz flux density must be greater than
20 mJy. This allowed imaging of the source with global VLBI at 5 GHz 
without recourse to phase referencing. Note that if no Greenbank flux density was 
available (noise level is about 5 mJy), the flux density was
estimated by interpolating the 1.4 and 8.4 GHz flux density points.
\end{itemize}

The resulting sample of 47 sources is listed in table \ref{GPS}.
One of the sources, B1807+5959, did not obey the criterion of decreasing 
flux density above the peak frequency, because the 5 GHz Greenbank flux density
flux point is too low. However it was kept in the sample because 
the fall off in flux density at both low and high frequencies suggests
that the low flux density point at 5 GHz is due to the different epoch of the
Greenbank observations. Additional observations showed this to be true.

The spectra of the selected GPS sources were fitted with the following 
function 
\begin{equation} \label{eq1}
S(\nu ) = S_{max} / (1-e^{-1}) \times \left( \frac{\nu }{\nu _{max}} \right)^k \times
(1  - e^{- \left( \frac{\nu }{\nu _{max}}\right) ^{l - k}})
\end{equation}
where $k$ is the optically thick spectral index, $l$ the optically
thin spectral index, and $S_{max}$ and $\nu _{max}$ respectively the 
peak flux density and peak frequency. This equation, which represents 
a homogeneous synchrotron self absorbed radio source for $k = 2.5$ 
(eg. Moffet 1975), fits 
the spectral peak well in most cases, 
however it did not always fit the flux density points at the lowest and
highest frequency frequency adequately, and therefore 
was only used to determine the 
peak flux density, peak frequency and Full Width Half Maximum of the 
spectra. The optically thick and thin spectral indices have been
 determined from
the two lowest and the two highest frequency data points respectively.
The fitted spectra are shown in figure \ref{spectra}.
Table \ref{GPS} gives the characteristics of the GPS sources:
column 1 gives the source name, columns 2 and 3 the peak frequency and 
peak flux density, columns 4 and 5 the optically thick and optically thin 
spectral indices, and column 6 the FWHM of the spectrum in logarithmic 
units.

Figure \ref{spectra} shows that three of the 47 sources initially 
selected probably do not have genuine GPS spectra, 
namely B0531+6121, B0748+6343 and B0755+6354. In these cases 
differences between MERLIN, Greenbank and WSRT  5 GHz flux density measurements
suggest that the measured spectra are contaminated by flux density
variability and it is not clear whether the spectra do indeed exhibit a peak.
Although we have included them in the sample, we omit them from the 
analysis below. Note that no sign of a turnover is seen in B1945+6024,
and that there are some sources in the sample which do not have a ``clean'' 
peaked spectrum, like B1954+6146 and B0535+6743.
To obtain a better determination of the spectral peak of B1954+6146, 
the 325 MHz flux density data point is not used to fit the spectrum.

\begin{table}
\begin{center}
\setlength{\tabcolsep}{1.05mm}
\renewcommand{\arraystretch}{0.9}
\begin{tabular}{|ccrrrc|}\hline
GPS& $\nu _{max}$ &  $S_{max}$ & $\ \ \ \alpha _{thick} $ &$\ \ \ \alpha _{thin}$ & FWHM\\
Source& (GHz) & (mJy) & & & log(freq)\\ \hline
B0400+6042&      1.0&         184&     0.52&    -1.48&      0.7\\
B0436+6152&      1.0&         237&     0.79&    -1.01&      0.7\\
B0441+5757&      6.4&         109&     0.30&    -0.50&      1.9\\
B0513+7129&      1.5&         242&     0.47&    -0.81&      0.9\\
B0531+6121&      5.9&          36&     0.09&    -1.12&      1.0\\
B0535+6743&      5.7&         192&     0.27&    -0.94&      1.1\\
B0537+6444&      2.3&          29&     0.31&    -1.10&      1.7\\
B0538+7131&      4.2&          85&     0.61&    -1.47&      0.7\\
B0539+6200&      1.9&         129&     0.67&    -0.33&      0.9\\
B0543+6523&      1.2&          69&     0.66&    -0.96&      1.0\\
B0544+5847&      1.4&          63&     0.45&    -1.16&      0.9\\
B0552+6017&      1.0&          50&     0.91&    -1.16&      0.5\\
B0557+5717&      1.1&          69&     0.85&    -1.20&      0.6\\
B0601+5753&      4.4&         187&     1.37&    -0.13&      0.9\\
B0748+6343&      6.6&          92&     0.37&     0.36&      1.1\\
B0752+6355&      6.4&         314&     1.79&    -0.10&      1.4\\
B0755+6354&      4.2&          28&     0.03&     0.10&      2.0\\
B0756+6647&      3.4&         127&     0.54&    -0.07&      0.8\\
B0758+5929&      2.0&         215&     0.51&    -0.40&      1.0\\
B0759+6557&      1.7&          46&     1.01&    -0.97&      0.6\\
B0826+7045&      3.5&         105&     0.79&    -0.20&      0.8\\
B0830+5813&      1.6&          65&     0.32&    -0.51&      1.1\\ 
B1525+6801&      1.8&         163&     0.38&    -1.07&      1.0\\
B1538+5920&      3.5&          64&     0.32&    -0.77&      1.0\\
B1550+5815&      4.6&         293&     0.41&    -0.02&      0.9\\
B1551+6822&      1.5&          52&     0.56&    -1.65&      0.8\\
B1557+6220&      2.3&          49&     0.40&    -1.89&      0.9\\
B1600+7131&      1.7&         346&     1.70&    -1.56&      0.3\\
B1620+6406&      2.2&          47&     0.17&    -1.56&      1.0\\
B1622+6630&      4.0&         363&     1.55&    -0.46&      0.5\\
B1639+6711&      1.0&          68&     0.92&    -0.61&      0.8\\
B1642+6701&      1.3&         124&     0.02&    -1.01&      2.0\\
B1647+6225&      0.9&          71&     1.03&    -2.53&      0.5\\
B1655+6446&      1.0&          69&     1.29&    -0.99&      0.5\\
B1657+5826&      0.5&          64&     0.19&    -0.56&      0.7\\
B1746+6921&      2.2&         164&     0.63&    -0.41&      1.1\\
B1807+5959&      1.0&          47&     1.56&    -0.90&      1.4\\
B1807+6742&      0.8&          54&     0.94&    -0.70&      0.6\\
B1808+6813&      1.3&          42&     0.20&    -0.55&      1.7\\
B1819+6707&      0.8&         338&     0.35&    -0.54&      1.0\\
B1841+6715&      2.1&         178&     1.53&    -0.63&      0.8\\
B1843+6305&      1.9&          75&     1.53&    -0.90&      0.7\\
B1942+7214&      1.4&         233&     0.72&    -0.50&      1.0\\
B1945+6024&    $>15$&      $>188$&     0.54&     0.70&        -\\
B1946+7048&      1.8&         929&     0.91&    -0.64&      0.6\\
B1954+6146&      8.4&         169&     0.00&    -0.31&      1.4\\
B1958+6158&      3.3&         142&     0.52&    -0.23&      0.9\\ \hline
\end{tabular}
\end{center}
\caption{\label{GPS}The resulting sample of GPS sources.}
\end{table}

\begin{figure*}
\vspace{-0.5cm}
\hbox{\hspace{-0.8cm}
\psfig{figure=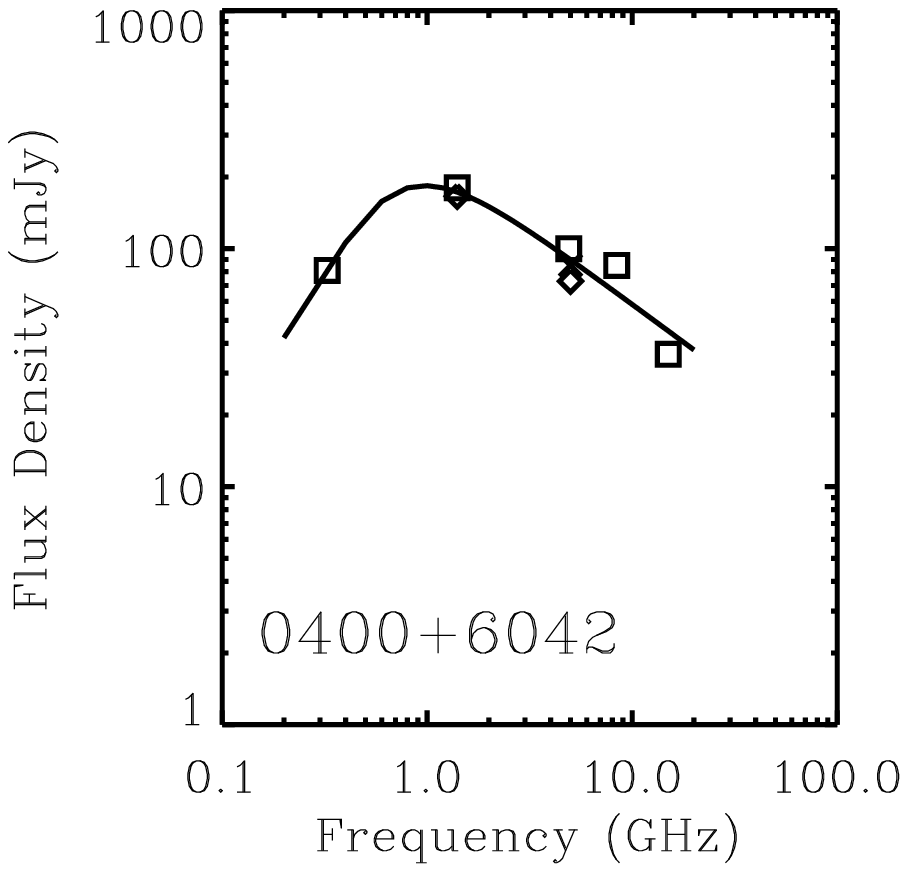,width=5.1cm}\hspace{-0.8cm} 
\psfig{figure=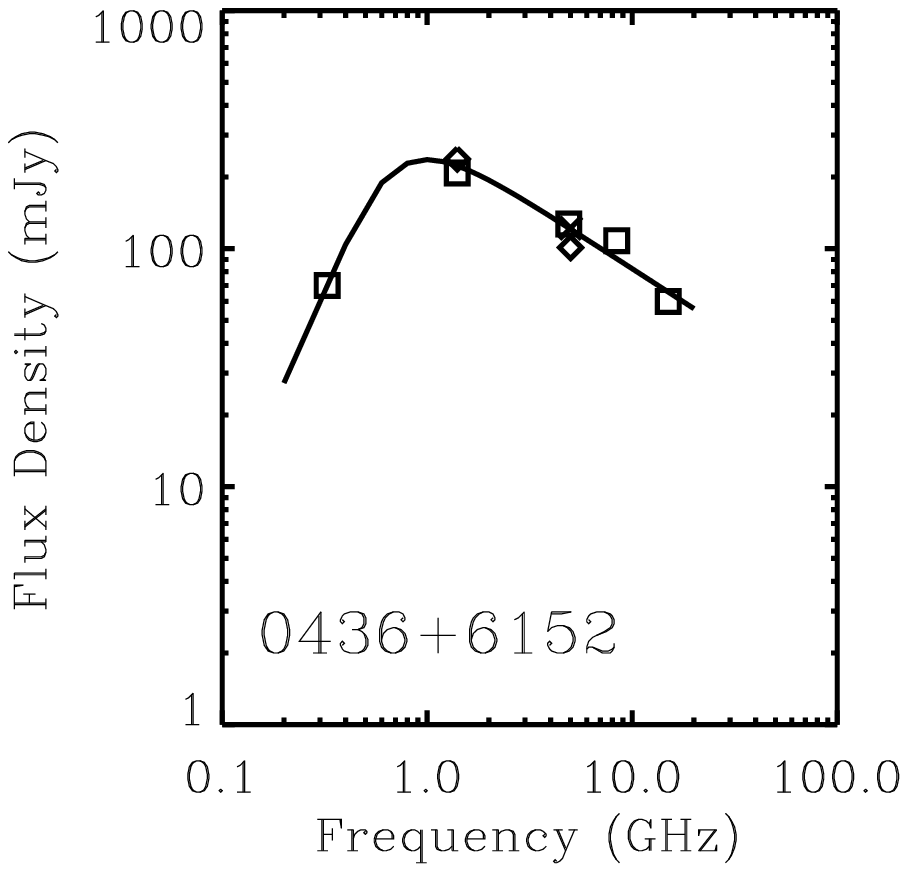,width=5.1cm}\hspace{-0.8cm} 
\psfig{figure=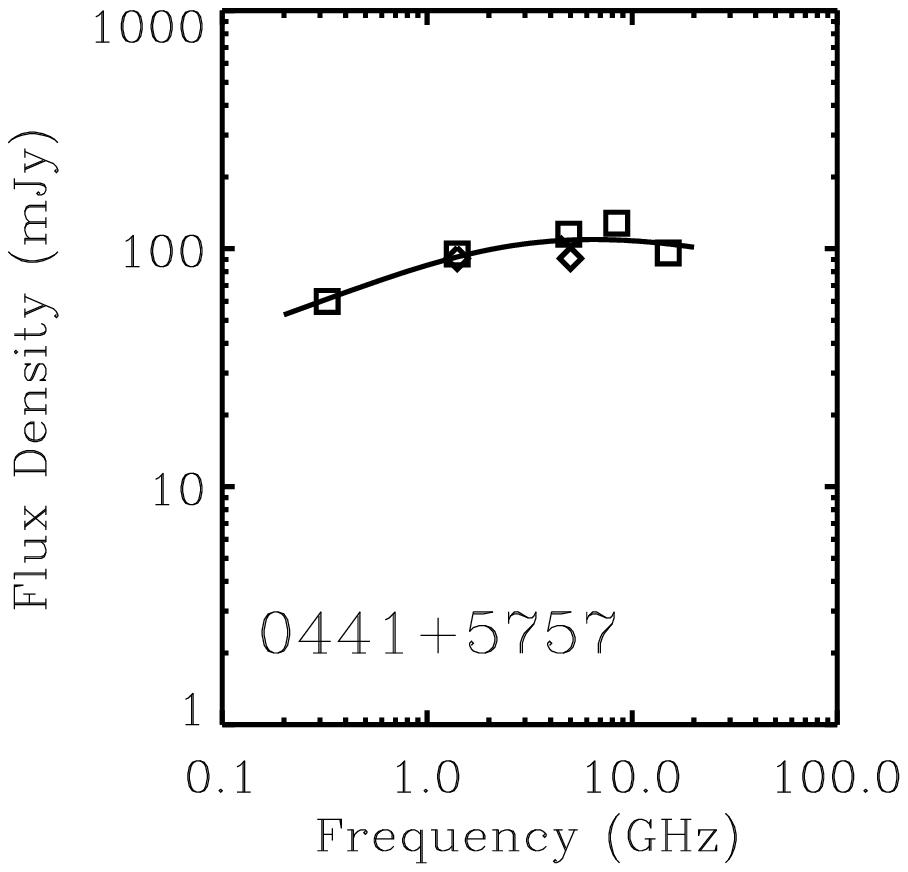,width=5.1cm}\hspace{-0.8cm}
\psfig{figure=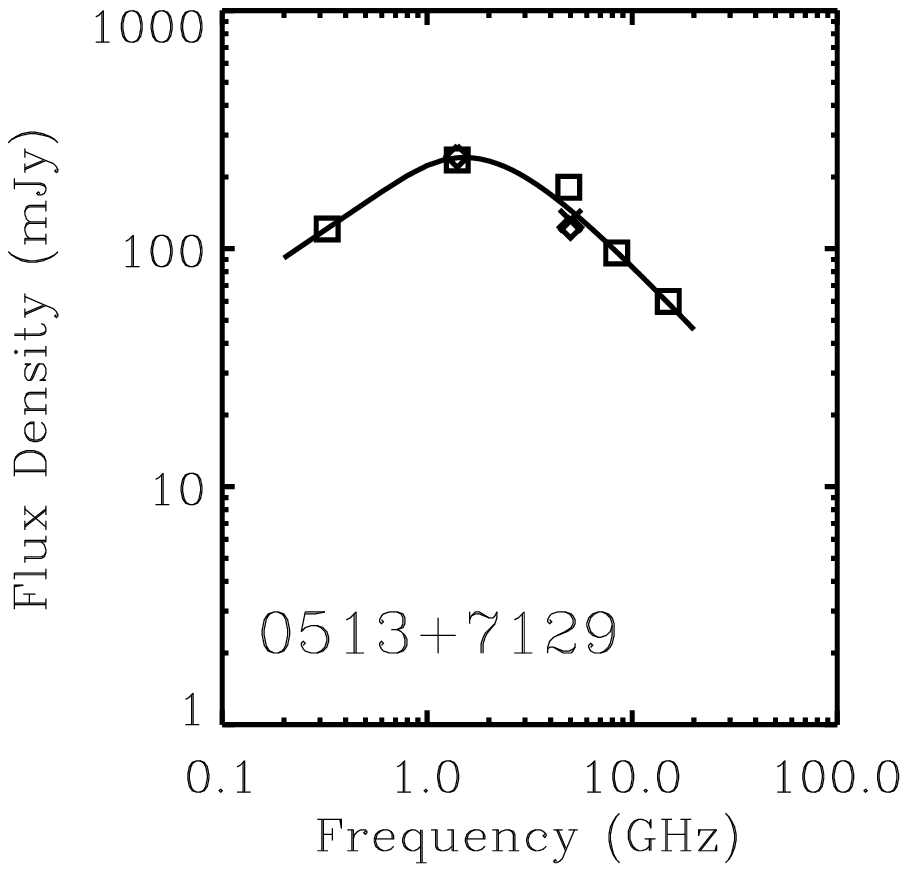,width=5.1cm}
}
\vspace{-0.5cm}
\hbox{\hspace{-0.8cm}
\psfig{figure=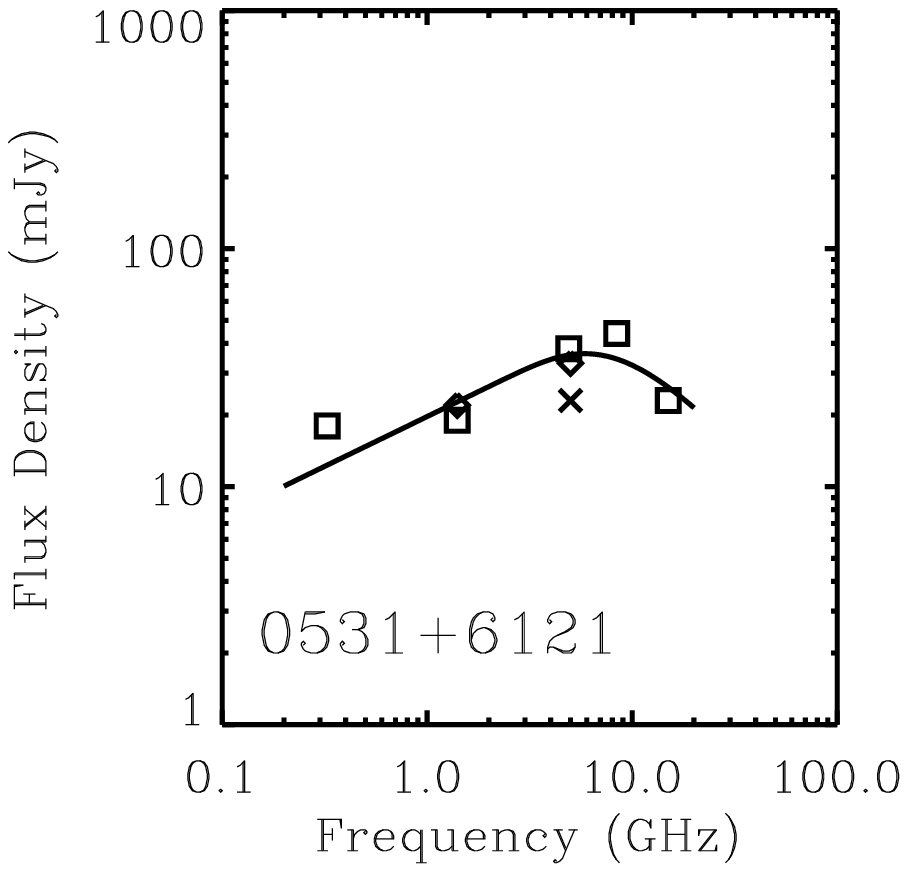,width=5.1cm}\hspace{-0.8cm}
\psfig{figure=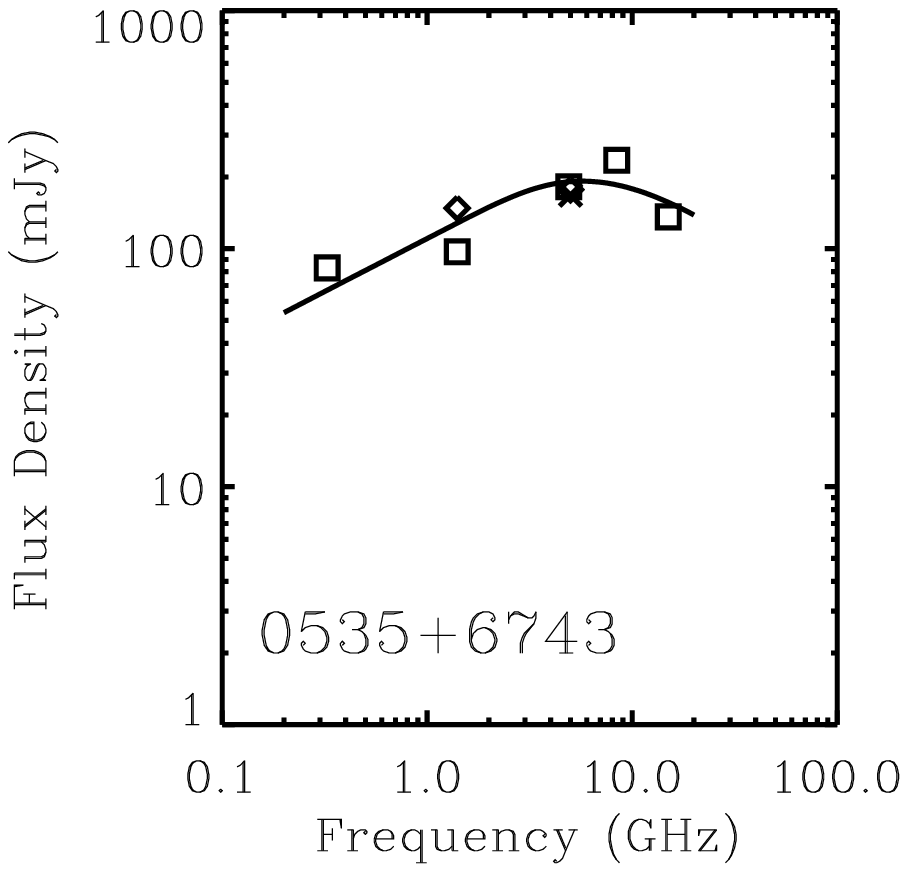,width=5.1cm}\hspace{-0.8cm}
\psfig{figure=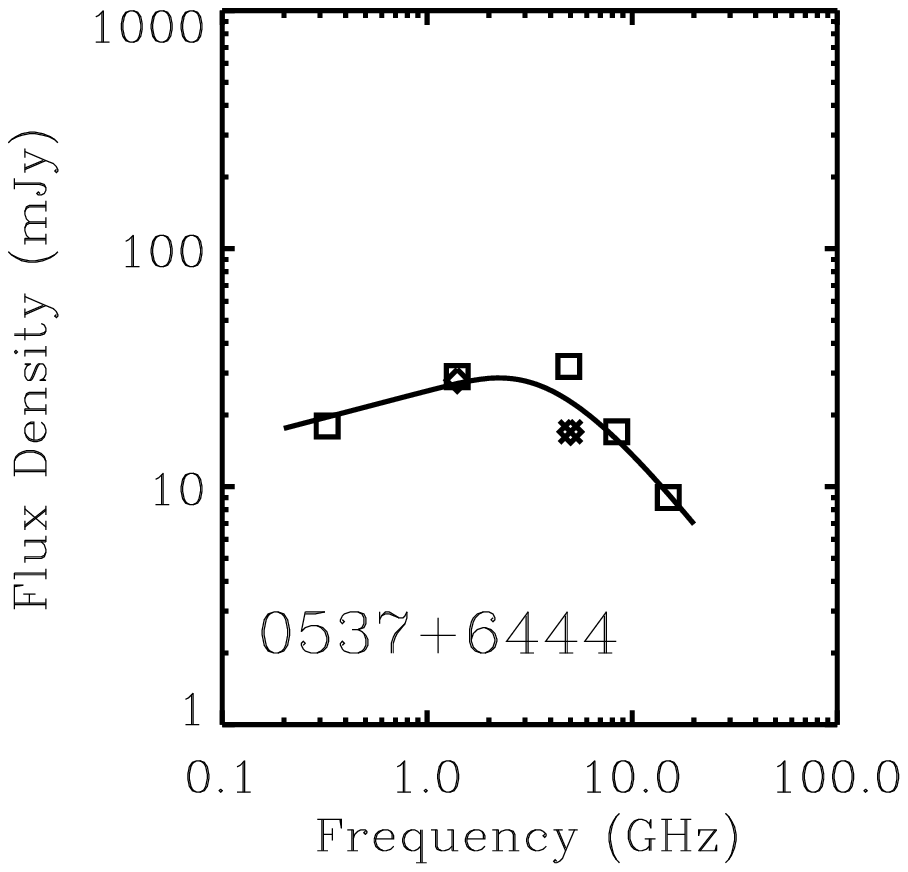,width=5.1cm}\hspace{-0.8cm}
\psfig{figure=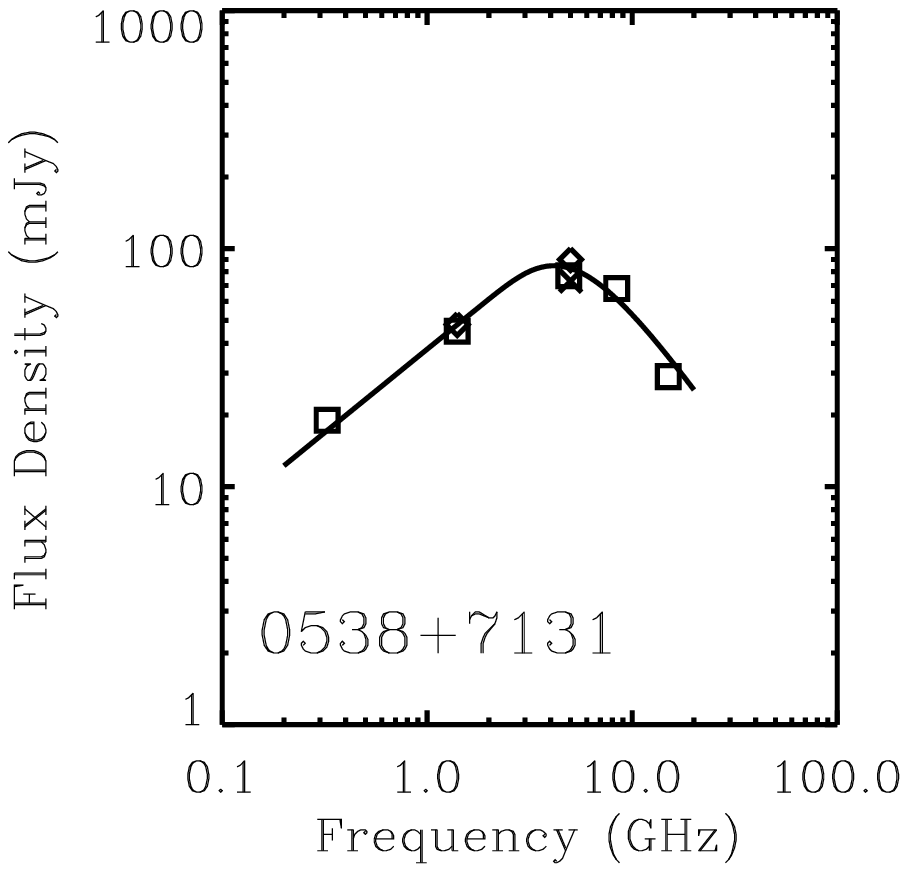,width=5.1cm}
}
\vspace{-0.5cm}
\hbox{\hspace{-0.8cm}
\psfig{figure=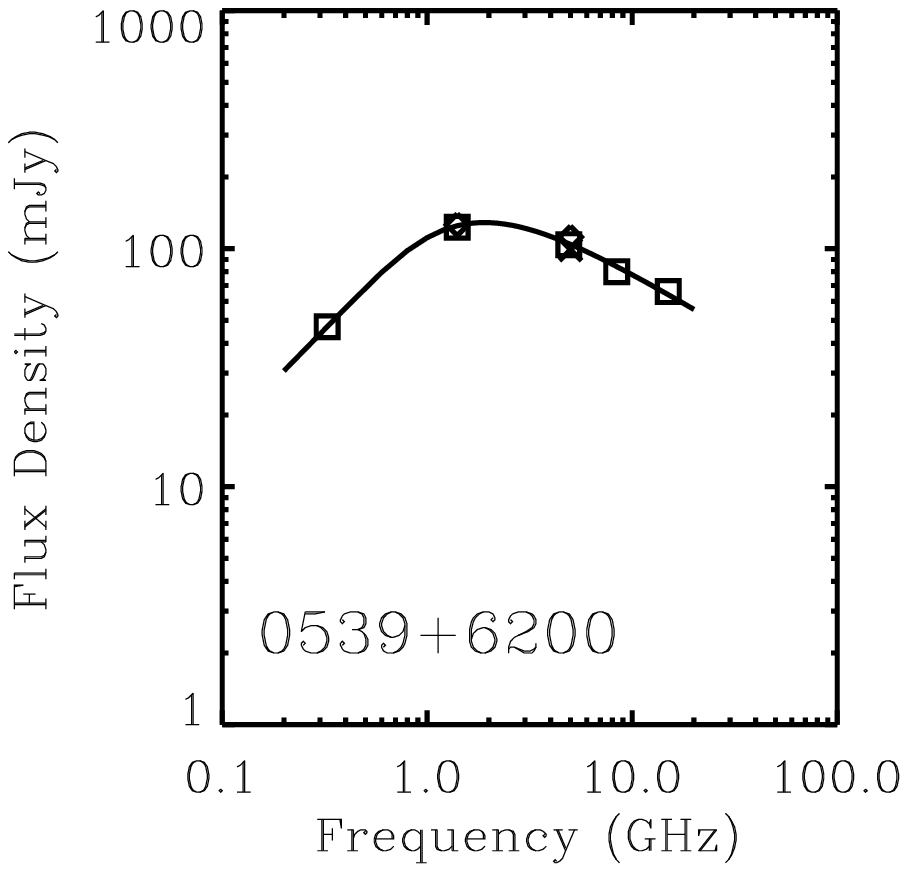,width=5.1cm}\hspace{-0.8cm}
\psfig{figure=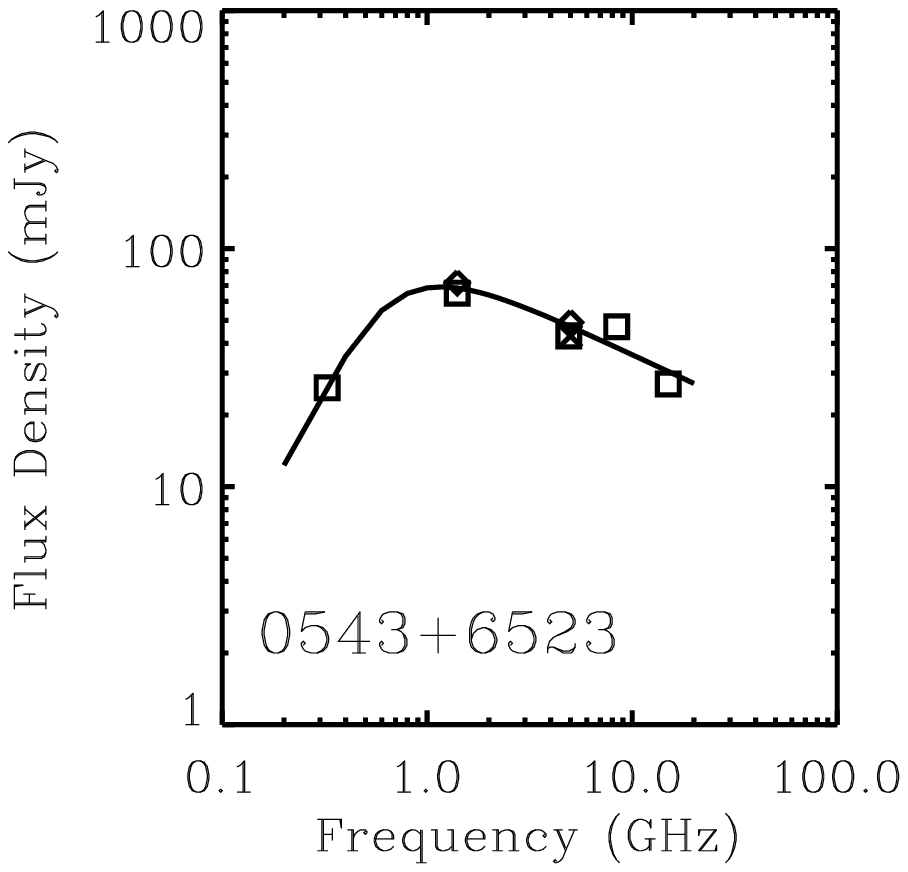,width=5.1cm}\hspace{-0.8cm}
\psfig{figure=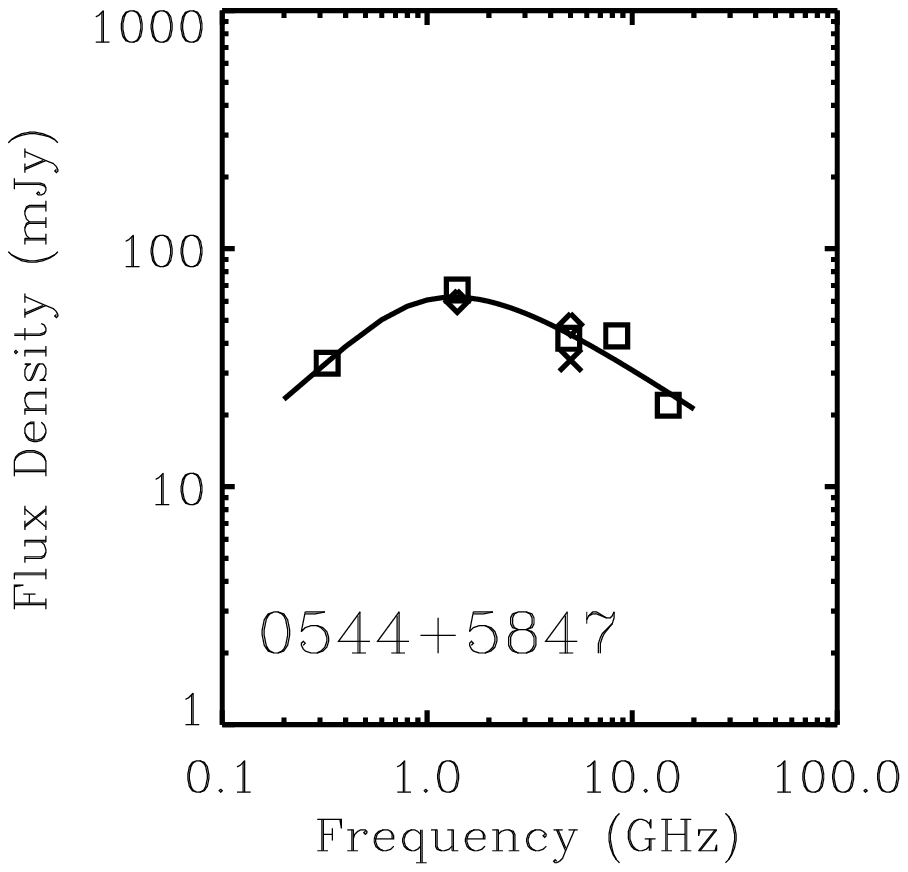,width=5.1cm}\hspace{-0.8cm}
\psfig{figure=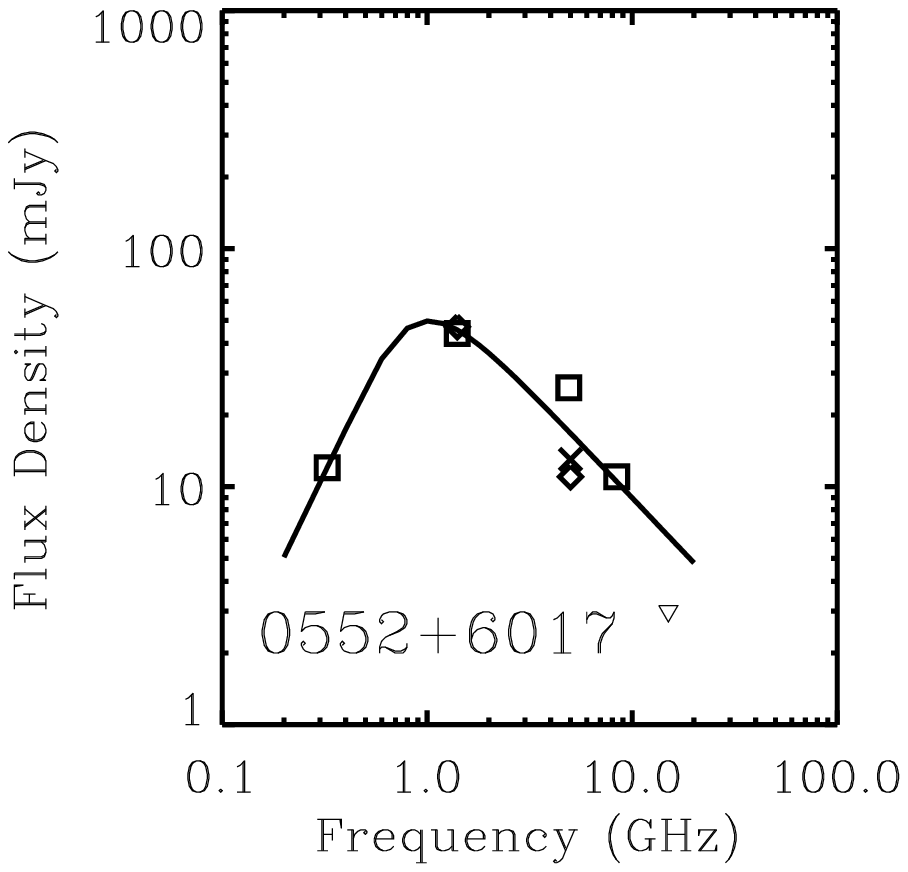,width=5.1cm}
}
\vspace{-0.5cm}
\hbox{\hspace{-0.8cm}
\psfig{figure=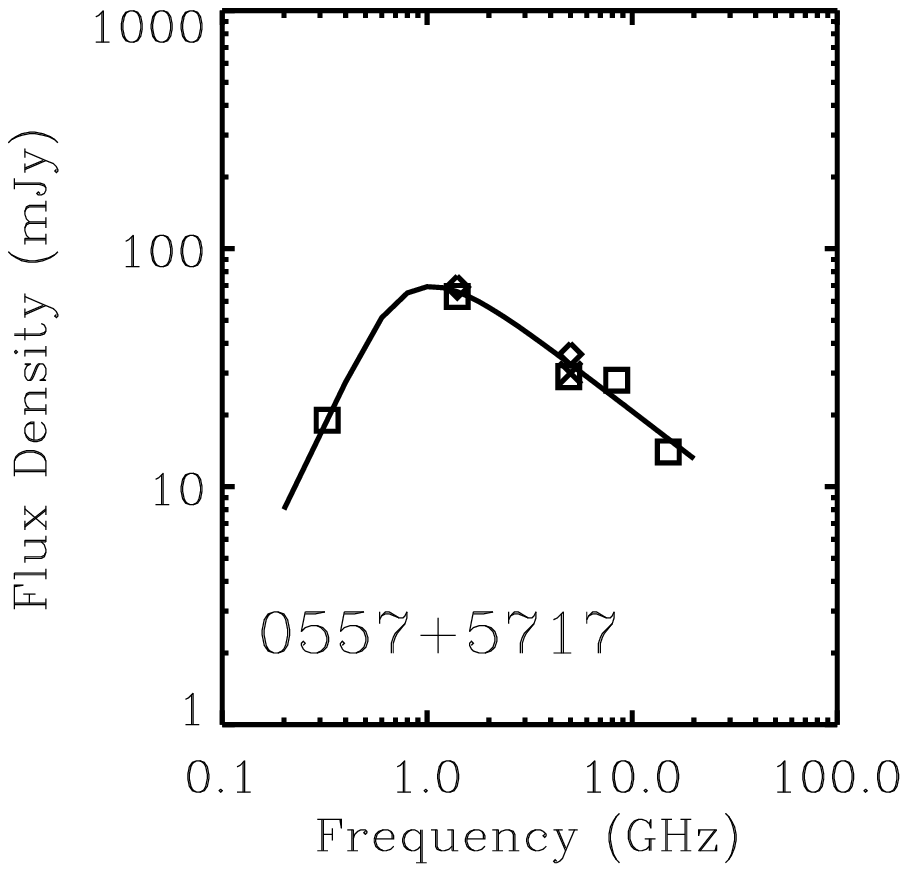,width=5.1cm}\hspace{-0.8cm}
\psfig{figure=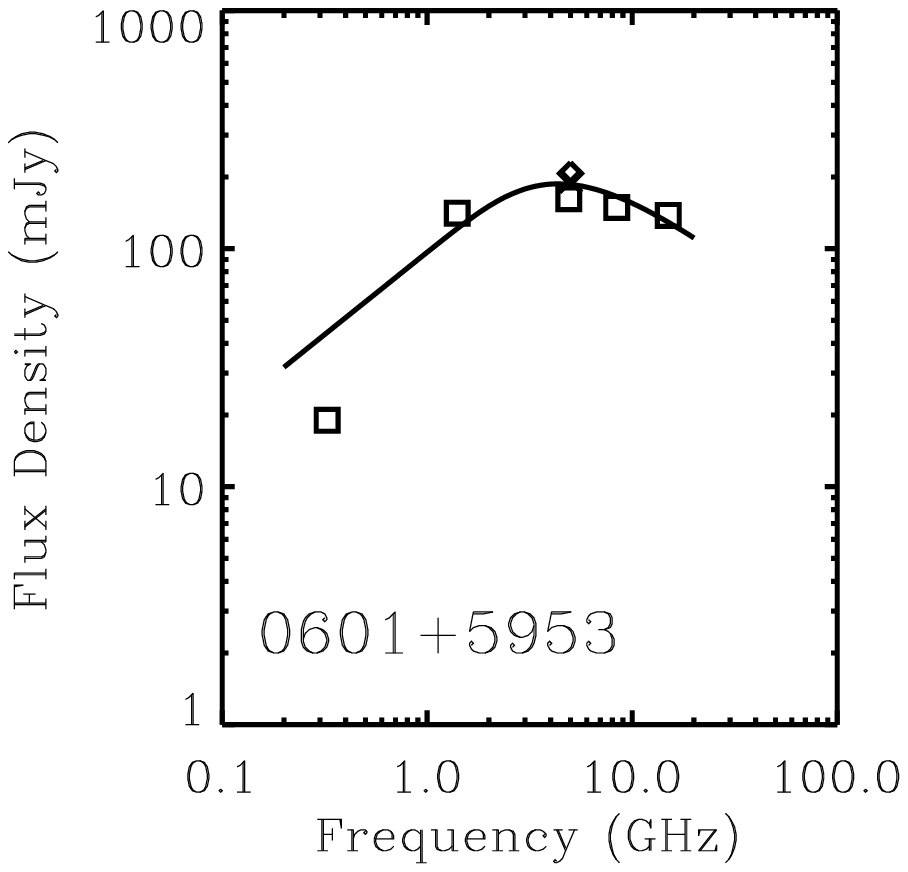,width=5.1cm}\hspace{-0.8cm}
\psfig{figure=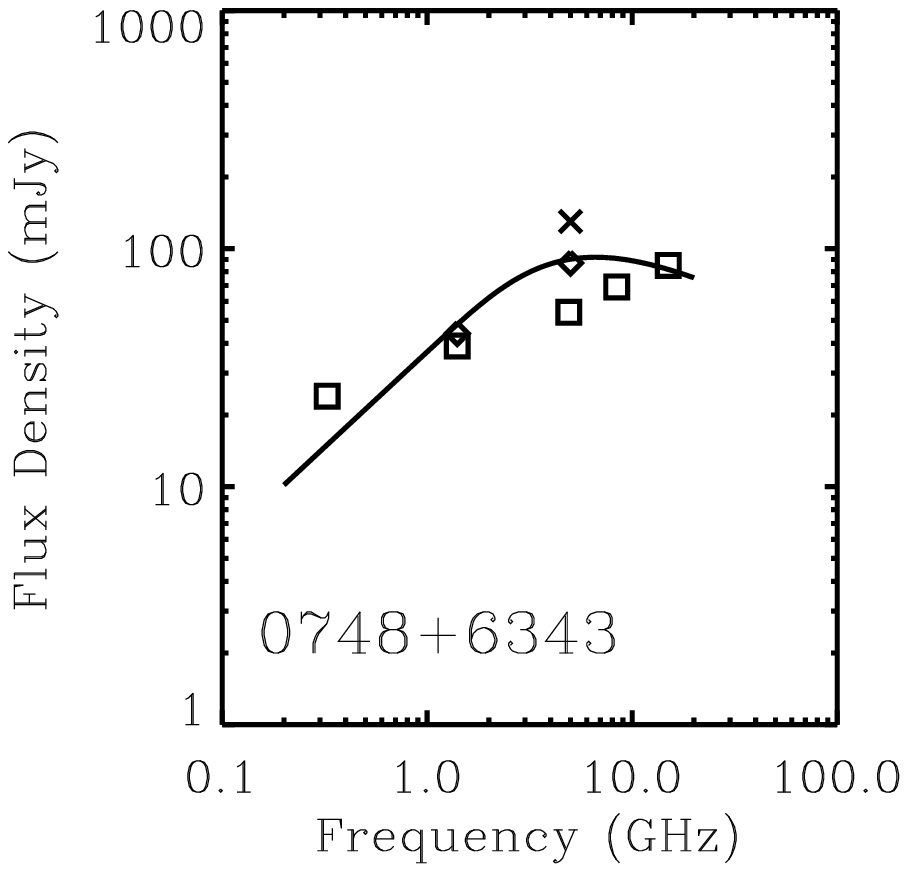,width=5.1cm}\hspace{-0.8cm}
\psfig{figure=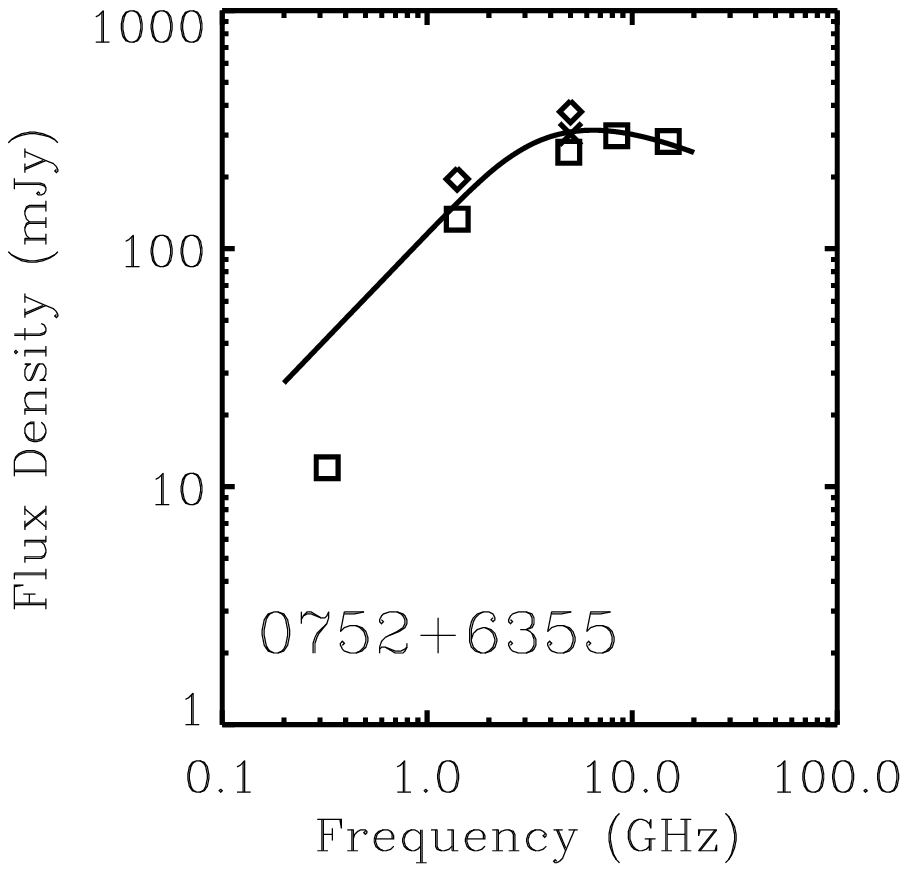,width=5.1cm}
}
\vspace{-0.5cm}
\hbox{\hspace{-0.8cm}
\psfig{figure=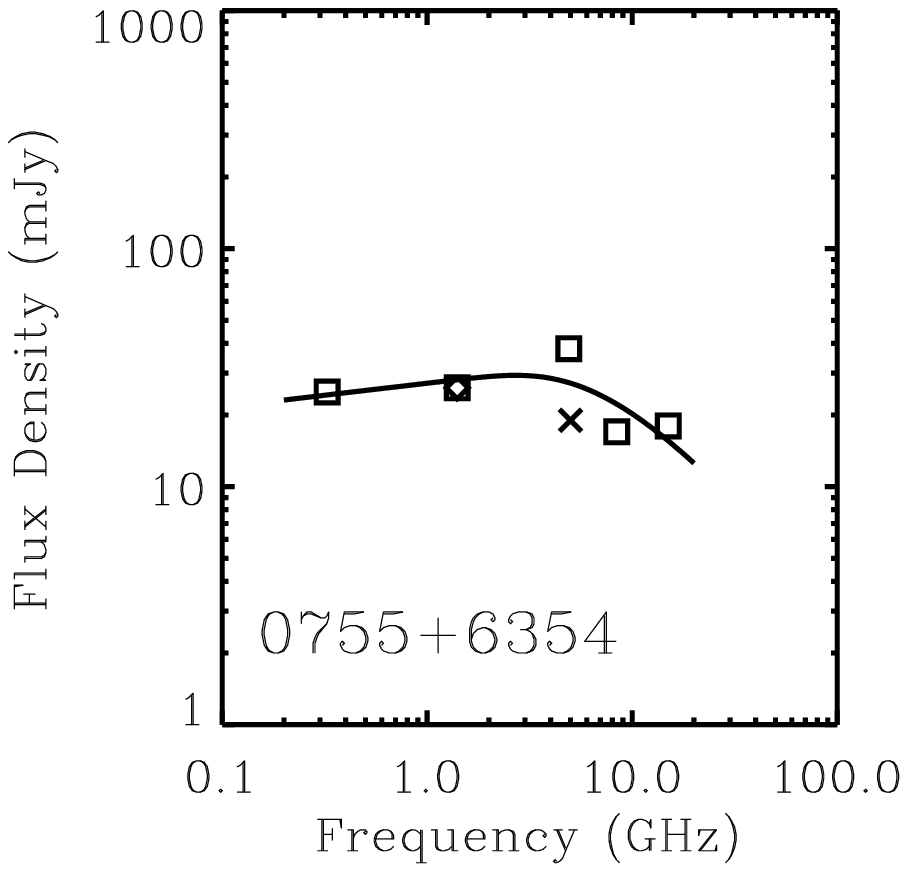,width=5.1cm}\hspace{-0.8cm}
\psfig{figure=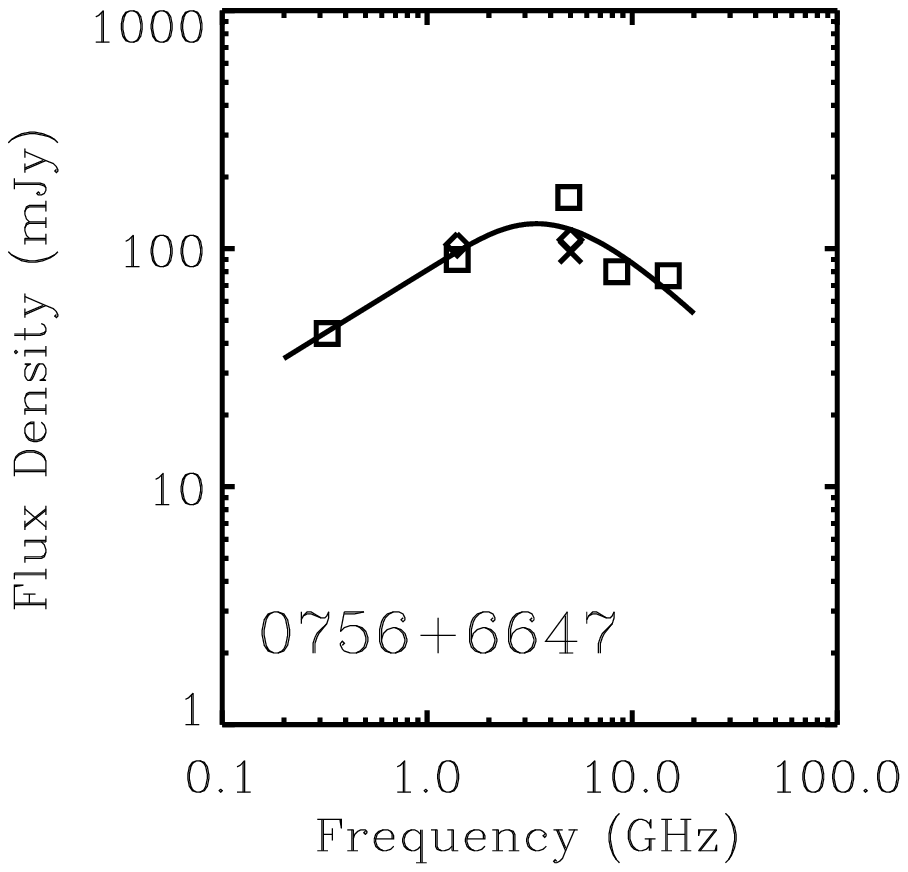,width=5.1cm}\hspace{-0.8cm}
\psfig{figure=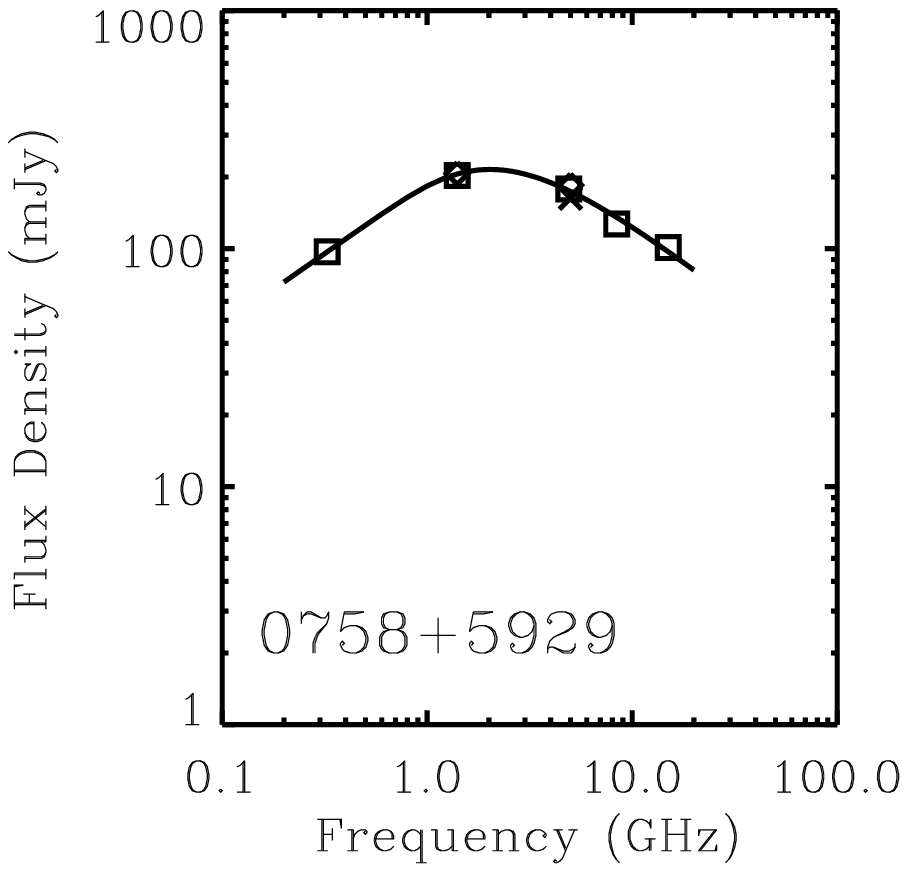,width=5.1cm}\hspace{-0.8cm}
\psfig{figure=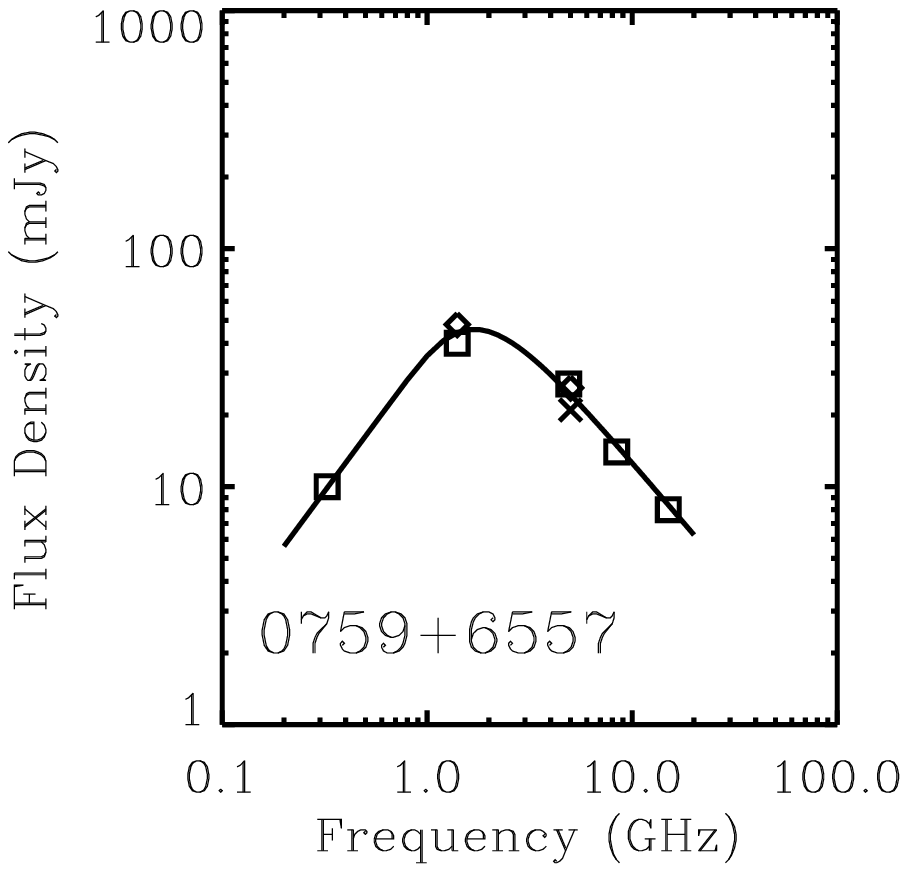,width=5.1cm}
}
\caption{\label{spectra} 
Radio spectra of individual sources. Crosses indicate MERLIN
data, diamonds indicate at 1.4 GHz NVSS data and at 5 GHz WSRT data.} 
\end{figure*}

\addtocounter{figure}{-1}

\begin{figure*}
\vspace{-0.5cm}
\hbox{\hspace{-0.8cm}
\psfig{figure=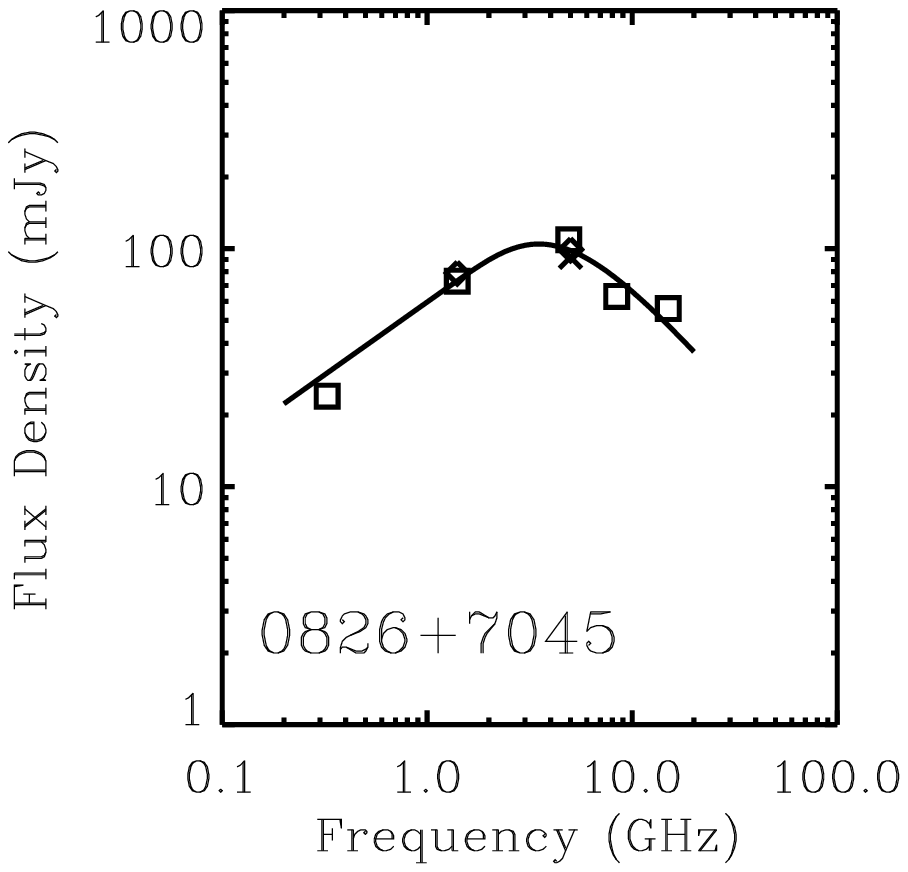,width=5.1cm}\hspace{-0.8cm}
\psfig{figure=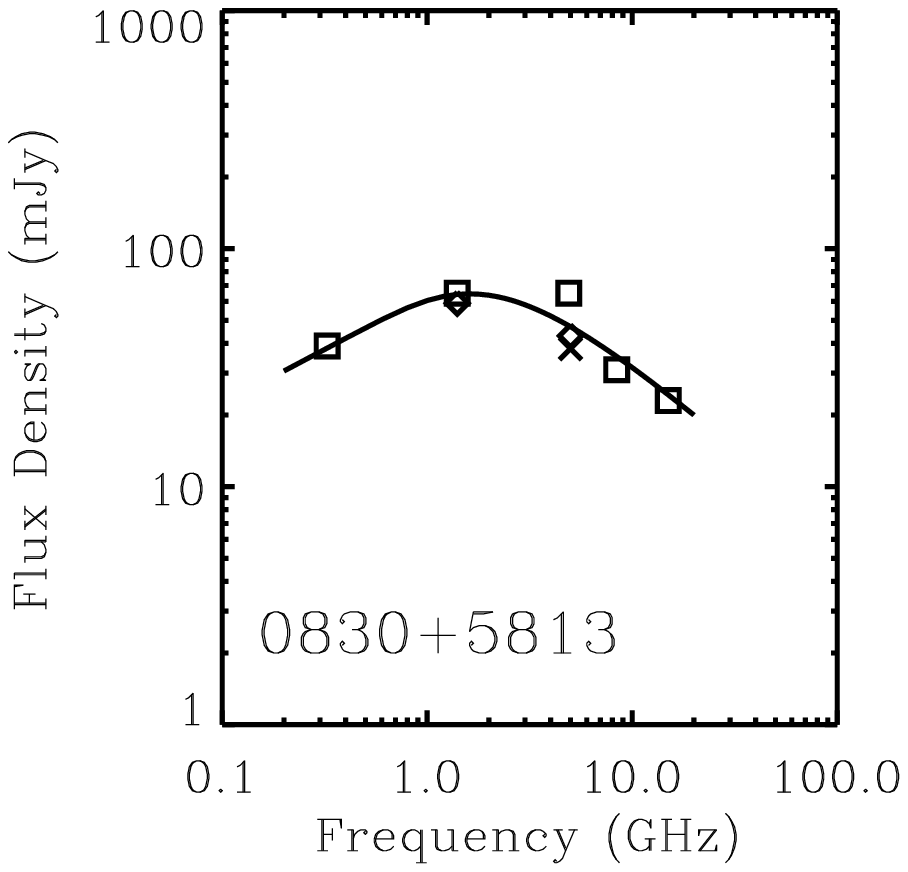,width=5.1cm}\hspace{-0.8cm}
\psfig{figure=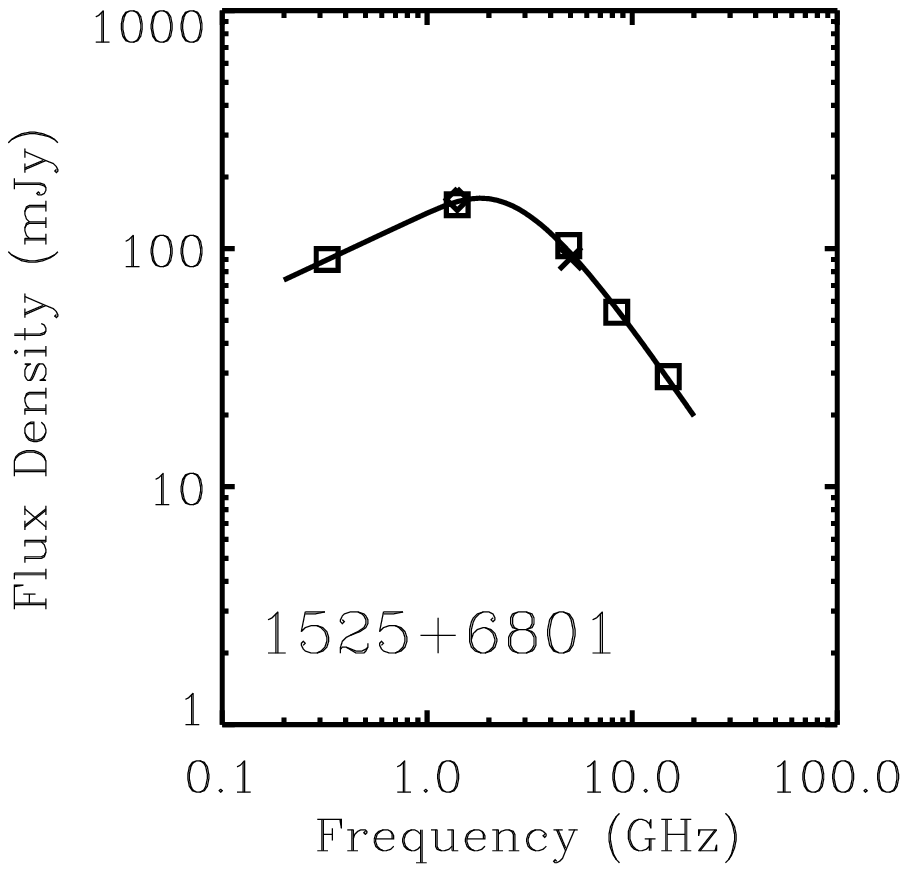,width=5.1cm}\hspace{-0.8cm}
\psfig{figure=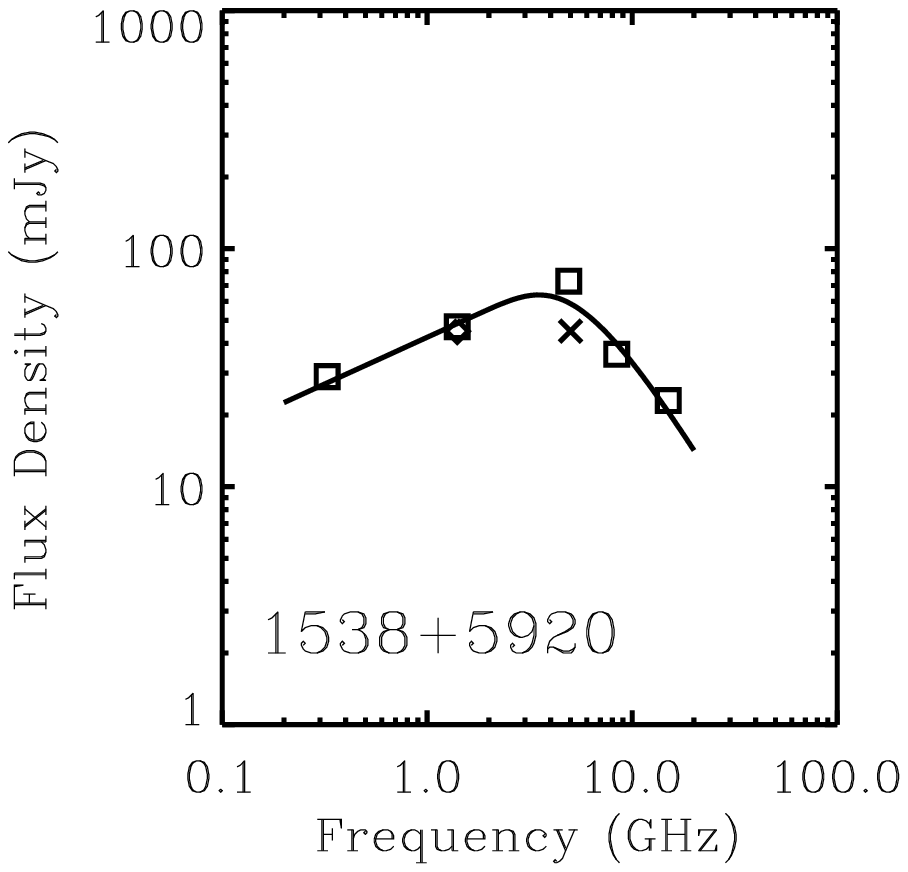,width=5.1cm}
}
\vspace{-0.5cm}
\hbox{\hspace{-0.8cm}
\psfig{figure=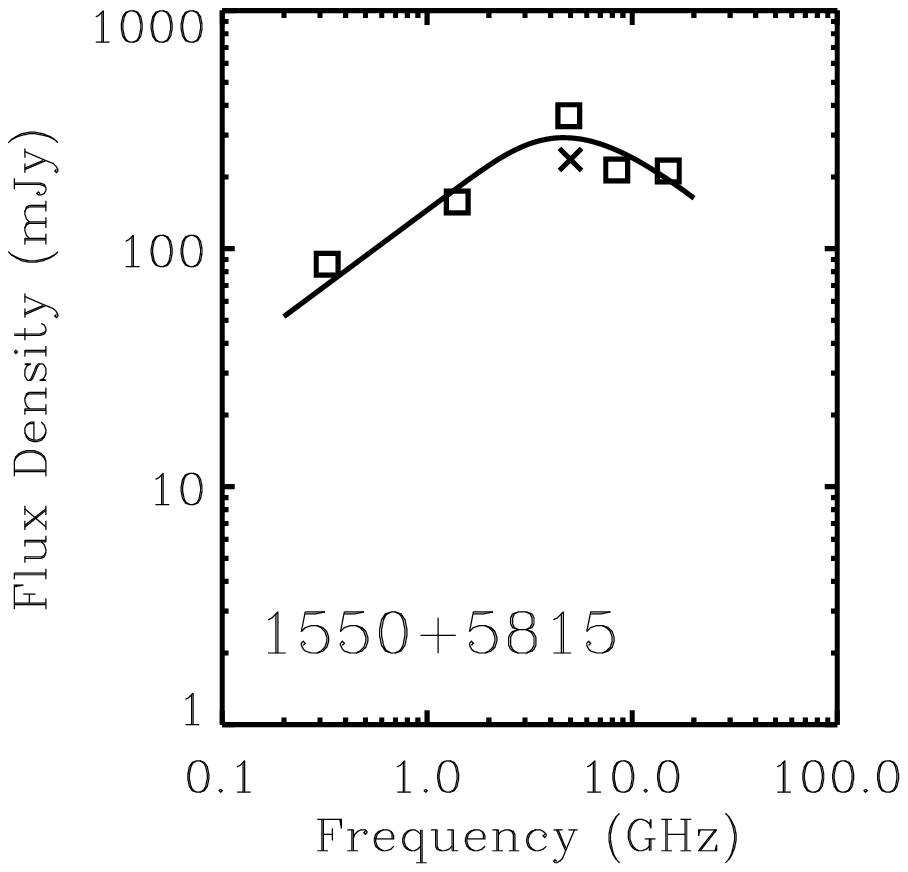,width=5.1cm}\hspace{-0.8cm}
\psfig{figure=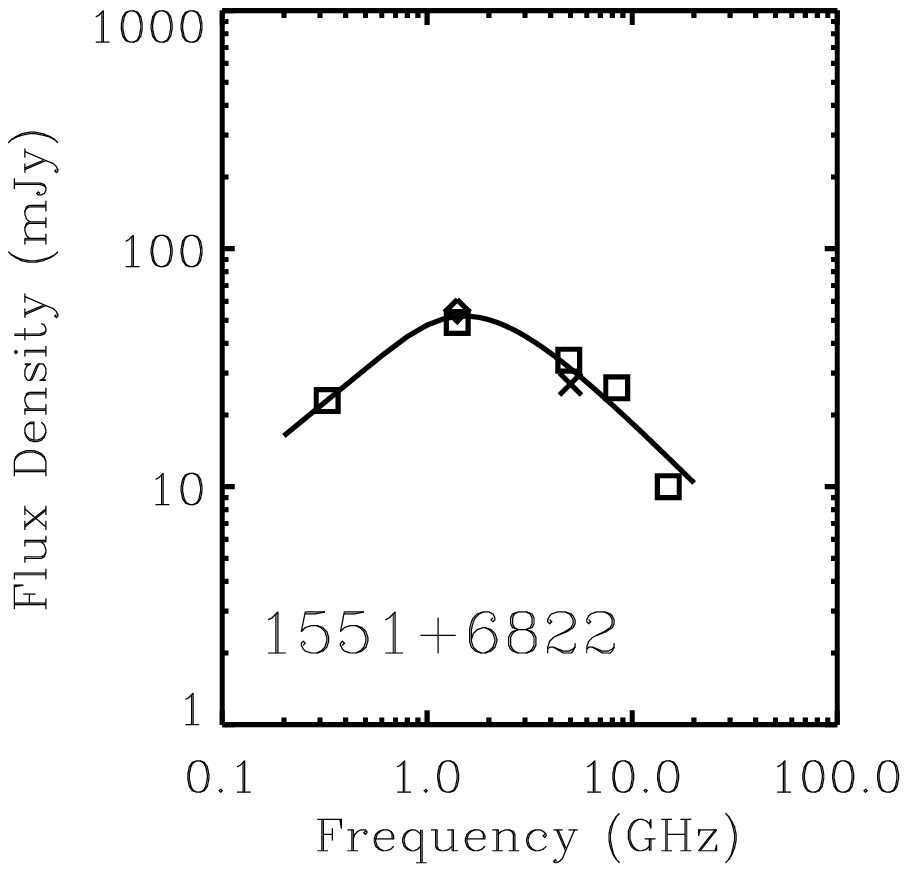,width=5.1cm}\hspace{-0.8cm}
\psfig{figure=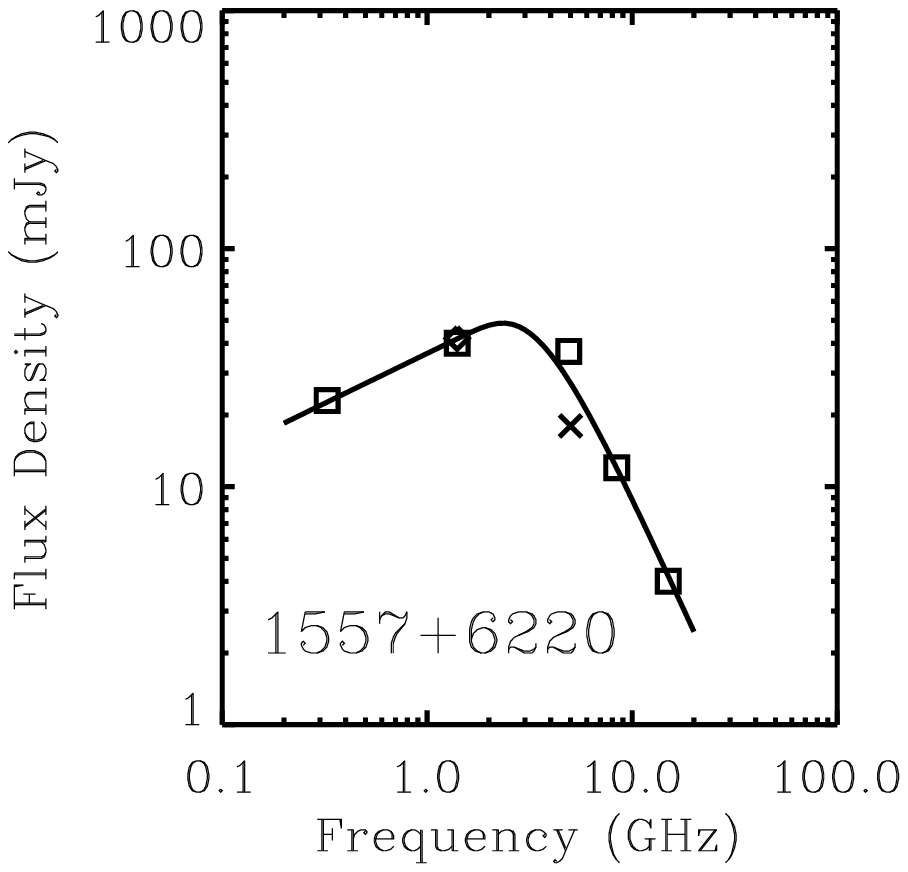,width=5.1cm}\hspace{-0.8cm}
\psfig{figure=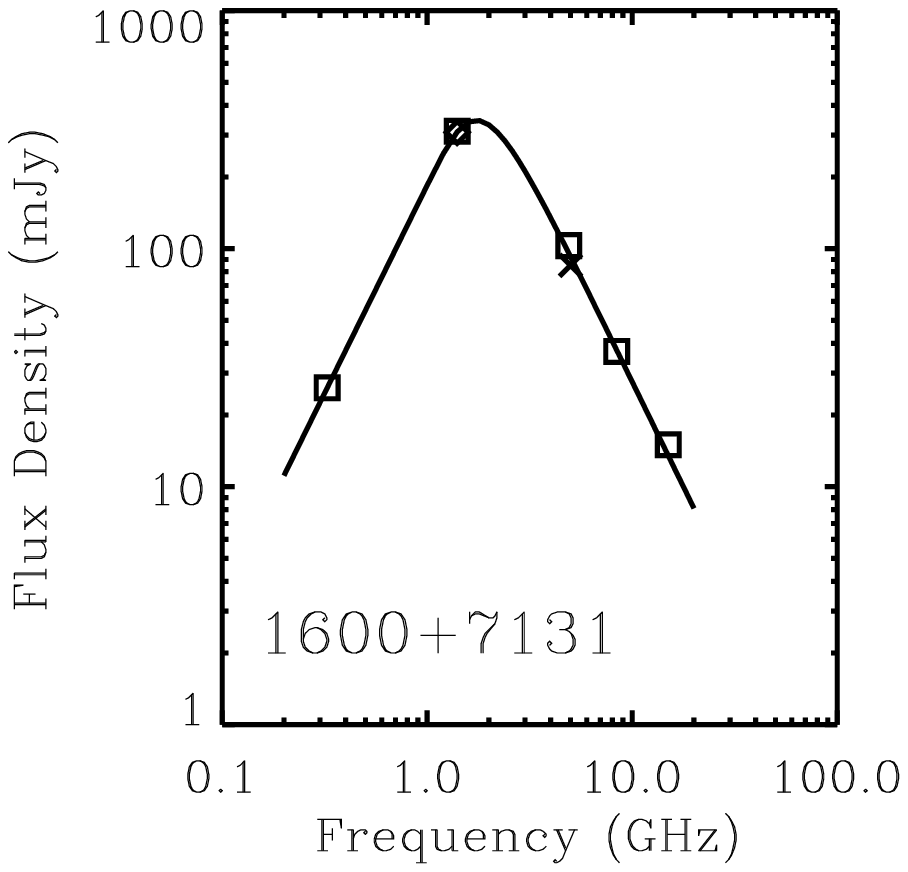,width=5.1cm}
}
\vspace{-0.5cm}
\hbox{\hspace{-0.8cm}
\psfig{figure=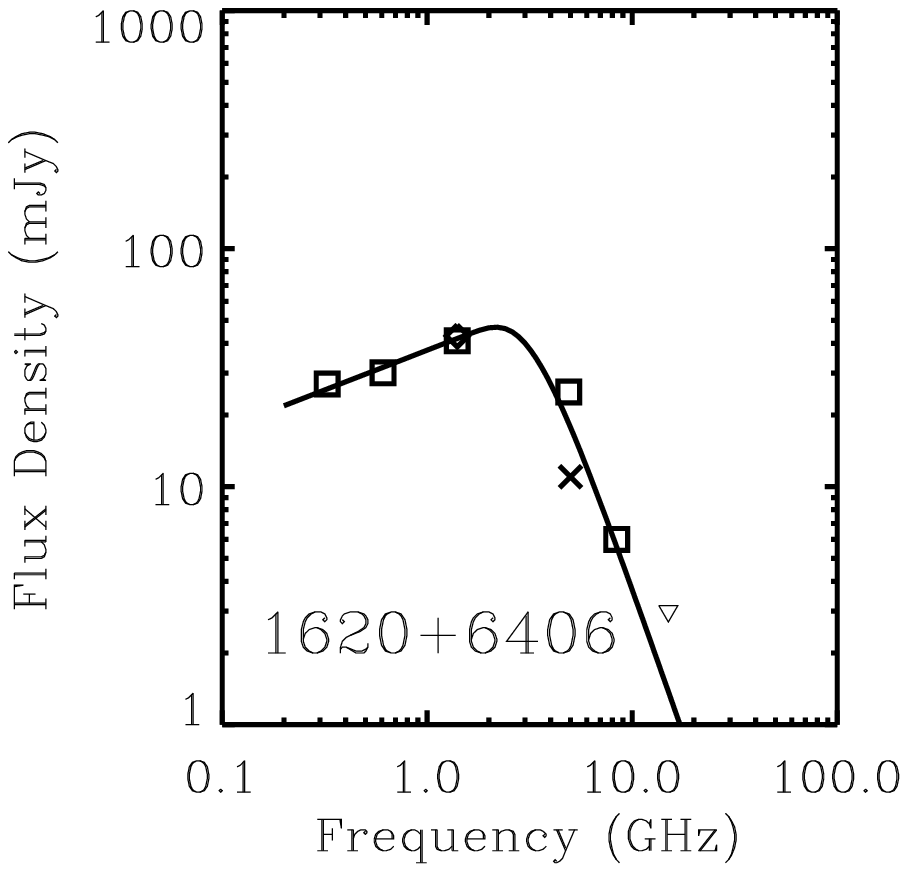,width=5.1cm}\hspace{-0.8cm}
\psfig{figure=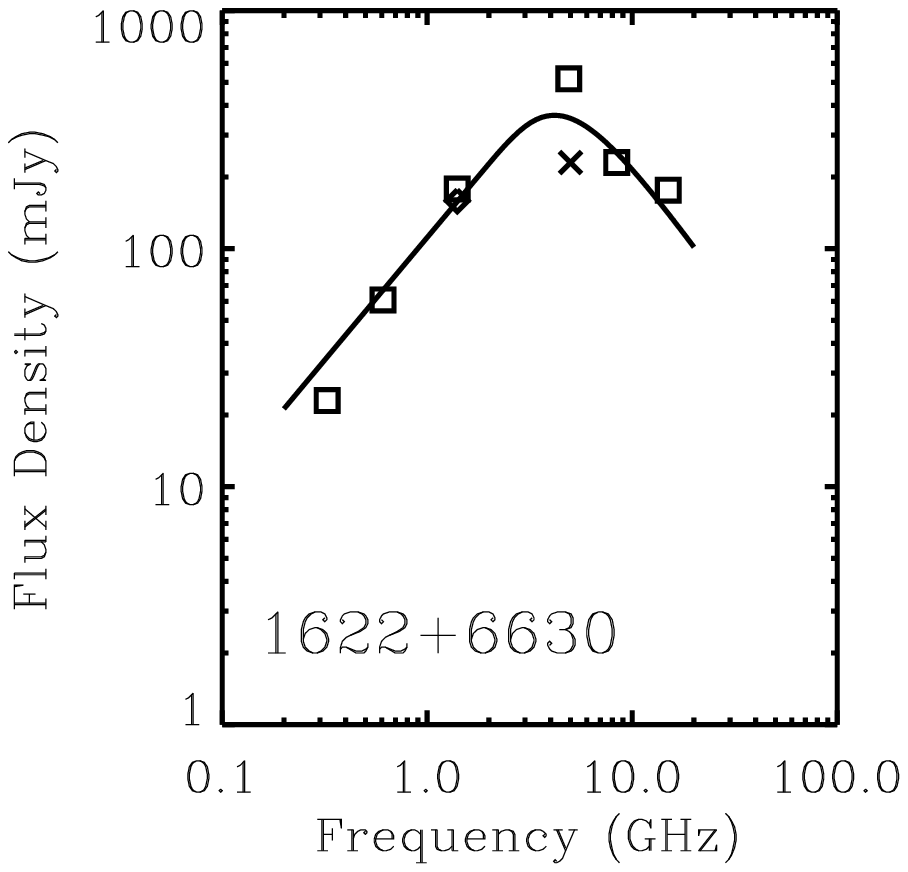,width=5.1cm}\hspace{-0.8cm}
\psfig{figure=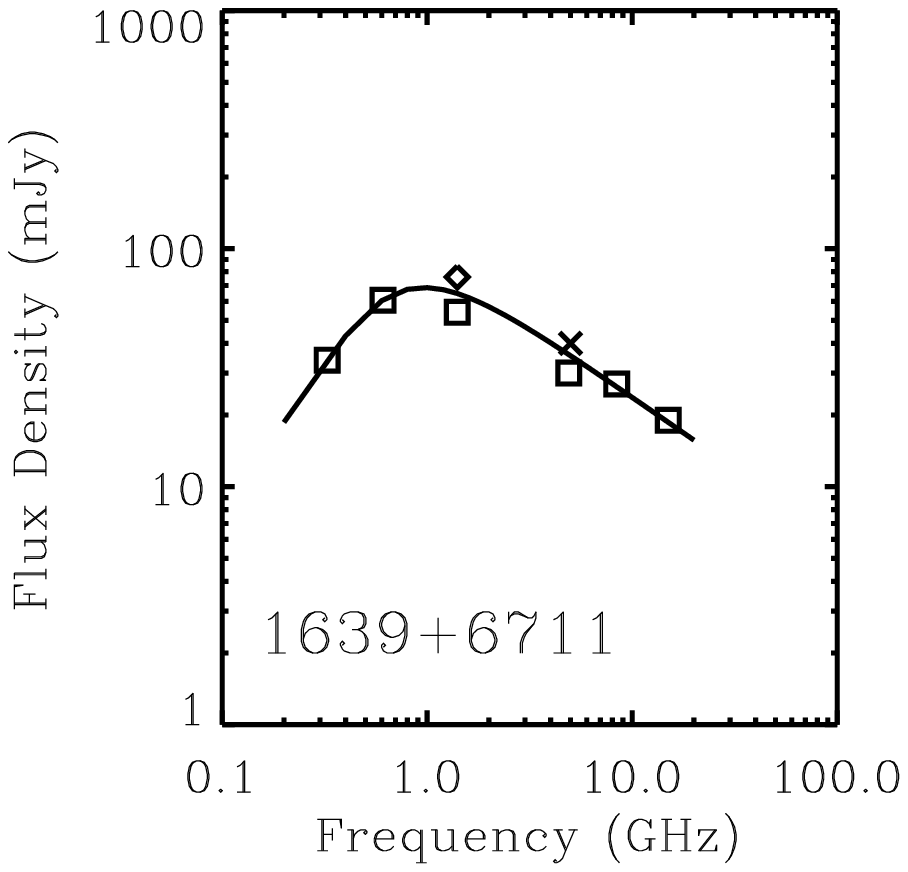,width=5.1cm}\hspace{-0.8cm}
\psfig{figure=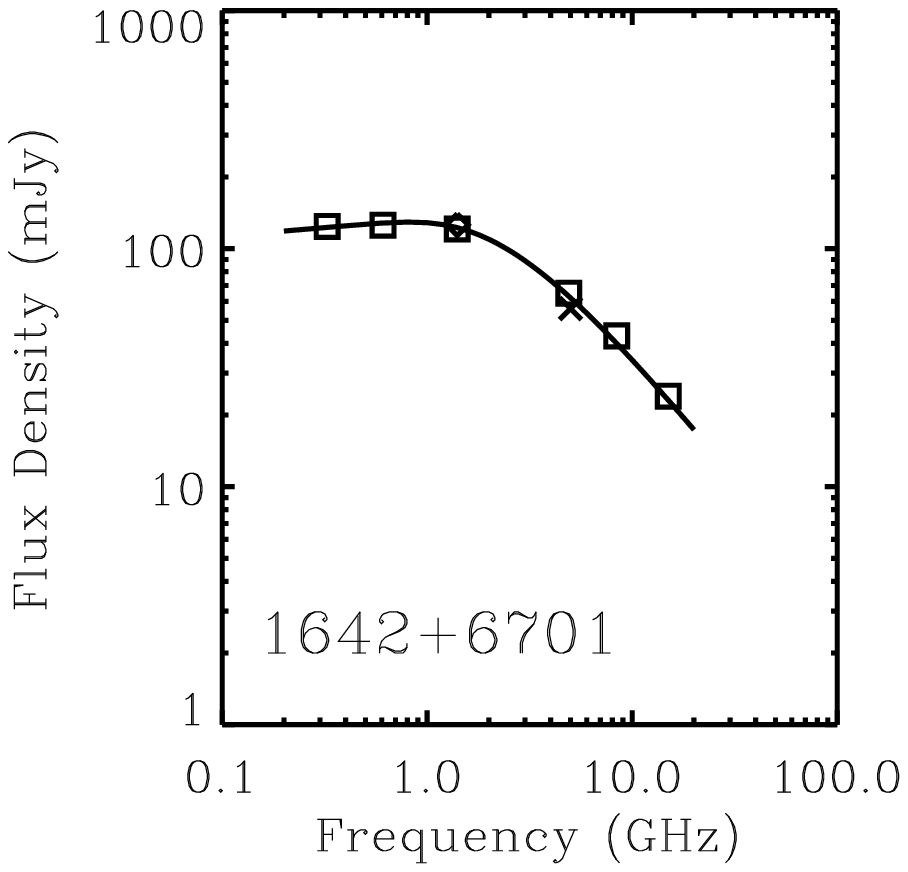,width=5.1cm}
}
\vspace{-0.5cm}
\hbox{\hspace{-0.8cm}
\psfig{figure=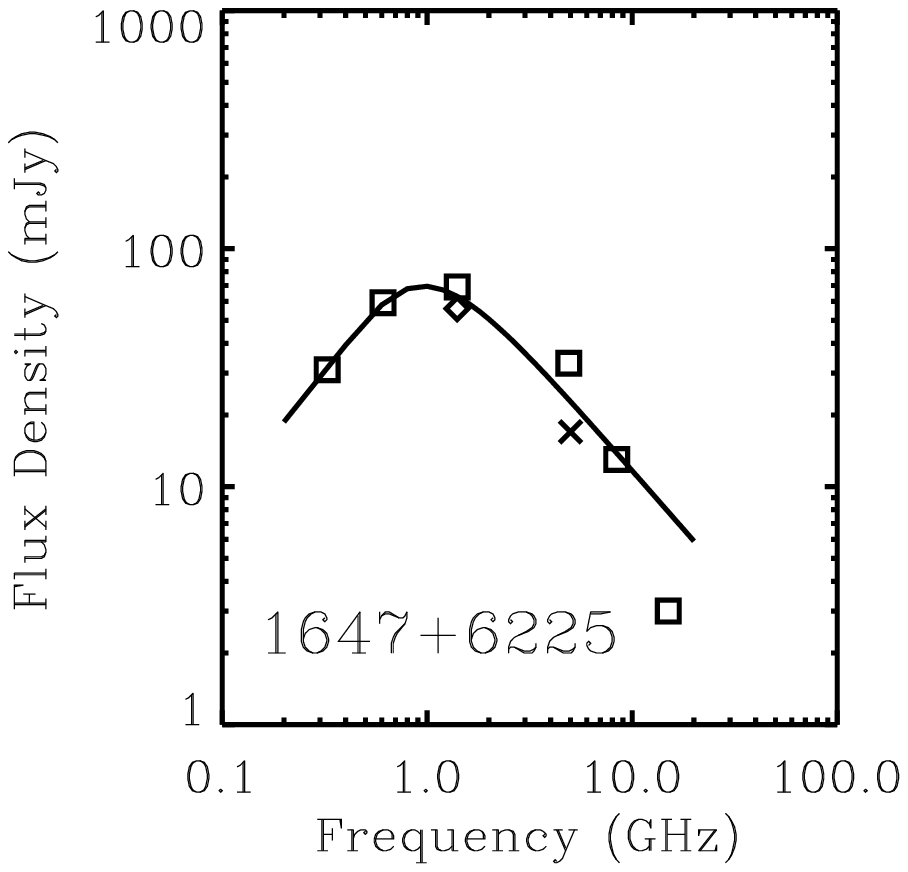,width=5.1cm}\hspace{-0.8cm}
\psfig{figure=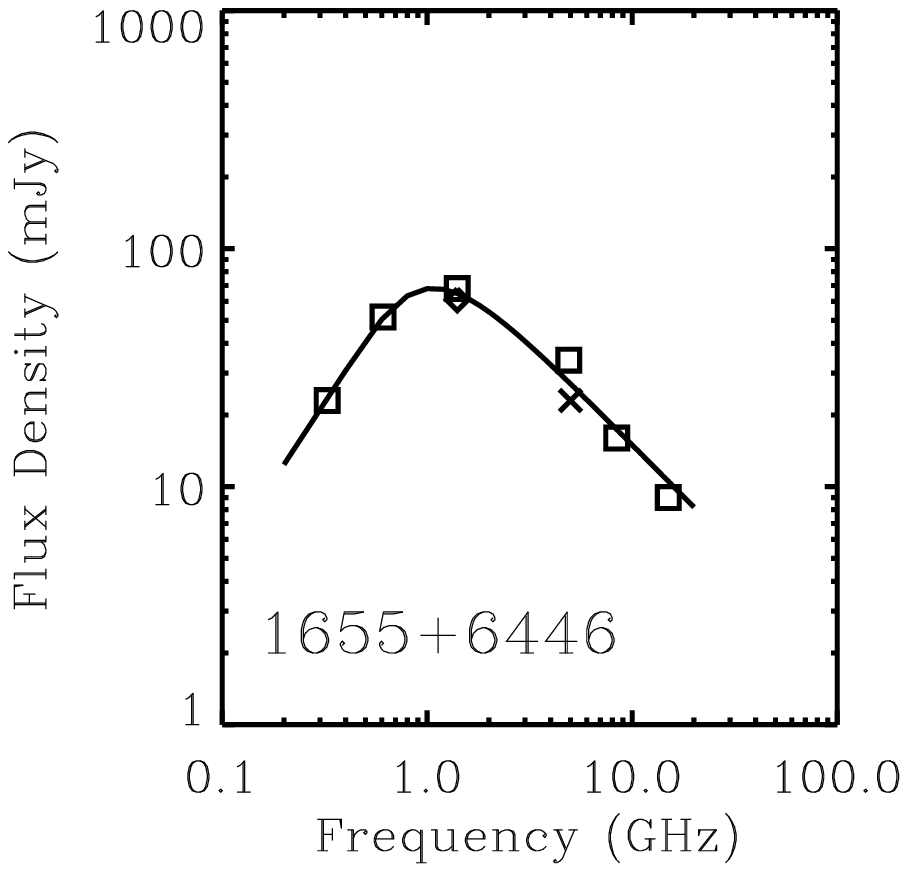,width=5.1cm}\hspace{-0.8cm}
\psfig{figure=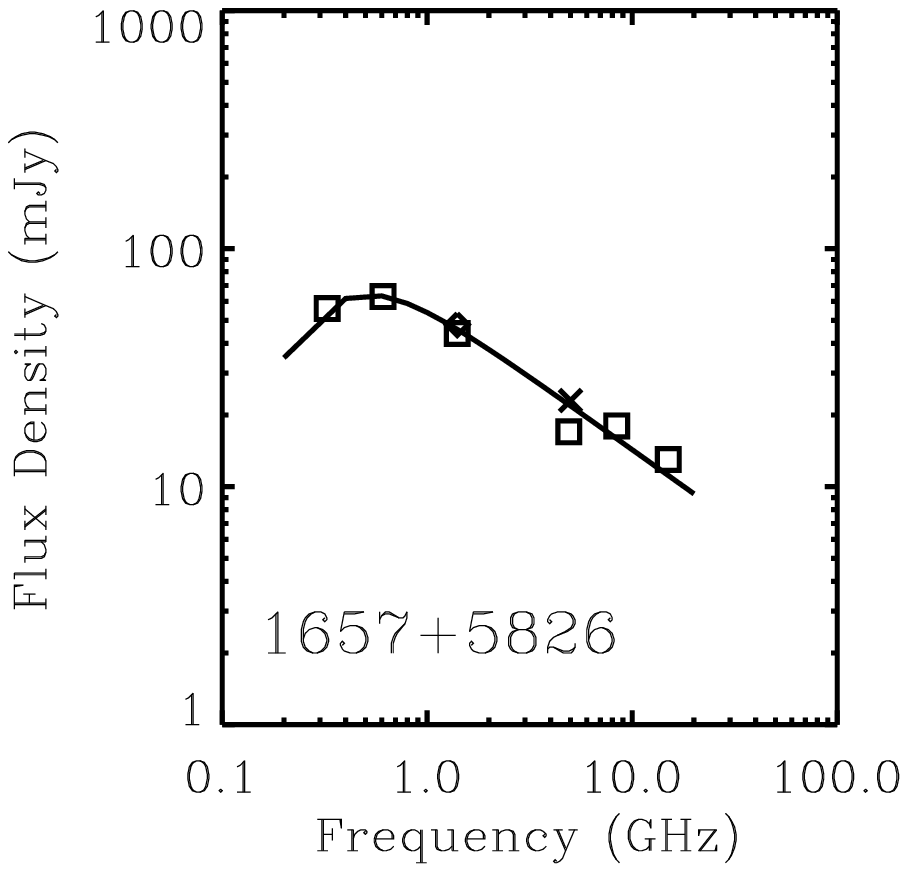,width=5.1cm}\hspace{-0.8cm}
\psfig{figure=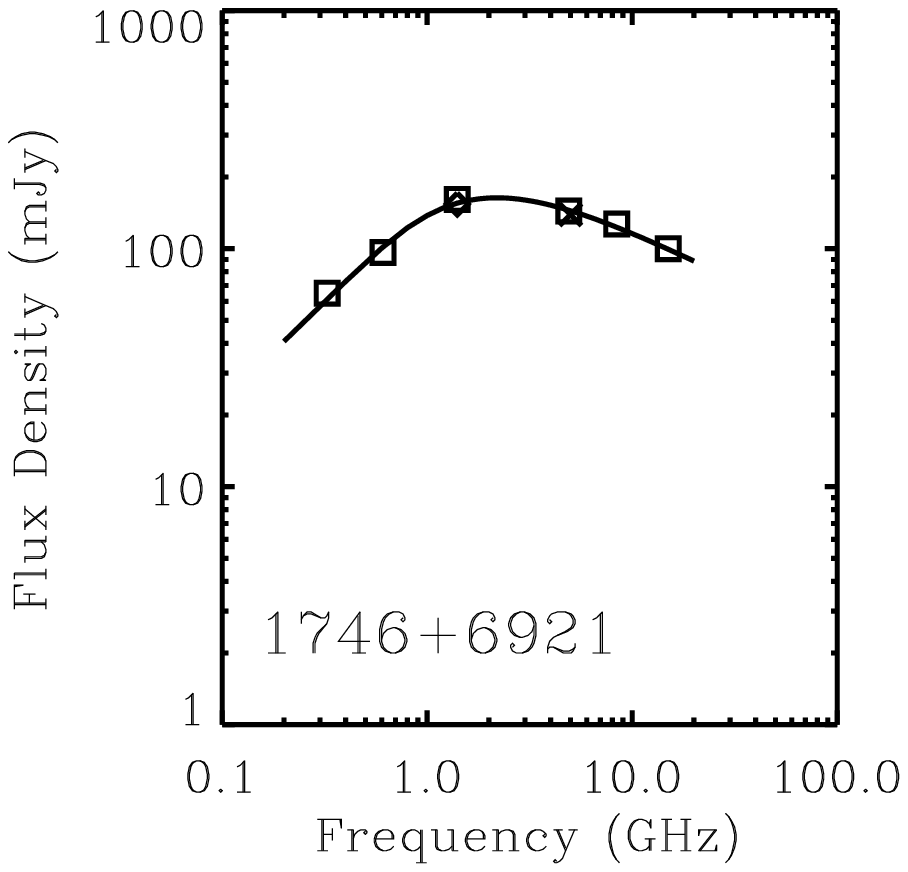,width=5.1cm}
}
\vspace{-0.5cm}
\hbox{\hspace{-0.8cm}
\psfig{figure=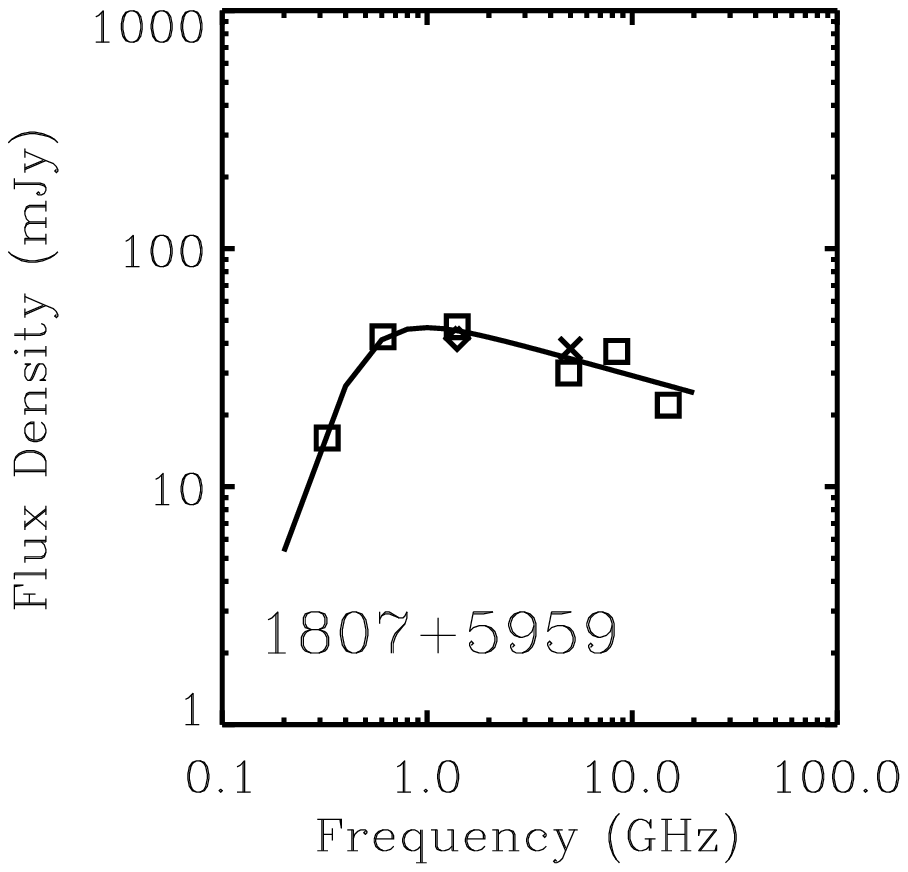,width=5.1cm}\hspace{-0.8cm}
\psfig{figure=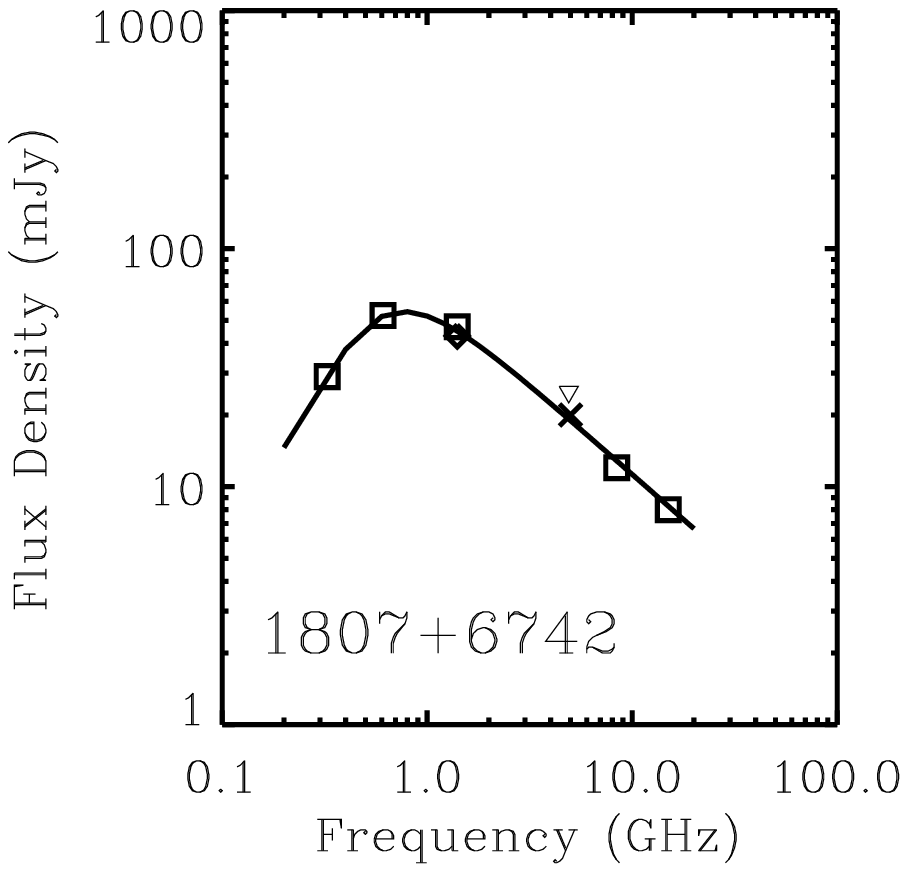,width=5.1cm}\hspace{-0.8cm}
\psfig{figure=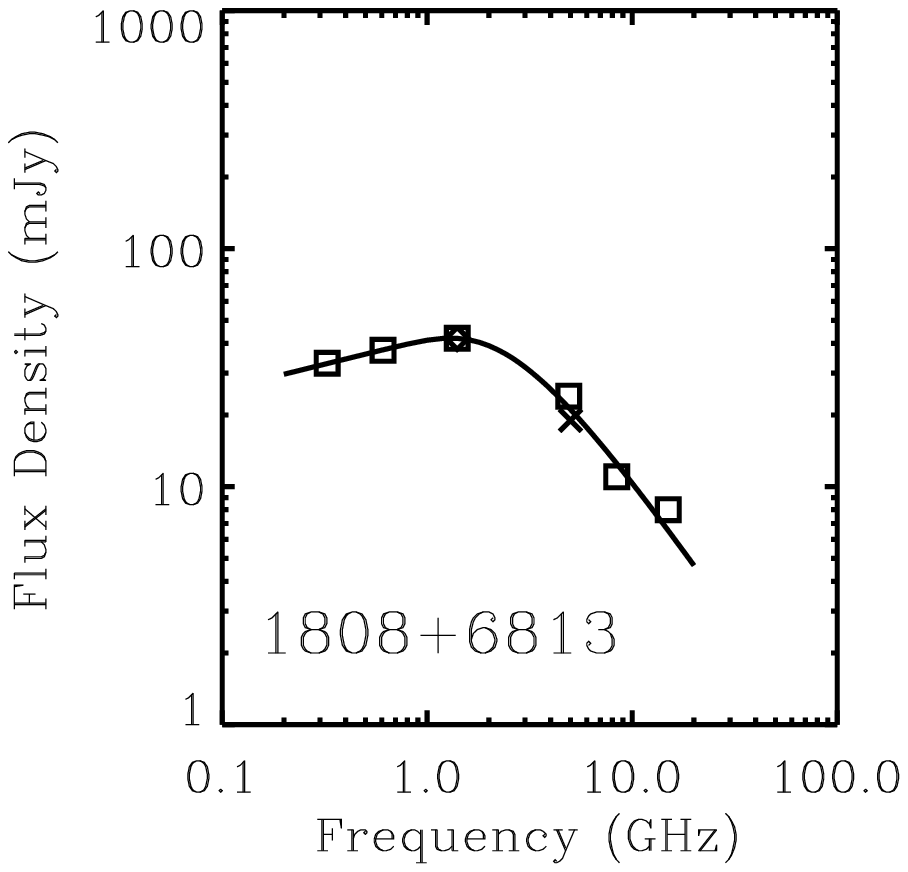,width=5.1cm}\hspace{-0.8cm}
\psfig{figure=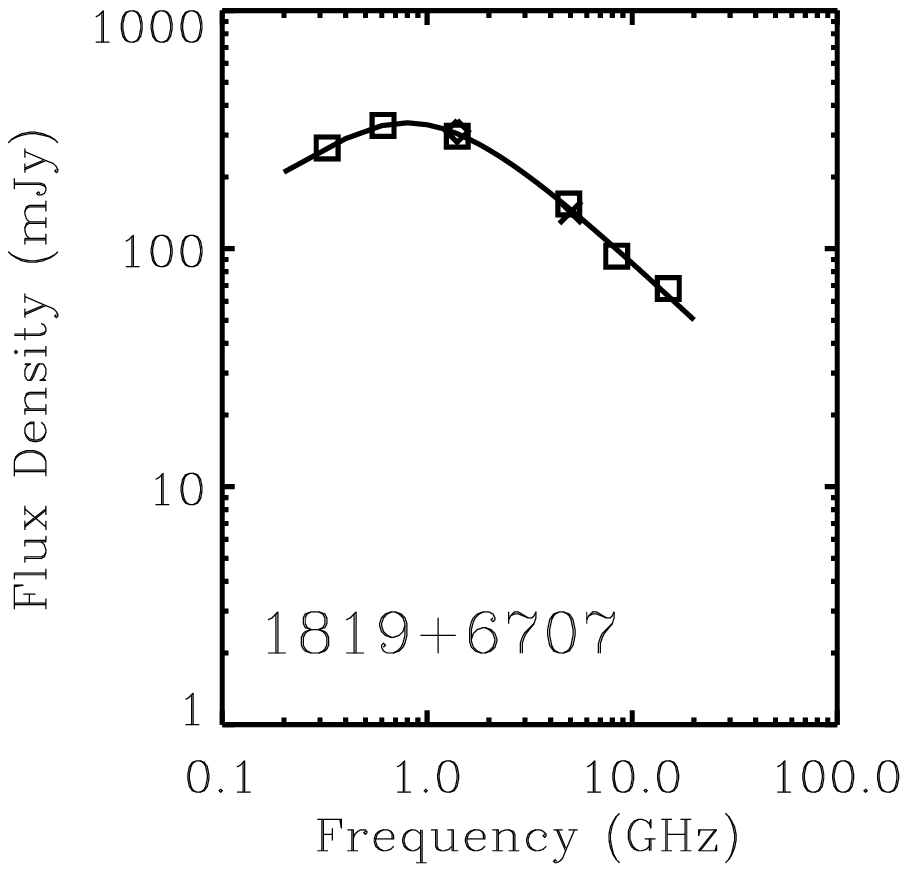,width=5.1cm}
}
\caption{{\it Continued...}}
\end{figure*}
\addtocounter{figure}{-1}

\begin{figure*}
\vspace{-0.5cm}
\hbox{\hspace{-0.8cm}
\psfig{figure=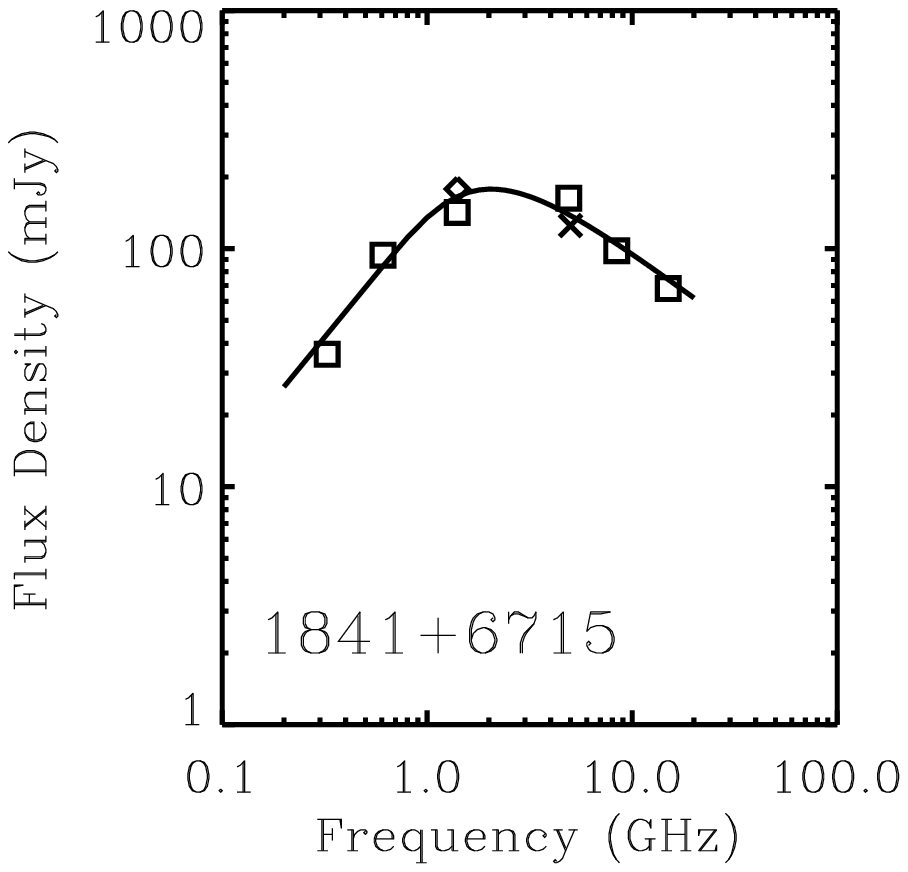,width=5.1cm}\hspace{-0.8cm}
\psfig{figure=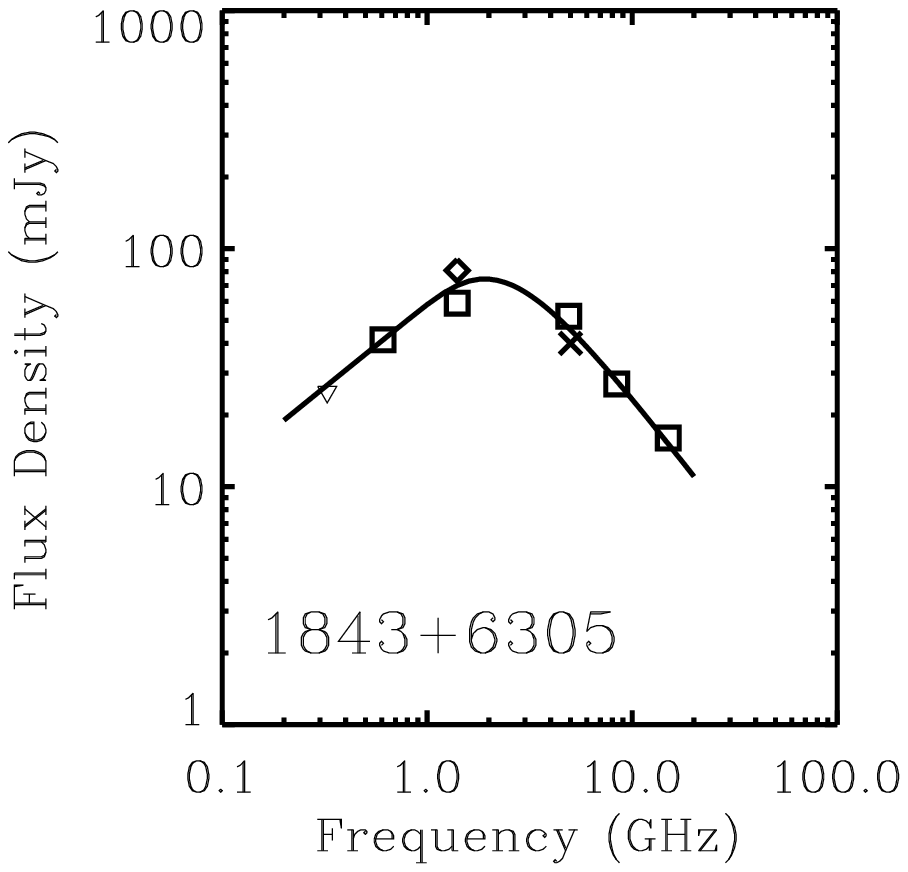,width=5.1cm}\hspace{-0.8cm}
\psfig{figure=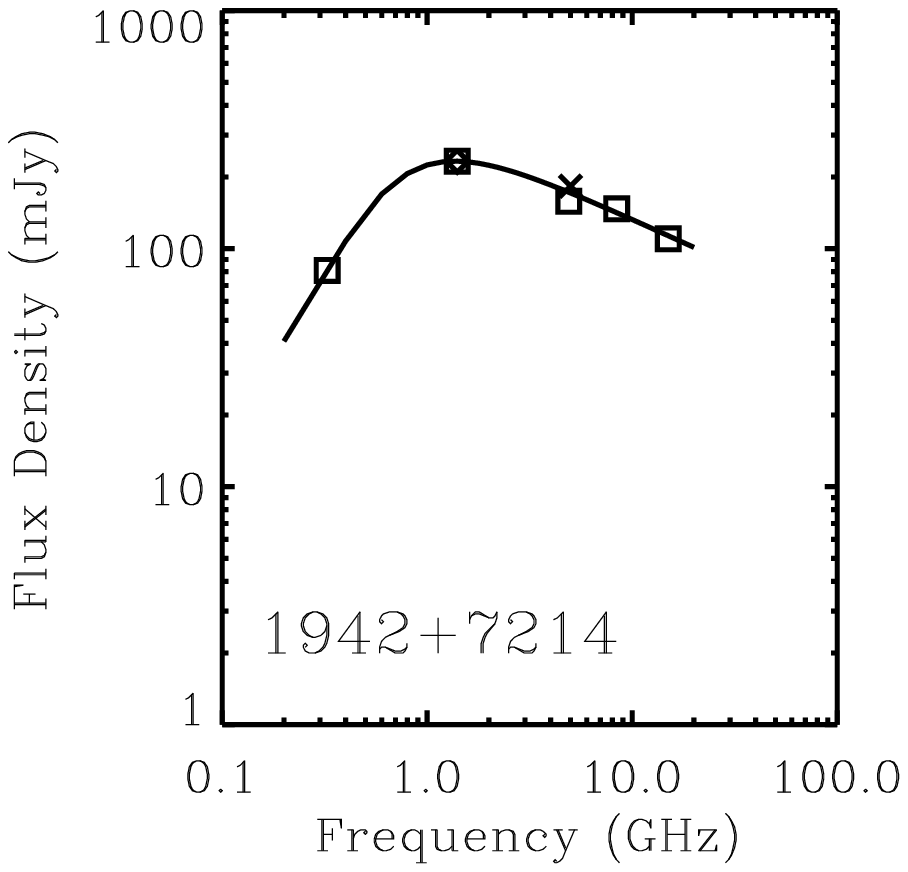,width=5.1cm}\hspace{-0.8cm}
\psfig{figure=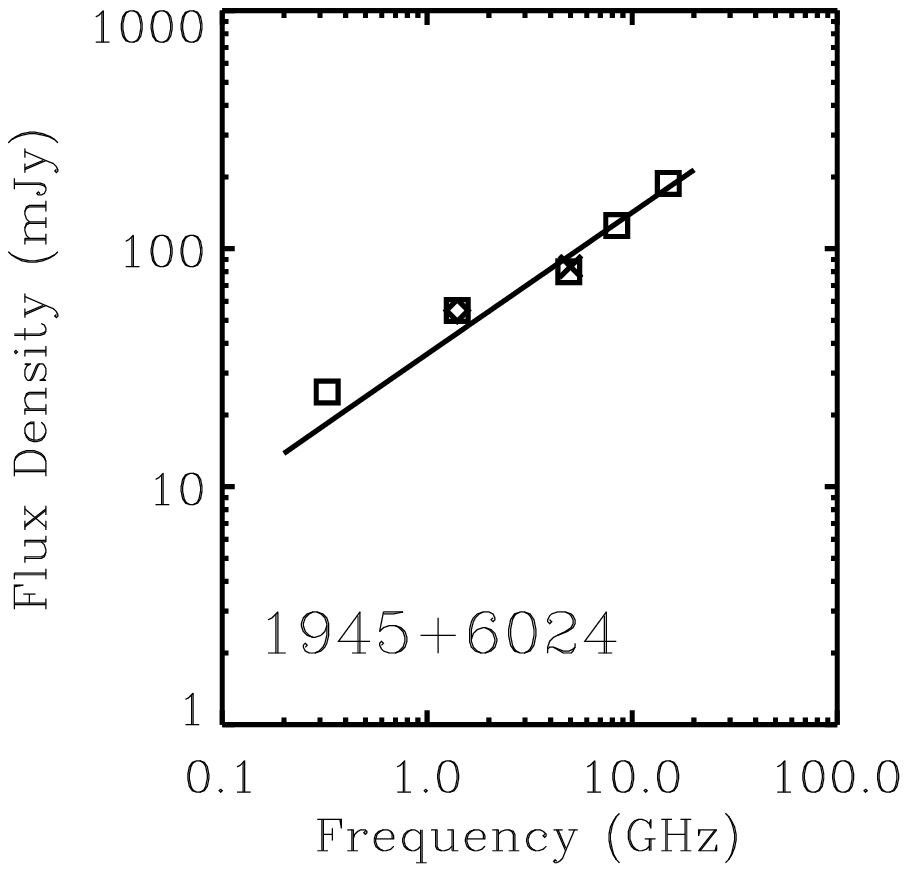,width=5.1cm}
}
\vspace{-0.5cm}
\hbox{\hspace{-0.8cm}
\psfig{figure=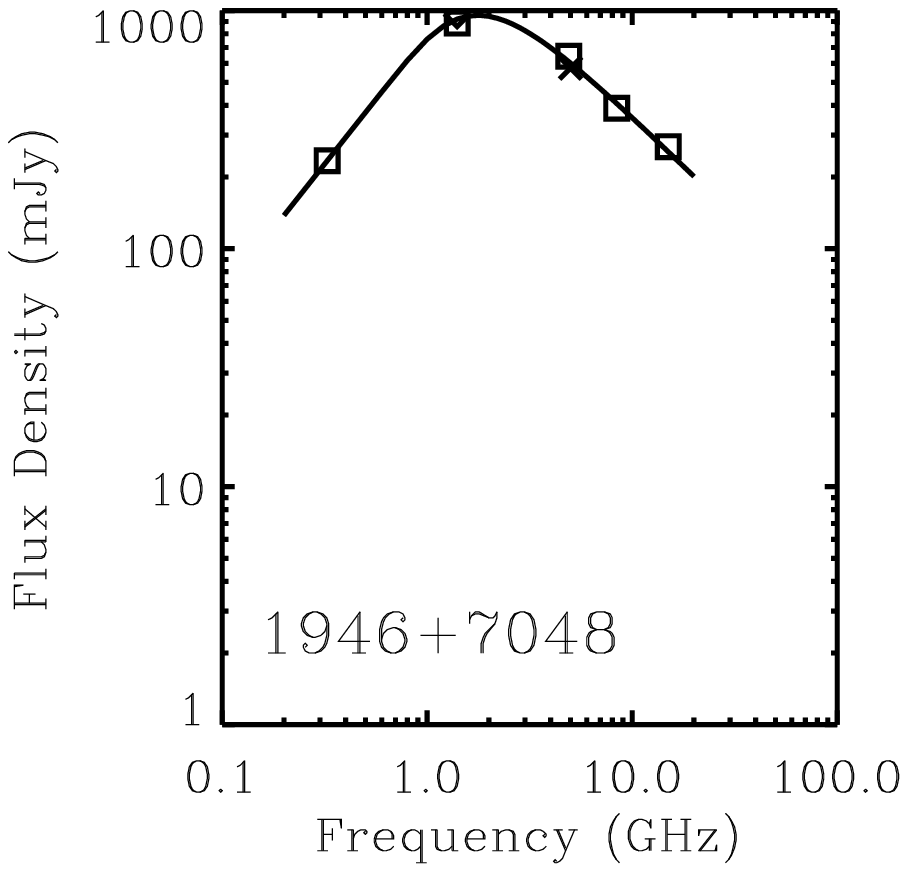,width=5.1cm}\hspace{-0.8cm}
\psfig{figure=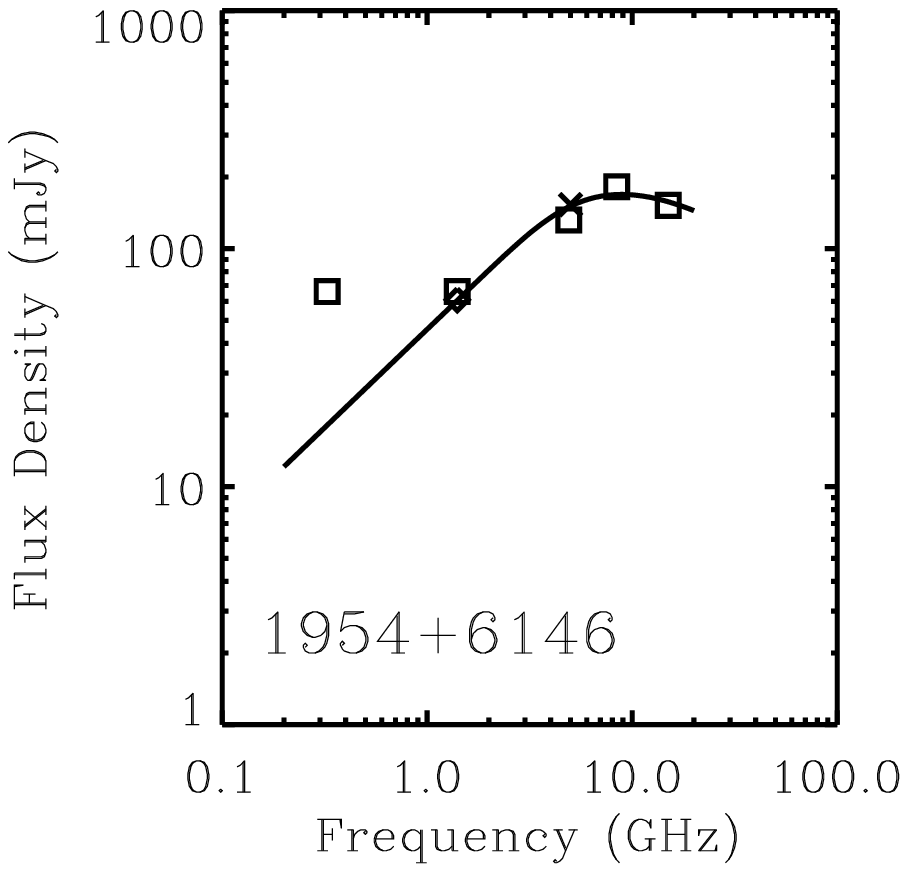,width=5.1cm}\hspace{-0.8cm}
\psfig{figure=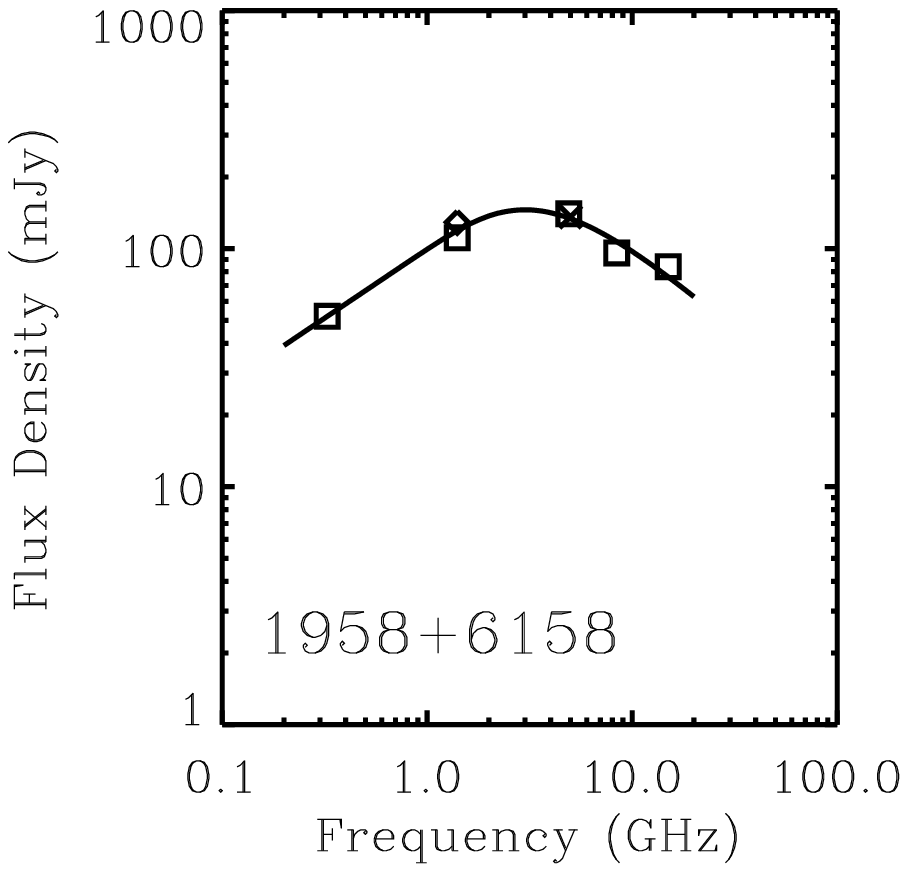,width=5.1cm}
}
\caption{{\it Continued...}}
\end{figure*}

\section{Flux Variability, Confusion and Extended Emission \label{varsec}}

The measurements at different epochs at 1.4 GHz 
(WSRT and NVSS) and at 5 GHz (Greenbank, WSRT and MERLIN) can be used 
to investigate the variability of the sources in our sample.
The measurements are taken with different resolutions, hence if 
extended emission is present this can contribute to a difference in
measured flux densities. In particular at 5 GHz, the Greenbank flux density 
measurements can be influenced by confusion of background sources in the $3.5'$
beam. However, in general this effect is small.

\begin{figure}
\centerline{
\psfig{figure=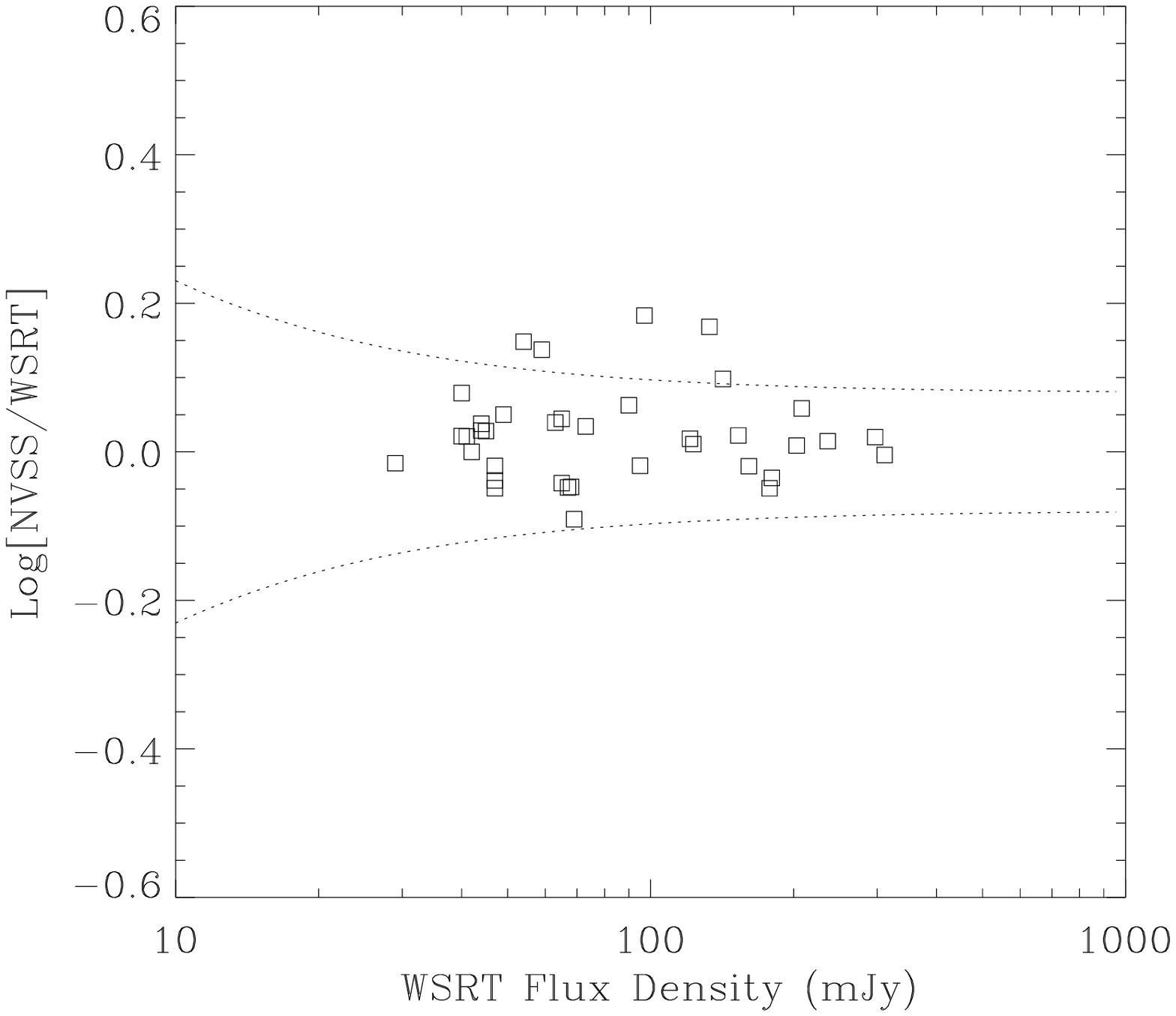,width=9cm}}
\caption{\label{ratio1.4}The logarithm of the ratio between the 1.4 GHz
NVSS and WSRT flux densities as function of NVSS flux density for the GPS
sources. The dotted lines indicate a difference in flux density of $20\%\pm5$
mJy.}
\end{figure}

\begin{figure}
\begin{tabular}{|c|}\hline
\\
{\large WENSS 325 MHz $\ \ \ $ NVSS 1.4 GHz}\\
\hbox{
\hspace{-0.5cm}
\psfig{figure=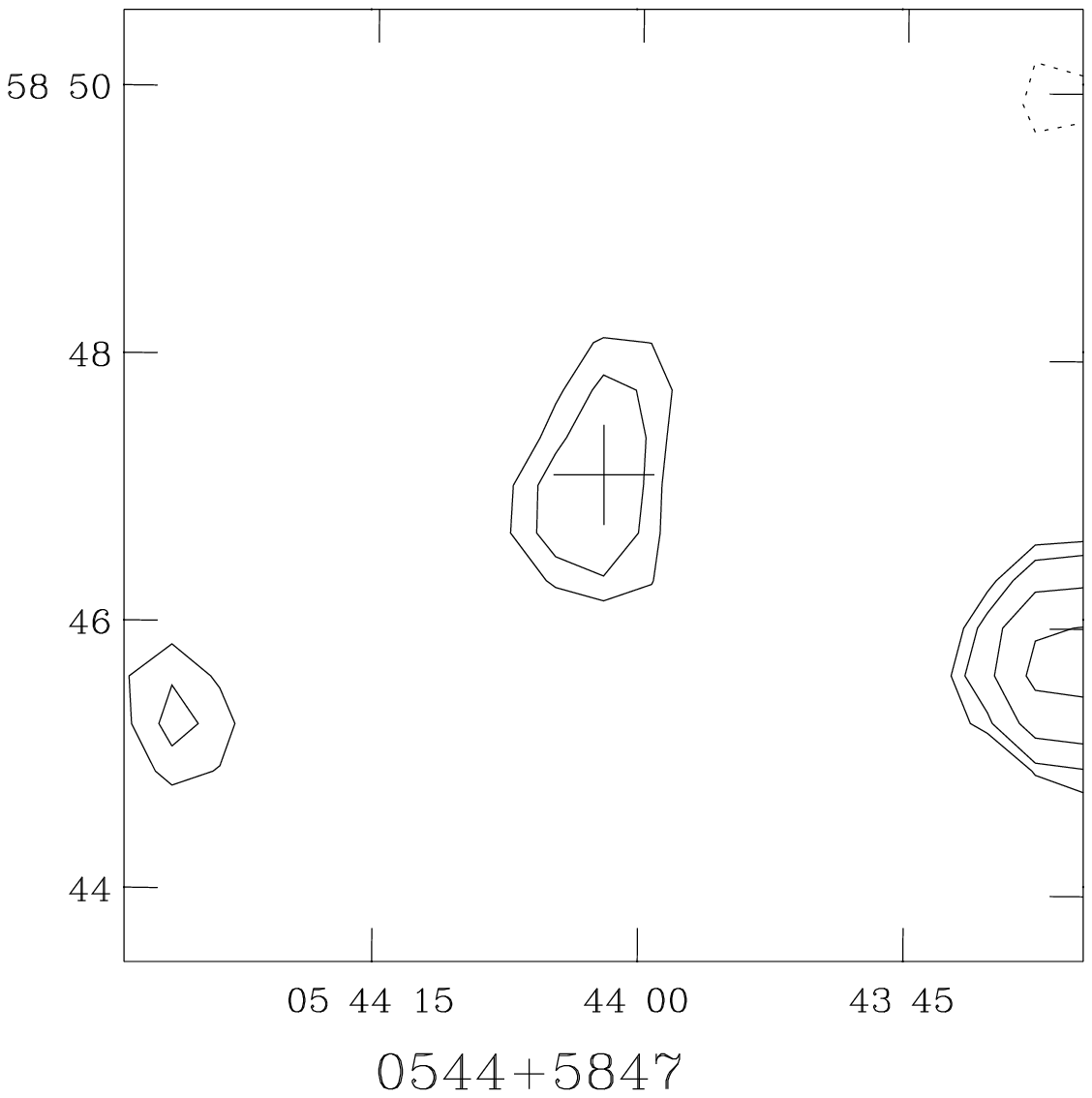,width=3.7cm} 
\psfig{figure=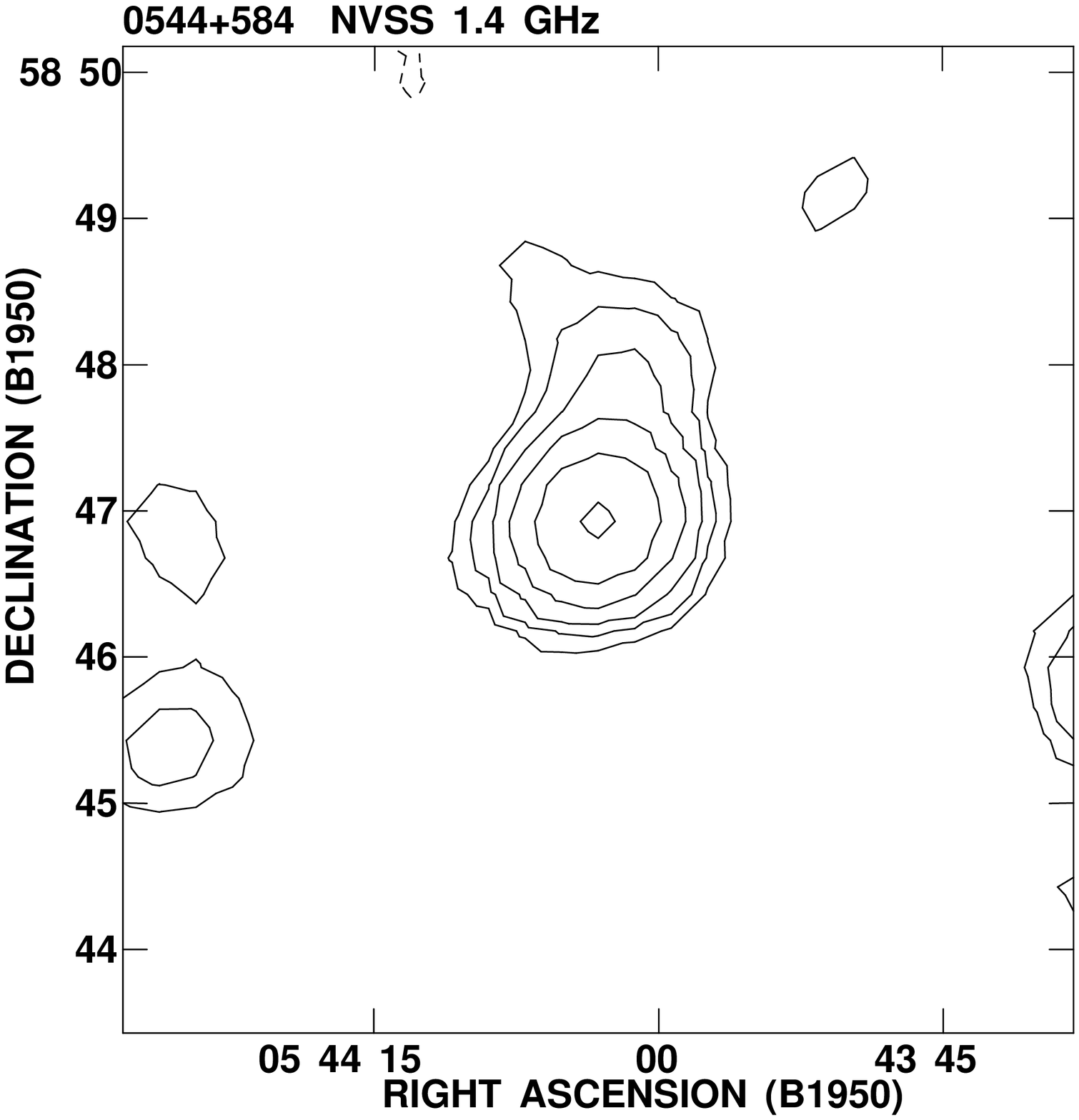,width=3.5cm}} \\ \hline

\hbox{
\hspace{-0.5cm}
\psfig{figure=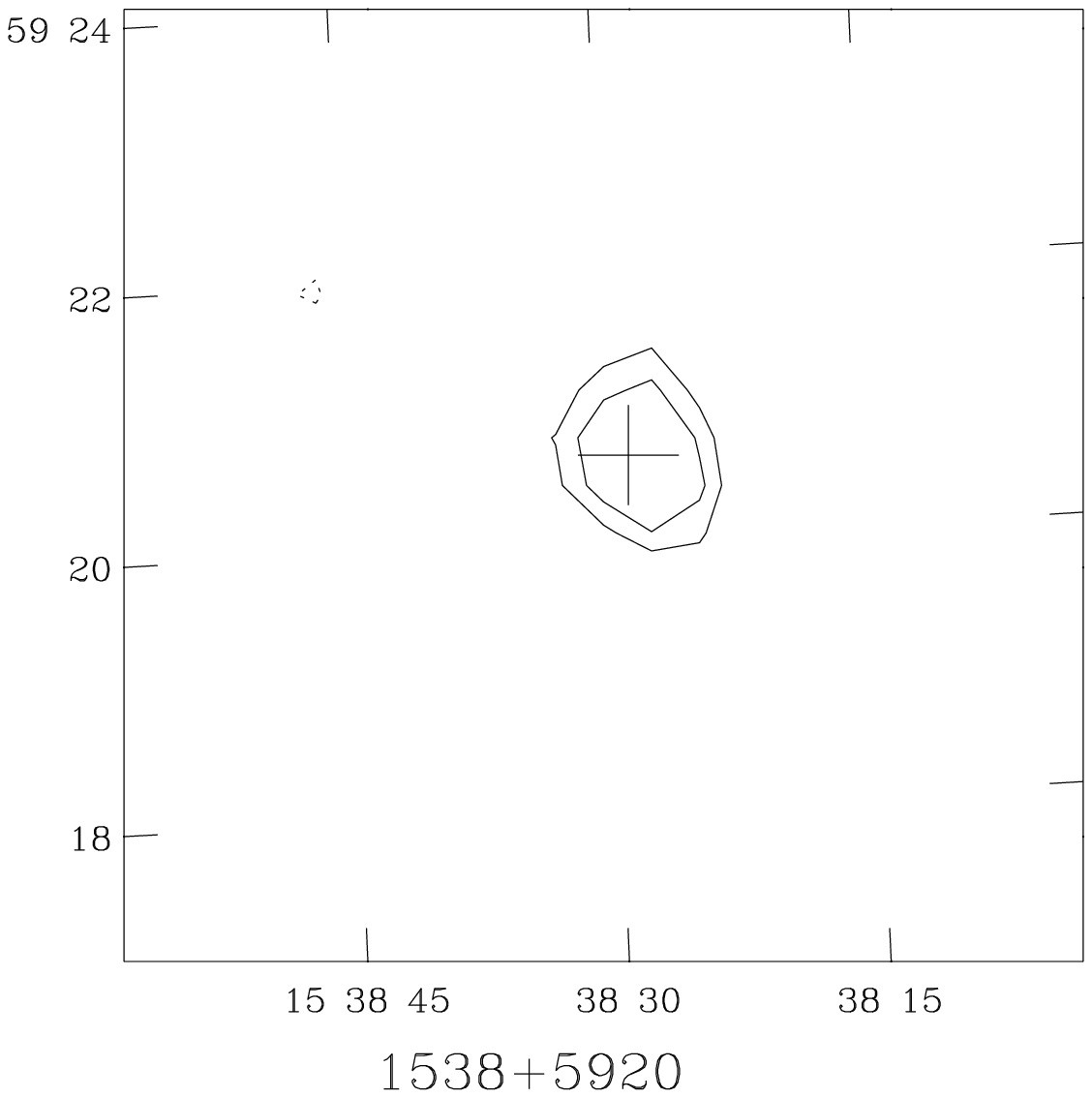,width=3.7cm}
\psfig{figure=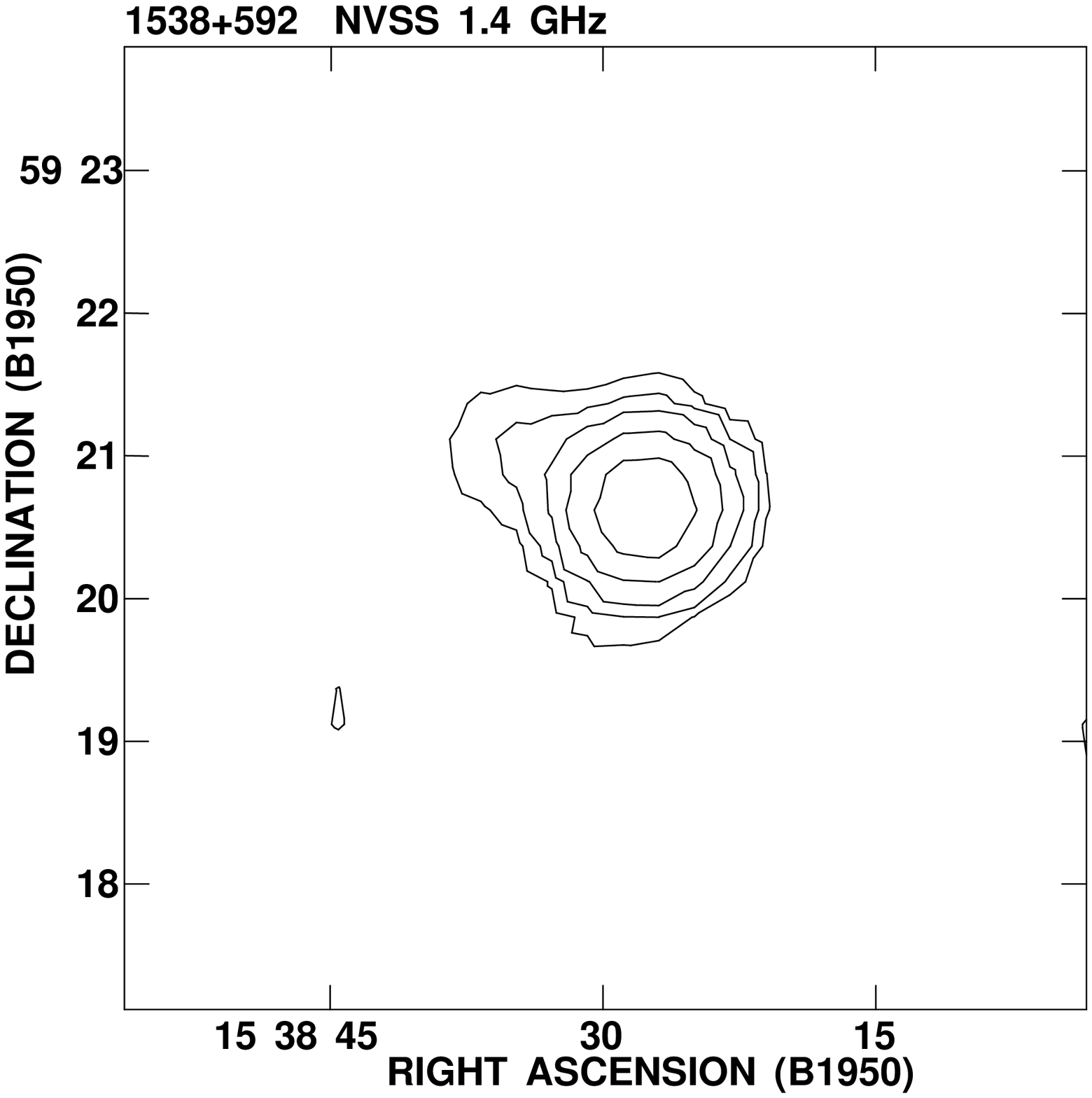,width=3.5cm}}\\ \hline

\hbox{
\hspace{-0.5cm}
\psfig{figure=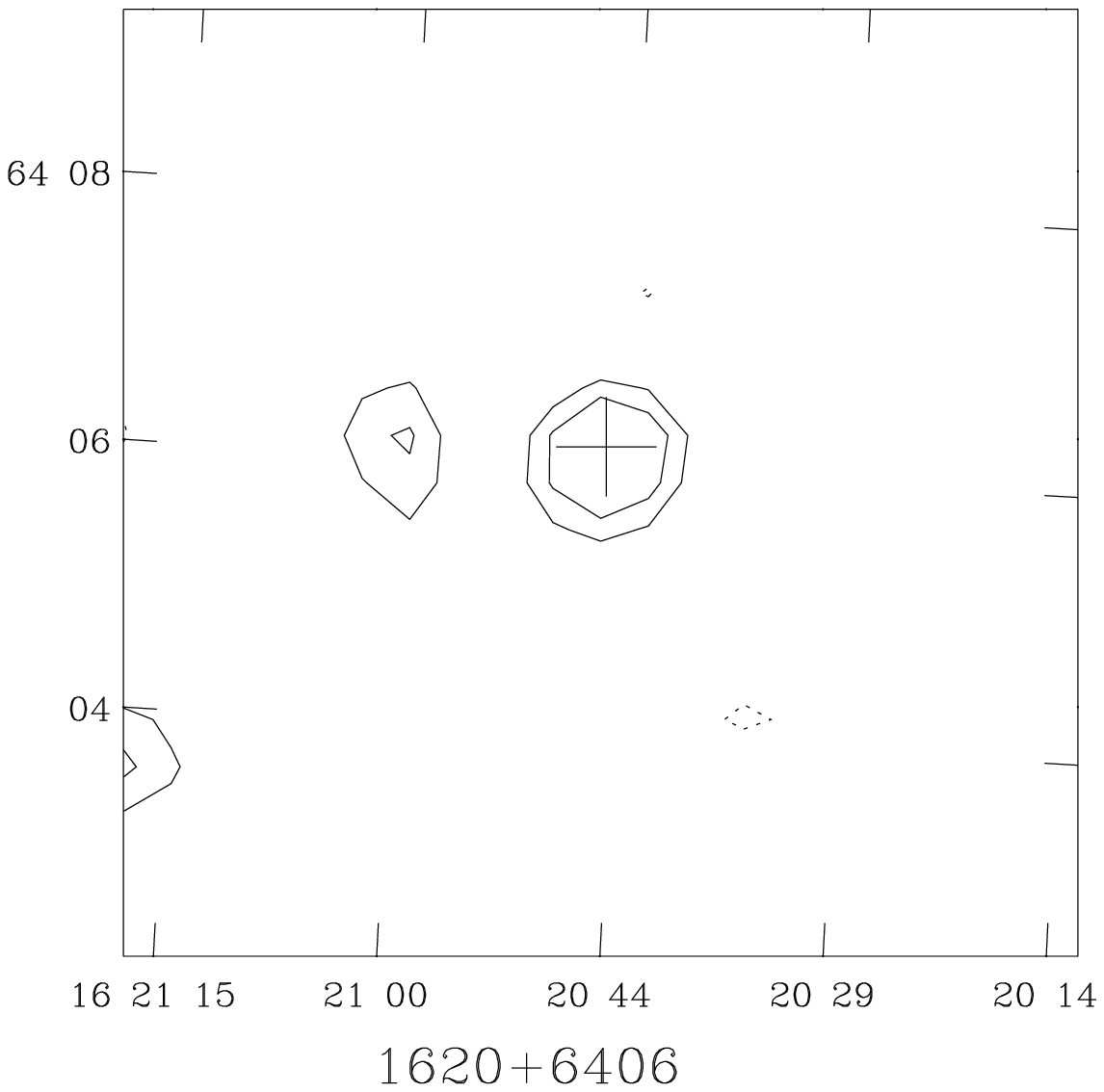,width=3.7cm}
\psfig{figure=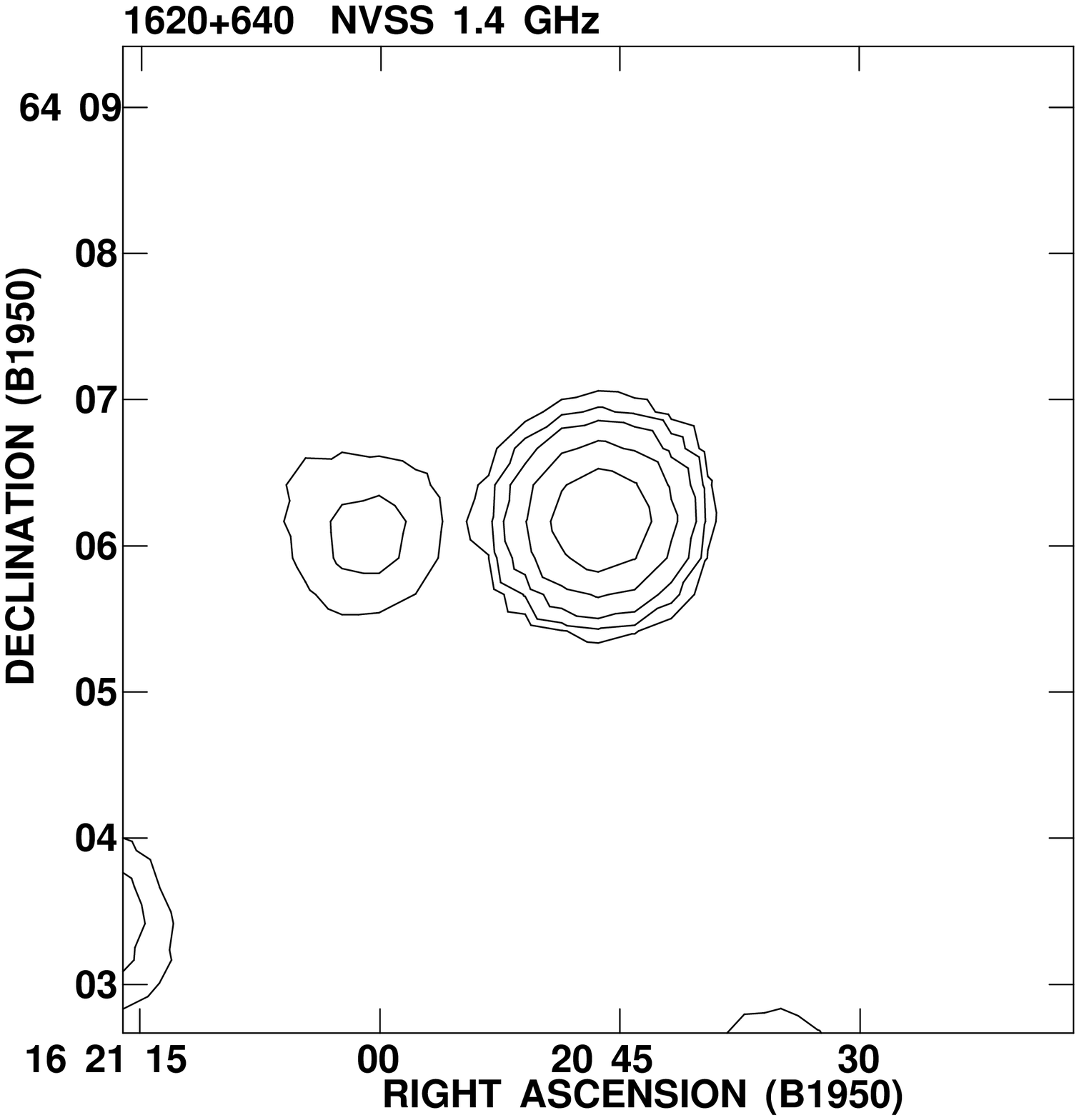,width=3.5cm} } \\ \hline

\hbox{
\hspace{-0.5cm}
\psfig{figure=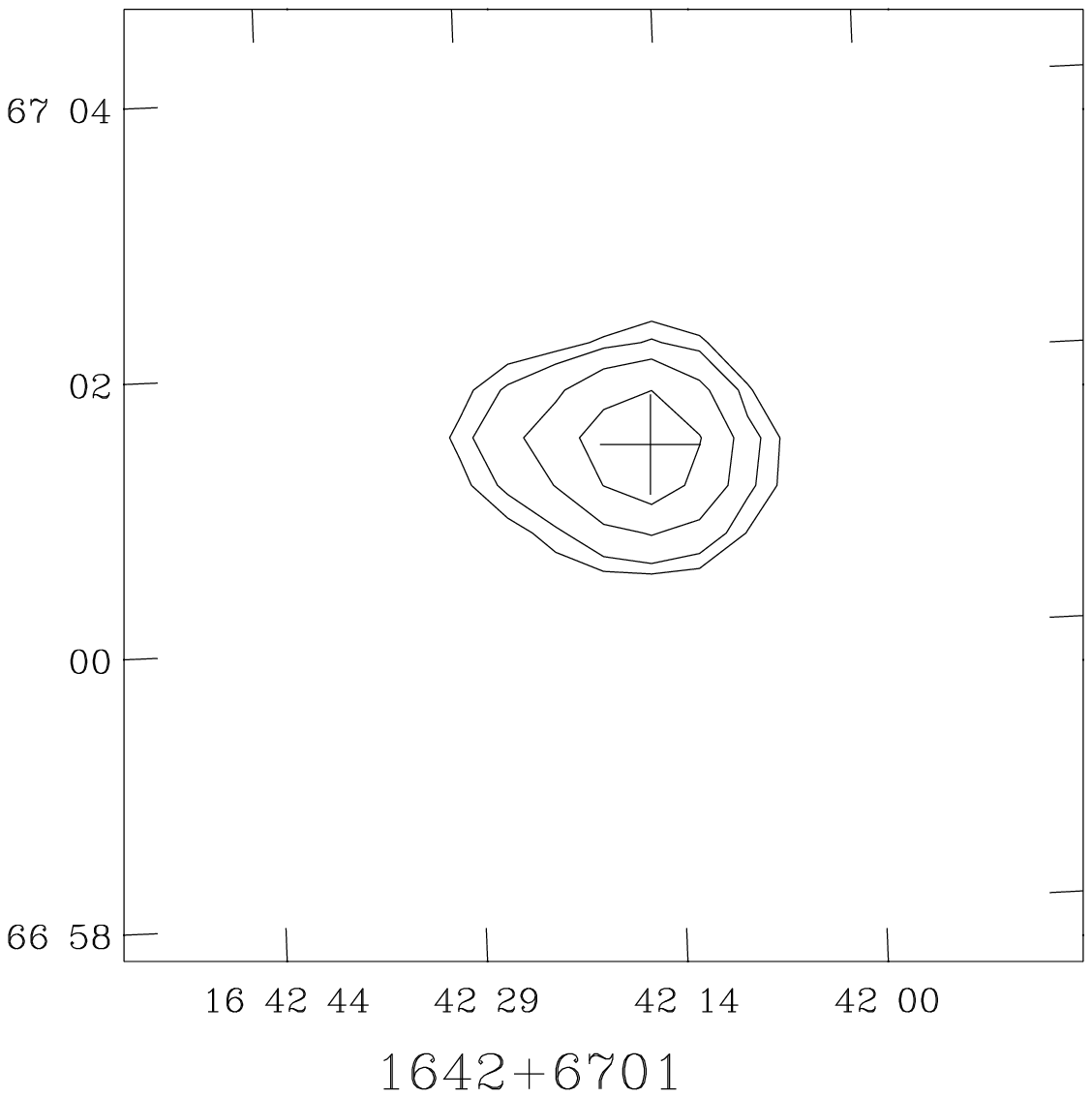,width=3.7cm} 
\psfig{figure=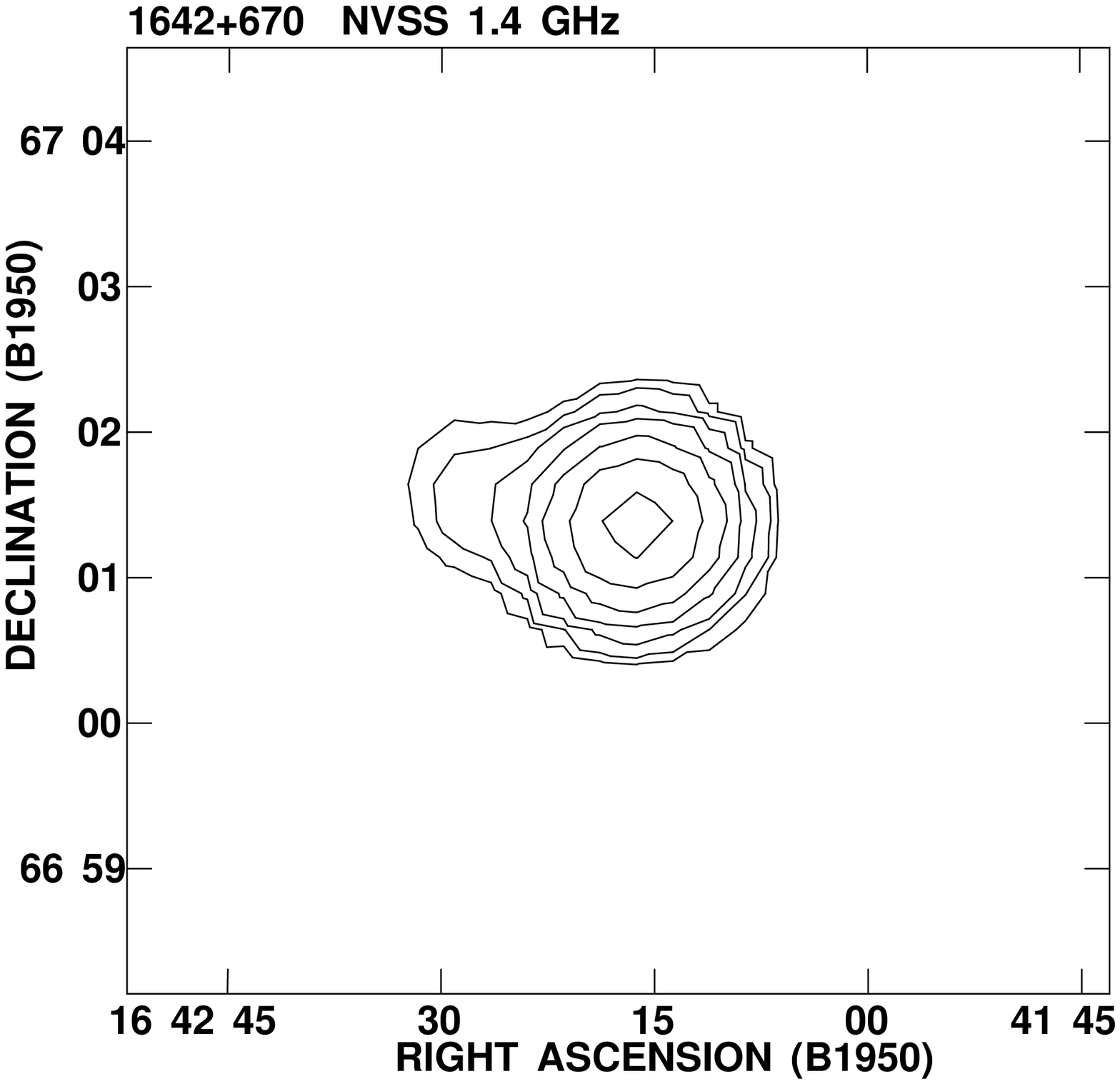,width=3.5cm} }\\ \hline

\hbox{
\hspace{-0.5cm}
\psfig{figure=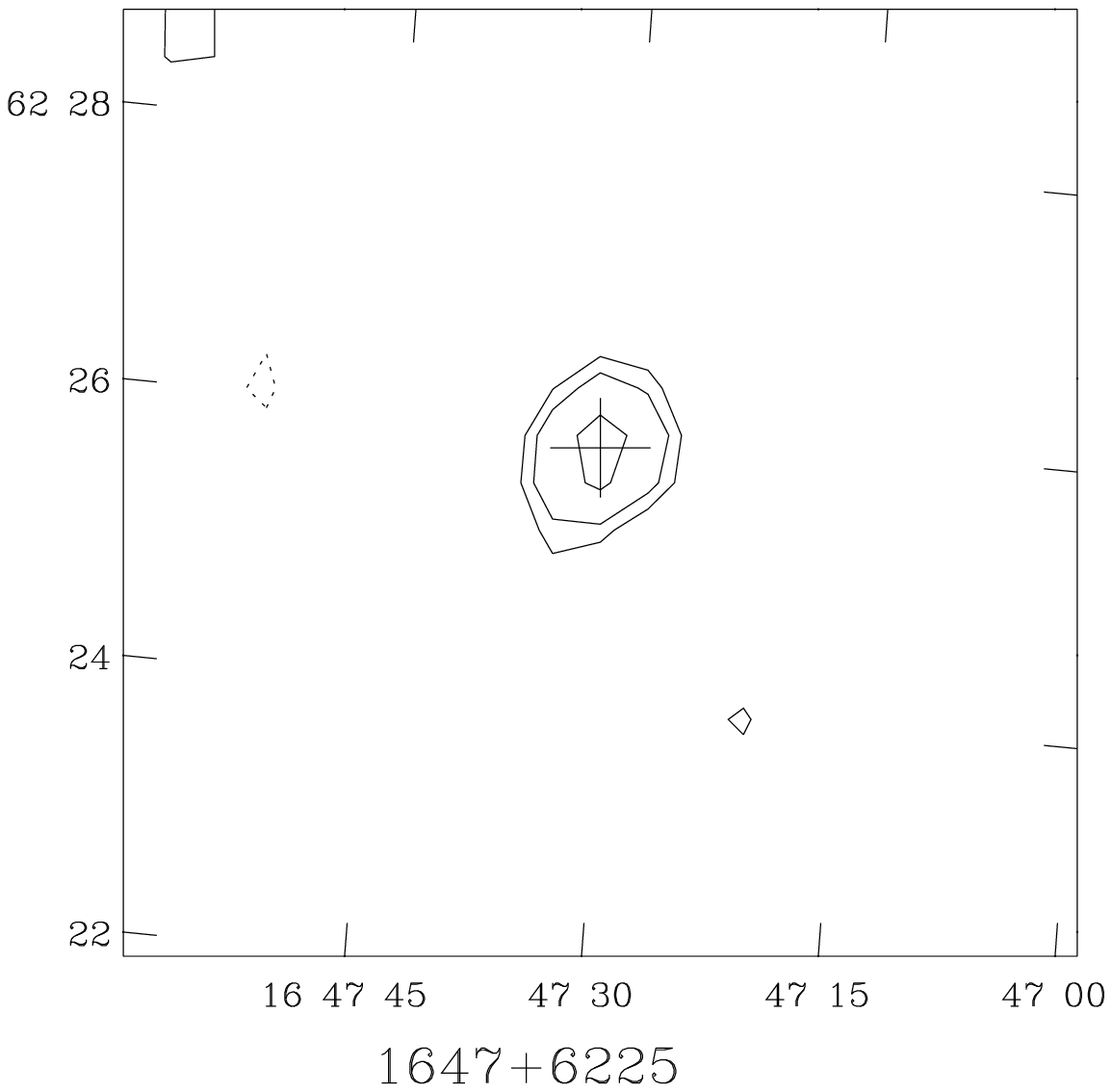,width=3.7cm} 
\psfig{figure=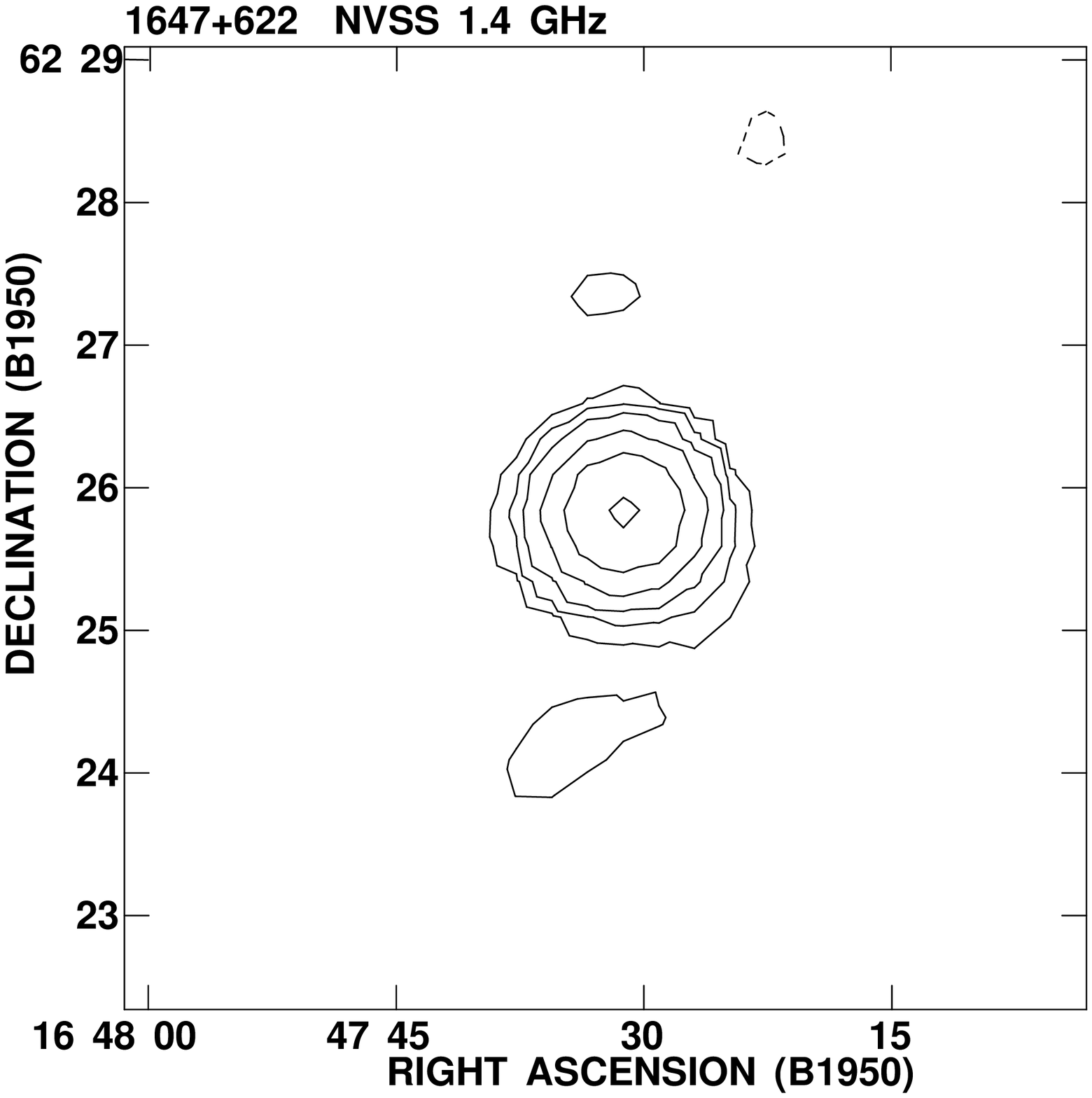,width=3.5cm} } \\ \hline
\end{tabular}
\caption{\label{extend}The WENSS 325 MHz (left) and the NVSS 1.4 GHz (right) 
maps of the five GPS sources for which nearby radio components 
are found in the NVSS.}
\end{figure}

Figure \ref{ratio1.4} shows the logarithm of the ratio between the 1.4 GHz
NVSS and WSRT flux densities as function of WSRT flux density for the GPS
sources. Note that for only 5 GPS sources (14 \%) this difference is larger 
than $20\% \pm 5$ mJy, which can be caused by either variability or extended 
emission on scales between $8''$ and $35''$, or both. 
For the sources which were not included in the final GPS sample, the 
``non-GPS sources'', the
percentage of objects outside $20\% \pm 5$ mJy is 26\%. 
None of the objects with a nearby radio source in the NVSS (see figure 4) has
a difference between their  1.4 GHz WSRT and NVSS measurements of $>20\%$.
Therefore it seems unlikely that extended emission is causing the 
discrepancy between the two 1.4 GHz flux density measurements.

\begin{figure*}[!t]
\centerline{\hbox{
\psfig{figure=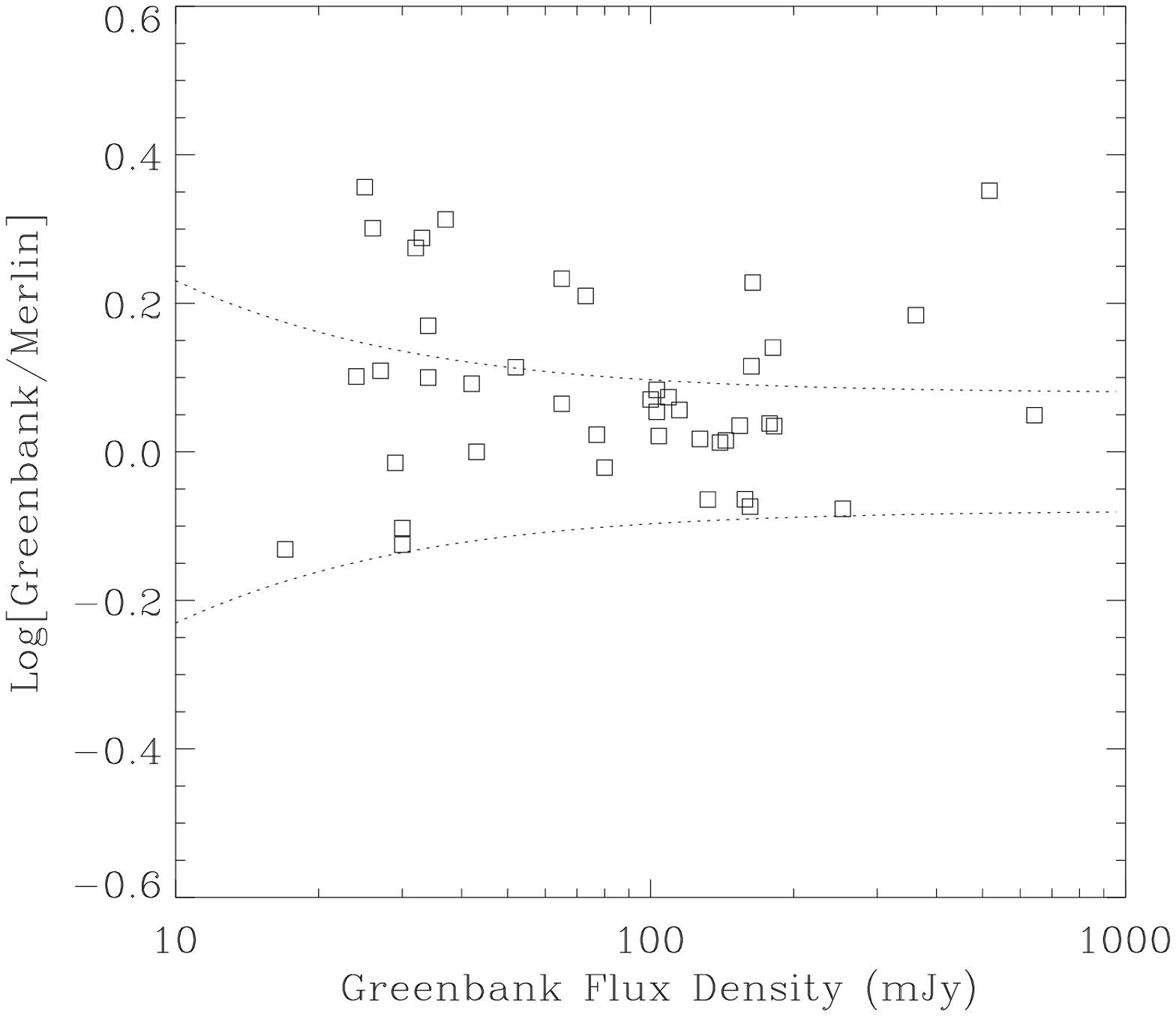,width=8cm}
\psfig{figure=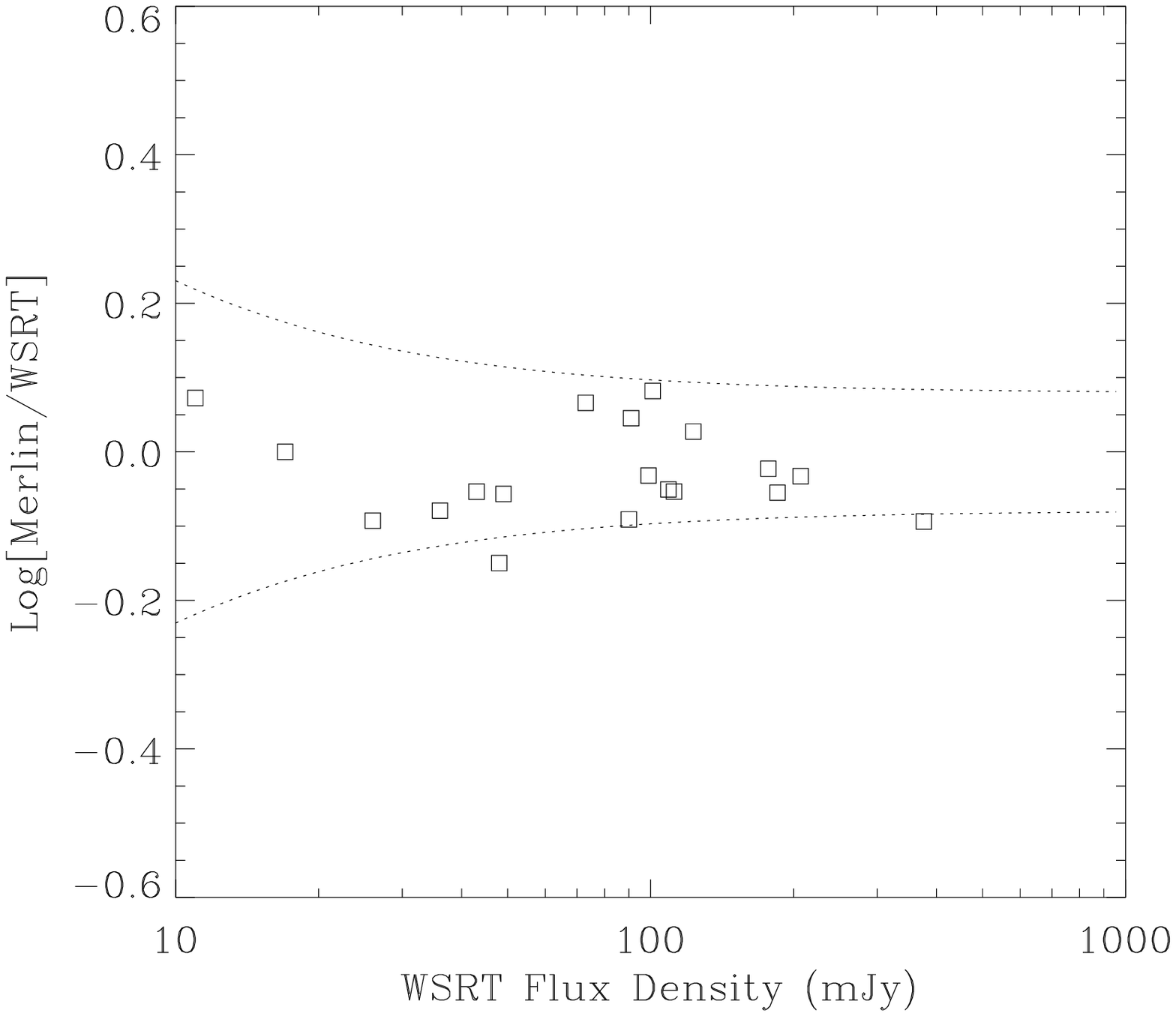,width=8cm}}}
\caption{\label{ratio5} The logarithm of the ratio of the 
Greenbank/MERLIN 5 GHz (left) and the WSRT/MERLIN 5 GHz (right) 
measurements as function of flux density.}
\end{figure*}

Figure  \ref{ratio5} shows the logarithm of the ratio of the Greenbank/MERLIN
5 GHz and the WSRT/MERLIN 5 GHz flux density measurements. For 13 out of 43
GPS sources (30 \%), the Greenbank-MERLIN flux density ratio is greater than
20\% + 5 mJy, while none of these ratios are smaller than 20\% - 5 mJy. All
except two of the WSRT-MERLIN 5 GHz flux density ratios are within $20 \% \pm
5$ mJy of each other. In contrast, the WSRT to Greenbank 5 GHz flux densities
of 71\% of the non-GPS sources differ more than 20\%. Hence the candidate GPS
sources which turned out not to be GPS sources are clearly more variable than
the GPS sources.

The Greenbank flux densities for the GPS sources in our sample are always
equal or higher than the MERLIN and WSRT flux densities. This suggests that
extended emission or confusion by background sources in the Greenbank
measurements plays an important role. Indeed, four of the five sources with
nearby components (see figure \ref{extend}) differ in their 5 GHz Greenbank 
and MERLIN
measurements by more than 20\%.
However, the flux densities of the nearby components are relatively low, and
therefore variability is more likely to be the dominant factor. The absence of
 ratios $<0.8$ is probably caused by a selection effect. The Greenbank
measurements are much older (1987) than all the other measurements. If a GPS
source has monotonically increased in flux density between 1987 and the first
half of this decade than the Greenbank flux density will be significantly
lower than the 5 GHz WSRT and the MERLIN flux density measurements. However
the chance that this source will appear in our sample is then very small,
since the spectral index between WENSS and Greenbank is probably not inverted.
Furthermore the Greenbank flux density would be too low with respect to the
newer WSRT 1.4 GHz and the VLA 8.4 GHz measurements to be included in our
sample as a GPS source. If a GPS source is monotonically decreasing with flux
density between 1987 and the first half of this decade then there is only a
problem if the peak frequency is greater than about 10 GHz, which could make
both the Greenbank 5 GHz and the VLA 15 GHz measurement greater than the 8.4
GHz flux density measurement. Thus only sources with stable or decreasing flux
densities are likely to be selected in our sample. Examples of GPS sources
with decreasing flux densities are B0537+6444, B0552+6017, B0756+6647, 
B0830+5813 and B1538+5920.

Unlike the NVSS observations, the snapshot WSRT observations are not suitable
for mapping possible low level extended emission around the GPS sources,
due to the poor UV-coverage of the WSRT observations. The
NVSS images have been used to look for extended emission around our GPS
sources on scales $<100''$. Extended emission was detected in 5 of our
objects, namely B0544+5846, B1538+5920, B1620+6406, B1642+6701 and B1647+6225.
Their corresponding maps are shown in figure \ref{extend}. It is not clear if
these additional sources are components associated with the GPS objects or
superimposed unrelated sources. 
From the source density in the NVSS survey, we estimate that there is 
a $\sim 6\%$ chance of finding a
radio source with a flux density of $>5$ mJy within a radius of $100''$ .
Hence in our sample of 47 sources, 3 GPS sources might be expected to have
unrelated sources within $100''$, while 5 are found. This is not a 
significant difference.

\section{Source Counts}

Since the 1960s it has been recognized that counts of radio sources can
provide important cosmological information, especially about the evolution of
radio sources. Identification and source counts statistics indicated that 
radio sources were preferentially located in early epochs of the universe 
(Longair 1966).
Recently, several authors (Readhead et al. 1996, Fanti et al.
1995, O'Dea et al. 1997) have shown that if GPS and CSS sources evolve into
large scale FRI/FRII radio sources, they must significantly 
decrease in radio luminosity (a factor 10 to 30)
 to account for the high number of GPS and CSS sources compared to
the number of FRI and FRII radio galaxies. In this section the GPS source
counts at faint flux densities derived from our faint sample are constructed
and discussed.

\begin{figure}[!t]
\psfig{figure=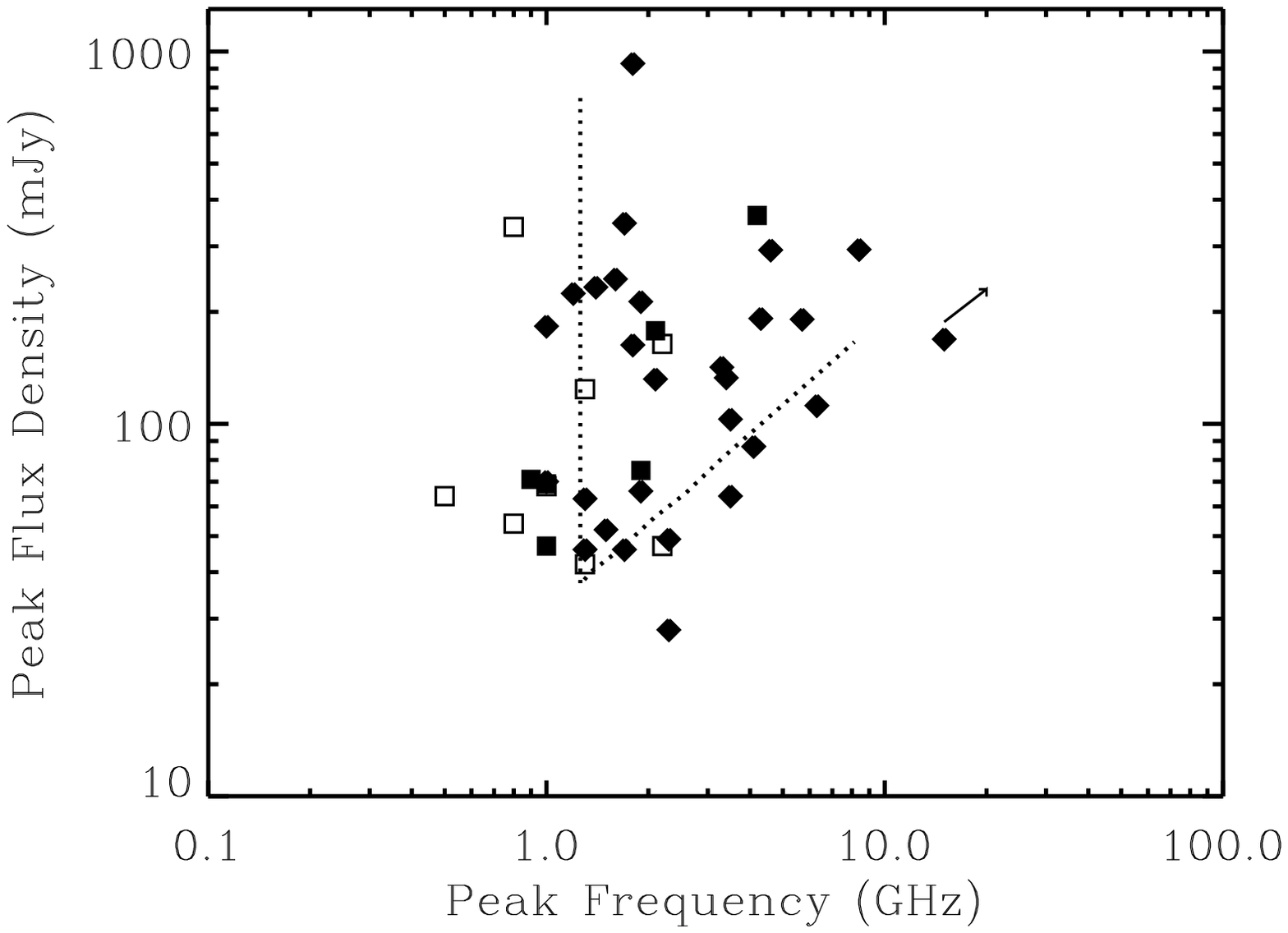,width=9cm}
\caption{\label{distr} The distribution of peak frequencies and peak flux
densities for the GPS sources in our sample. The diamonds represent the sources
selected on their 325-5000 MHz spectral index; the squares represent
the sources selected on their 325-609 MHz spectral index; the
filled squares represent the sources selected from the 
325-609 MHz comparison which also would have been in the sample
if selected on their 325-5000 MHz spectral index. The dotted lines represent
the limits for which a GPS source with optically thin and optical thick
spectral indices of $-$0.75 and +0.80 respectively would be selected on the 
basis of its 325-5000 MHz spectral index. The arrow indicates the 
lower limit for the peak frequency and peak flux density of B1945+6024.}
\end{figure}

To be able to derive the GPS source counts from the sample, the intrinsic
distribution of GPS sources with peak frequency and peak flux density has to
be determined. The observed distribution of peak flux densities and peak
frequencies of the GPS sources in our sample is shown in figure \ref{distr}.
The diamonds represent the sources initially selected on the basis of 
their 325-5000 MHz
spectral index, the squares represent the sources selected on their 325-609
MHz spectral index, while the filled squares represent the sources
selected from the 325-609 MHz comparison which also would have been in the
sample if selected on their 325-5000 MHz spectral index. It should be noted
that figure \ref{distr} is the observed distribution, 
and that  the selection criteria are a function of peak frequency, peak
flux density, and optically thin and thick spectral indices of the sources,
making statistical studies complicated.

Whether a GPS source with a certain peak flux density and peak frequency will
appear in our sample depends on its optically thin and thick spectral
indices. If the optically thick spectral index is too inverted, the 325 MHz
flux density is too low to be in the sample. If the optically thin spectral
index is too steep, then the 325-5000 MHz spectral index will not be inverted.
The range of optically thick and thin spectral indices allowed for a GPS
source to appear in the sample is a strong function of peak frequency and peak
flux density. To be able to derive the parent distribution of peak flux
densities and peak frequencies of GPS sources from the observed distribution
in our sample, the fraction of GPS sources selected as a function of peak flux
density and peak frequency has to be estimated.

\begin{figure}[!t]
\psfig{figure=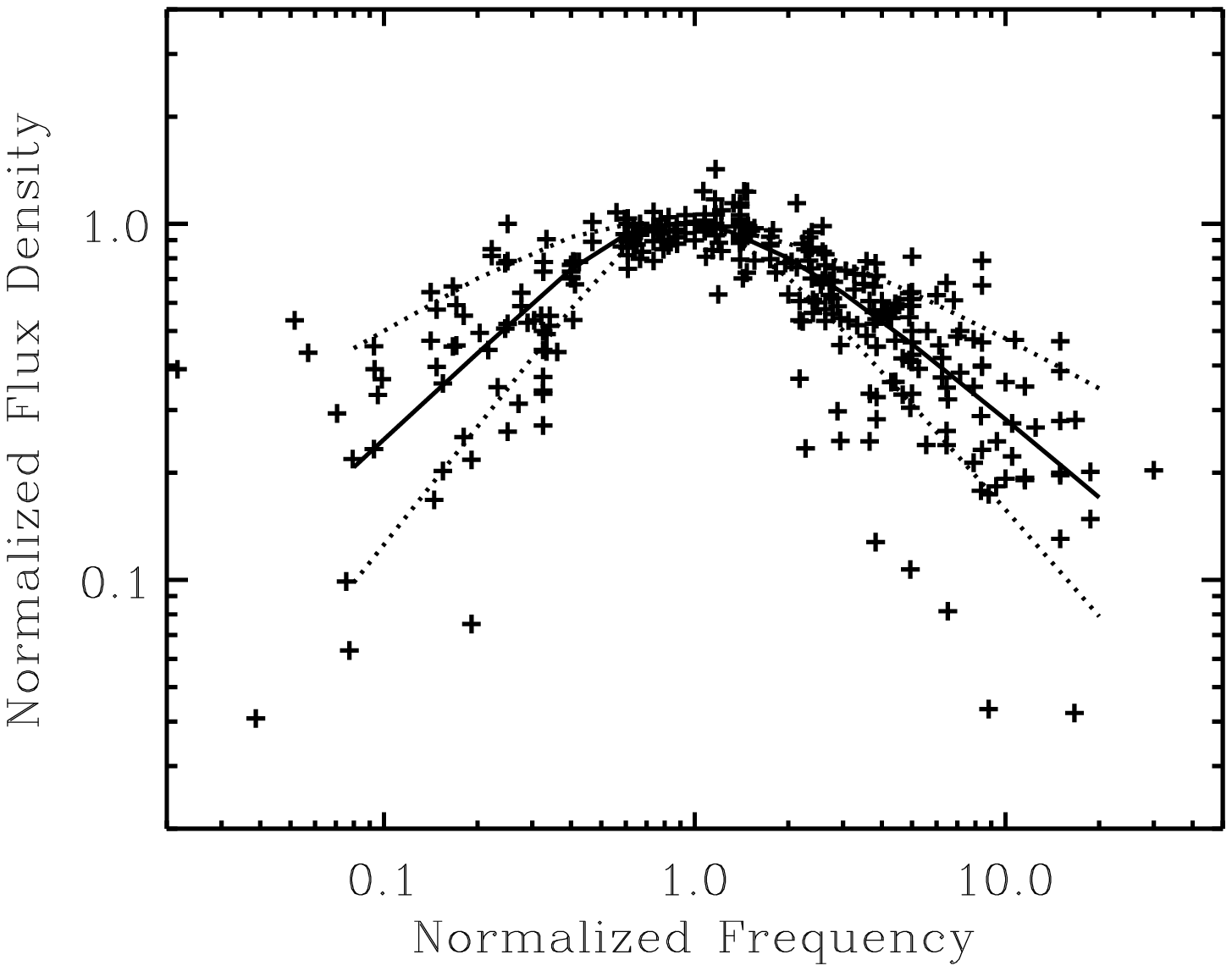,width=9cm}
\caption{\label{shape} The spectra of the GPS sources normalized (scaled) 
in both
frequency and flux density. The solid line represents the best fit of equation
\ref{eq1} to the data, which gives optically thick and thin spectral indices
of respectively +0.80 and -0.75. The dashed lines represent optical thick and
thin spectral indices of $+0.5$, $+1.1$, $-0.5$ and $-1.0$}
\end{figure}

We assume that the peak frequency, peak flux density and the optically thin
and thick spectral indices are independent of each other. The intrinsic
distribution of optically thin and thick spectral indices have to be
determined from their observed distributions. The normalized spectra 
(in both frequency and flux density) of the
sources in our sample are plotted in figure \ref{shape}. The solid line
represents the best fit of equation \ref{eq1} to the data, which gives
optically thin and thick spectral indices of $-0.75$ and $+0.80$ respectively.
If we assume the intrinsic or parent spectral index distributions 
to be Gaussian
functions with means and standard deviations of $-0.75$ and 0.15 (thin) and
+0.80 and 0.18 (thick), then the observed spectral index distributions are 
recovered after applying
the selection effects to the parent spectral index distributions. 
Although the outcome of this is not very accurate, this is not too important
since our main concern is too show that no
significant fraction of GPS sources was missed; sources with highly inverted
spectral indices of 2, say, are unlikely to be included unless the peak flux density is
high and the peak frequency is low enough. Sufficient numbers of these 
sources could have an effect on the
parameters of the optically thick spectral index distribution. To obtain an
indication of the number of sources with very inverted spectra
($\alpha_{thick} >1.1$), we looked for highly inverted GPS sources in our sample
with peak
frequencies lower than 2 GHz and peak flux densities greater than 100 mJy,
because these will be the least influenced by selection effects on the optically
thick spectral index. Only one object out of ten (10\%) was found to have an
optically thick spectral index greater than 1.1, while 5\% would be expected
from the distribution of optically thick spectral indices. Hence we regard any
missed population of GPS sources with very steep optically thick spectral
indices as small and negligible.

The GPS sources in our sample which have a flux density at 325 MHz 
 $>25$ mJy and an inverted spectral index between 325 and 5000
MHz have been used to determine the peak frequency and peak flux density 
number distributions. The sample is divided into bins of peak flux density of 
50-100 mJy, 100-200 mJy and 200-400 mJy and bins of peak frequency of
1-2 GHz, 2-4 GHz and 4-8 GHz. If it is assumed that the parent population
of GPS sources has spectral index distributions as described above, then
for each bin the percentage of GPS sources selected in our sample can
be determined. For each bin, the number of sources in the parent population 
of GPS sources was estimated by dividing the number of GPS sources in the 
sample which fall in the bin by these percentages which were 
between 40\% and 100\%.

\begin{figure}[!t]
\psfig{figure=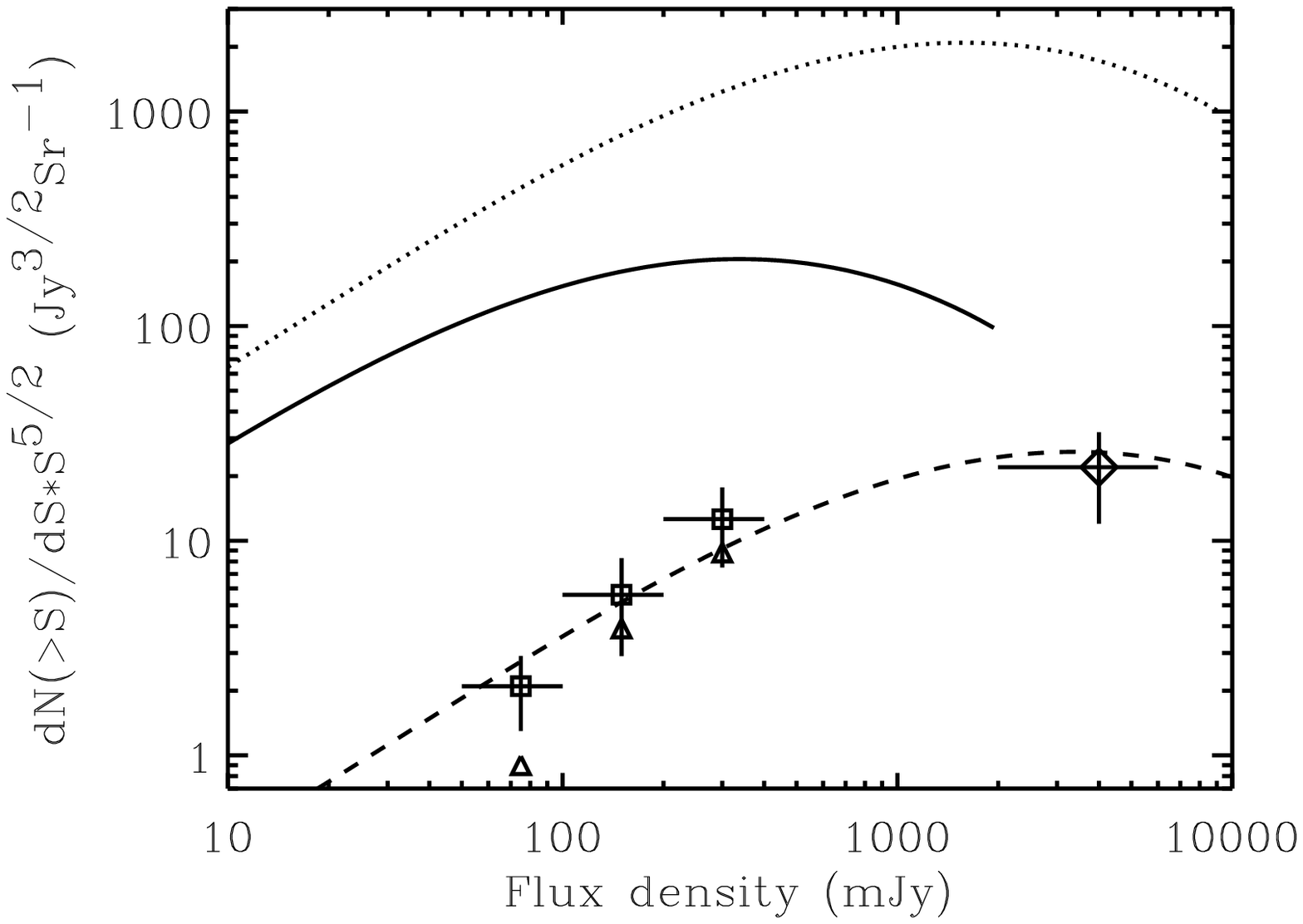,width=9cm}
\caption{\label{counts} The number counts of GPS sources as function of peak
flux density. The squares are from the data in this paper. The triangles are
the observed, not corrected, source counts. The point indicated by the diamond
is derived from de Vries et al. (1997). The dotted line represents the WENSS
radio source counts at 325 MHz in the mini-survey region, and the solid line
are these source counts corrected to 2 GHz using a spectral index of $-0.85$
(see text). The dashed line is the curve for 1/250 of the 2 GHz
counts shifted a factor 10 upward in flux density. This is consistent with
the data.}
\end{figure}

The observed number of GPS sources, corrected for selection effects as above,
have been summed in peak frequency space to determine the surface density of
GPS sources as function of flux density. The  source counts  for the three
flux density bins, normalized by the flux density to the power 2.5 are plotted
as squares in figure \ref{counts} (a horizontal relation is expected for a
uniformly distributed Euclidean space). The triangles represent the observed
surface density counts, not corrected for selection effects, which is a good
lower limit. We have used the well defined sample of De Vries et al. (1997)
compiled from the working sample of GPS sources from O'Dea et al. (1991) to
determine the surface density of GPS sources at high flux density. We
determined the number of GPS sources in the de Vries et al. sample which lie
within the Pearson and Readhead survey region (Pearson and Readhead 1988,
$\delta > 35 ^{\circ}$, $|b > 10^{\circ}|$, 2.0 sr), having peak frequencies
between 1 and 8 GHz and flux densities between 2 and 6 Jy. There are five
objects satisfying these criteria, which leads to a normalized surface density
count of $22\pm10$ $Jy^{-3/2} Sr^{-1}$ in this flux density range, assuming
that the spectra of all the radio sources in this region brighter than 2 Jy
are well known and that therefore this sample of GPS sources is complete. This
measurement is indicated by the diamond symbol in figure \ref{counts}.

Several authors (eg. Fanti et al. 1995, Readhead et al. 1996, 
O'Dea et al 1997) have proposed that GPS sources may evolve into large size
FRI/FRII radio sources. Therefore it is interesting to compare 
the GPS source counts with source counts of FRI/FRII radio sources. 
It is not very useful to compare them directly to the total radio source counts
at comparably high frequency, because these are dominated by compact flat 
spectrum sources, which are probably not related to GPS sources in an 
evolutionairy way. 
However, it can be assumed that the WENSS source counts at 325
MHz are dominated by the large size radio source population, because
they have in general steep ($\alpha <-0.5$) spectra. The radio source
counts at 325 MHz from the WENSS mini-survey region (Rengelink et al. 1997) is
shown in figure \ref{counts} by the dotted line. Note the resemblance between
the shape of this curve and the data for the GPS sources (squares + diamond).
The median spectral index is about $-0.85$, which we used to estimate
the source counts for large scale radio sources at higher frequencies
comparable to the peak frequencies of the GPS sources. Note that the source
counts of the GPS sources are not at a certain fixed frequency, but resemble
the distribution of peak flux densities. However, we assume that within the
errors the distribution of the peak flux densities of the GPS sources in our
sample is the same as for the flux densities at the median peak frequency,
which is 2 GHz. We determined the large size radio source number counts at 2
GHz from the 325 MHz counts of the WENSS mini-survey region, assuming a fixed
spectral index of $-0.85$. This is represented by the solid curve in figure
\ref{counts}.

How can the differences in number counts between the GPS sources and large
scale radio sources be interpreted? If GPS sources evolve into large scale
radio sources, it is reasonable to assume that they undergo the same
cosmological evolution, because the typical lifetime of a radio source is
significantly smaller than cosmological time-scales. Assuming that the slope of 
the luminosity function of GPS sources
is identical to the slope of the luminosity function of large size radio 
sources and that {\it all} GPS sources evolve into large size radio sources, 
one could obtain the ratio of the life time of the two classes of radio source
and the luminosity evolution of the GPS sources.
If a radio source is 10 times
brighter in its GPS phase than in its FRI/FRII phase, and  if the time scale
of the GPS phase is 250 times shorter than the age of large scale radio
sources, then the dashed line is expected for the source counts of GPS
sources. Namely in that case the curve for FRI/FRII radio sources moves a 
factor 250 down due to the age difference, and a factor of 10 to the right
and a factor factor $10^{3/2}$ upward due 
to the luminosity evolution.  
This agrees quite well with the observed GPS source counts.

However, this straightforward interpretation is probably too simplistic.
Firstly, the redshift distributions of GPS galaxies and
large size radio sources are not the same, which indicates that the slopes of
their luminosity functions are different (see Snellen 1997). 
This has 
consequences for the interpretation of the radio source counts of GPS sources.
It should be investigated if a simple radio source evolution
model, as for example presented in Snellen (1997), 
combined with the cosmological 
evolution of the  radio luminosity function for large size radio sources,
is consistent with the GPS source counts presented here. This is beyond
the scope of this paper. Secondly, the GPS source counts presented here 
also include the GPS
quasars, although it is not clear that they are related to the GPS galaxies.
Excluding the GPS quasars, which contribute about one third of the members 
of both the bright and
faint GPS samples, shifts the squares and the diamond in figure \ref{counts}
down by a factor 1.5. In
this case, the number counts are consistent with radio galaxies in the GPS
phase being 10 times brighter for a period $\sim 400$ times shorter than the
FRI/FRII phase.

\section{Conclusions}
A sample of GPS sources has been
selected from the Westerbork Northern Sky Survey, with flux densities one to 
two orders of magnitude
lower than bright GPS sources investigated in earlier studies. 
Sources with inverted
spectra at frequencies $>325$ MHz have been observed with the WSRT at 1.4 and
5 GHz and with the VLA at 8.6 and 15 GHz to select genuine GPS sources. This
has resulted in a sample of 47 GPS sources with peak frequencies ranging from
$\sim$500 MHz to $>$15 GHz, and peak flux densities ranging from $\sim$40 to
$\sim$900 mJy.

Five GPS sources in our sample show extended emission or nearby components
in the NVSS maps at 1.4 GHz. However it is not clear if these components 
are related to the GPS sources.

About 30\% of the objects show flux density differences greater than 20\% 
between the Greenbank and MERLIN 5 GHz measurements, with the Greenbank
data points all higher than the MERLIN observations. We believe this is due to
variability, and that the lack of sources with reverse variability (the 
MERLIN flux density greater than the Greenbank flux density) is 
due to a selection effect caused by the ``old'' epoch (1987) of the Greenbank
observations.

GPS source counts are comparable to 1/250 of the 2 GHz 
source counts for large scale radio sources, if the latter sources were to
have 10 times their measured flux densities. 
Unfortunately, apparent differences in redshift distributions between 
GPS and large scale radio sources hamper a direct and straightforward
interpretation of the source counts.
Potentially, the comparison of GPS source counts with that of large
scale radio sources can provide clues about the age of GPS sources and their 
luminosity evolution. If it is assumed that the redshift distributions
are the same for GPS and large size radio sources, the source counts indicate 
that GPS sources have to decrease in luminosity by a factor of $\sim 10$ 
if they all evolve into large scale radio sources.

\section*{Acknowledgements}
This research was partly supported by the European Commission, TMR Programme,
Research Network Contract ERBFMRXCT96-0034 ``CERES''

\end{document}